\documentclass{article}
\usepackage[a4paper, total={6in, 8in}]{geometry}

\usepackage{enumitem}
\usepackage{graphicx} 

\usepackage{amsfonts}
\usepackage{amssymb}
\usepackage{amsmath}
\usepackage{amsthm}
\usepackage{mathrsfs}
\usepackage{breqn}
\allowdisplaybreaks
\usepackage{bm}

\usepackage{natbib}
\usepackage{authblk}

\newtheorem{theorem}{Theorem}
\newtheorem{lemma}{Lemma}

\newtheorem*{theorem*}{Theorem}

\newtheorem{assumption}{Assumption}
\newtheorem*{assumption*}{Assumption}
\newtheorem{definition}{Definition}
\newtheorem*{definition*}{Definition}

\usepackage{thmtools} 
\usepackage{thm-restate}

\date{}

\usepackage[ruled,linesnumbered]{algorithm2e}
\SetKwComment{Comment}{/* }{ */}
\SetKwInput{KwInput}{Input}

\usepackage{subcaption}
\usepackage{float}

\newcommand{\E}[1]{\mathbb{E}\left\{#1\right\}} 
\newcommand{\EE}[2]{\mathbb{E}_{#1}\left\{#2\right\}} 

\newcommand{\prob}[1]{\mathbb{P}\left(#1\right)} 
\newcommand{\pprob}[2]{\mathbb{P}_{#1}\left(#2\right)} 
\newcommand{\cprob}[2]{\mathbb{P}\left(\left.#1\right|#2\right)} 
\newcommand{\cpprob}[3]{\mathbb{P}_{#1}\left(\left.#2\right|#3\right)} 

\newcommand{\eprob}[1]{\hat{p}\left(#1\right)} 

\newcommand{\ind}[1]{\mathbf{1}\left(#1\right)} 

\newcommand{\argmin}{\operatorname*{arg\,min}}

\newcommand{\loss}[2]{\mathcal{L}\left(#1,#2\right)} 

\newcommand{\iidsim}{\ensuremath{\stackrel{\text{\tiny i.i.d.}}{\sim}}}

\title{Calibration of answer probabilities in verbal autopsies by internal consistency}

\author[1]{James Liley}
\author[1]{Nathan Higgins}
\author[2]{Eilidh Cowan}
\affil[1]{Department of Mathematical Sciences, Durham University, Durham, UK}
\affil[3]{Health Determinants Research Collaboration, Gateshead Council, Gateshead, UK; Population Health Sciences Institute, Newcastle University, Newcastle, UK; Institute of Medical Sciences, Aberdeen University, Aberdeen, UK}

\begin{document}

\maketitle

\begin{abstract}

We consider the problem of calibrating a posterior estimator of a categorical latent variable given a fixed-length ternary string using only unlabelled observations and without a specified likelihood. 
We primarily consider a setting in which the estimator is parametrised by estimated conditional probabilities of elements of the string given the latent variable, with calibration if the estimates are correct.  

Our motivating application is the `Verbal Autopsy' procedure, whereby a cause of death is probabilistically inferred following a structured interview with associates of the deceased. 
More generally, our setting applies to circumstances where experts can more readily describe posterior beliefs than likelihoods, due to similarity with diagnostic practices. 

We argue combinatorially that in general the problem is intractable without a simplifying assumption on the data distribution, though some posterior estimators can be ruled out as incompatible.
We propose an assumption of block-conditional independence on substrings, allowing calibration procedures based on substring frequency, imputation, and pairwise and three-way distributions of string elements. 
We give theoretical results on identifiability, on consistency for distributions of either fixed support or fixed entropy, and on robustness to assumptions, finding essentially that three conditional independence blocks of size at least the number of latent categories are necessary and sufficient for calibration.
We empirically evaluate methods on data simulated to resemble realistic verbal autopsy questionnaires, and find substantial promise for the approach in the practical problem of calibrating posterior estimates for causes of death.

\end{abstract}

\section{Introduction}

Suppose that each member of a population is associated with an unknown latent variable $D \in \{d_1,d_2,\dots,d_r\}$, and a corresponding string $A \in \{1,0,-1\}^s$. We observe the string $A$ and wish to estimate $D$. We are equipped with knowledge of the marginal distribution of $D$, and partial knowledge of the posterior probabilities $\cprob{D=d}{A=a}$ (of which there are $3^s \times d$); for instance, up to some number of parameters much smaller than $3^s \times d$. We are most interested in the setting in which the posterior estimator is parametrised by the values $\{q_{kj}\}_{k \in [s],j \in [r]}$, and is correctly calibrated if $q_{kj}=\cprob{A_k=1}{D=d_j}$, where $A_k$ is the $k$th element of $A$. We presume that we do not, however, have access to other information; in particular, we have no labelled samples of $(A,D)$. 

This goal is motivated by a practical problem in the field of Verbal Autopsies (VA), a method to infer a cause of death in a recently-deceased individual using a standardised post-mortem interview with the family of the deceased~\citep{byass10,byass19}. 
The interview consists of a series of binary questions, not all of which will necessarily be answered. 
Given the answers to an interview, a probability is estimated for a set of potential causes of death (CoD), using one of a set of methods we will collectively call `VA algorithms'. 
The method is widely used in global health and has been extended in various ways~\citep{ambruoso22,byass10}. 

In most methods for estimation, the probabilities are interpretable as a (multinomial) posterior distribution over causes of death~\citep{li14,byass03,byass19,miasnikof15}.
Verbal autopsies share a statistical characteristic with general medical diagnosis in that, although most easily described a Bayesian process in which a probability distribution over CoDs is maintained, practitioners tend to be have a better intrinsic knowledge of the posterior (given symptoms, what affected the patient?) than they do the likelihood (given an affliction, what symptoms are expected?), since it is the posterior they generally estimate in day-to-day work. 
Consequently, practitioners aiming to estimate a posterior over diagnoses or causes of death (including in the development of VA algorithms) tend to use methods which circumvent the use a fully specified likelihood function. 

A common parametrisation of VA algorithms is a matrix of estimated probabilities of answering a question a particular way given causes of death (corresponding to $q$ above)~\citep{li14,byass19,miasnikof15}. 
Although a natural interpretation of this matrix is as an incompletely specified likelihood~\citep{byass19,miasnikof15}, it may also be taken as only a suggestive approximation of true probabilities correct up to ordering~\citep{mccormick16}, or an informative prior on the true probabilities~\citep{li19}.
This matrix is generally estimated from expert opinion from medical practitioners, due to the absence of labelled empirical data, and is often called a `probbase': we will use that term here, even when considering the problem abstractly. 

The true probbase matrix tends to differ across populations~\citep{clark18}, making it particularly laborious to estimate. 
In addition, in many settings where VA is used, training data labelled with causes-of-death is unavailable, meaning that there is no way to directly verify the accuracy of a VA algorithm (or its parameters). 
Evaluation of probbase entries is consequently a problem of interest, with various approaches.  
Some training data is available for specific geographic settings~\citep{groenewald23} in which physicians review VA answers and assign likely causes of death, enabling a probbase to be learned directly.
Some modern approaches to verbal autopsy analysis reduce reliance on the probbase, by treating it as a prior estimate of a probability to be estimated~\citep{li19,li24} or by allowing an open interpretation of the questionnaire~\citep{chu25}. 
A potentially appealing approach is to consider the probbase as a parameter, and maximise the likelihood of an observed VA dataset (or use the likelihood to develop a Bayesian model~\citep{li19}), but VA methods do not always lead to tractable likelihoods, particularly in black-box approaches~\citep{chu25}.

VA questionnaires are regularly updated. In order to ensure reliable cause-of-death coding continues across these updates, VA algorithms must be updated to match current questionnaire standards~\citep{byass03,byass19}, which correspondingly require updating of probbases.  
Previous revision has generally relied on physician coded VAs (as above) to validate a new probbase. 
In 2022 the World Health Organisation (WHO) made considerable changes to the global standard verbal autopsy questionnaire~\citep{who22} but the updating of the algorithm has been severely constrained. 
Although an algorithm is now available, it remains unvalidated due to lack of physician coded VAs using the 2022 questionnaire.

We take the distribution of $D$ to be known. 
Typically, an estimate of CoD frequencies in the population is also generally specified as an input to the estimation procedure~\citep{fottrell11}, which may be interpreted as a prior over cause-of-death probabilities~\citep{byass19,miasnikof15}  and which is often called a `cause-specific mortality fraction' (CSMF). 
An important secondary use of VA algorithms is updating CSMF estimates in light of information from VA interviews; we do not consider this application in this paper, and take the CSMF to be known and fixed.

In this work, we propose a series of approaches to `learn' a probbase $\{q_{kj}\}_{k \in [s],j \in [r]}$ and potentially a complete set of posterior probabilities $\cprob{D=d}{A=a}$ using only knowledge of the CSMF and unlabelled questionnaire data. 
Our methods comprise a set of objective functions admitting consistent estimators which are minimised if, and in some cases only if, the values $\{q_{kj}\}_{k \in [s],j \in [r]}$ and/or the posterior estimates of CoD probability are correct. 
While we consider it generally impractical to learn a VA algorithm or probbase from scratch using only observed questionnaires in this way, there is potential to use such methods for `tuning' existing estimates. 
In particular, if our methods indicate a marked improvement by a small change in some probabilities in a working version of a probbase table, then these can be prioritised for expert checking. 
This could potentially reduce the burden of probbase re-estimation, particularly when a probbase needs to be re-tuned to a new population. 
The intent of this paper is to establish the soundness and applicability of our proposed methods. 
A complete development to the point of functionality to end-users is beyond the scope of this paper, but we intend to describe this in a follow-up work.

This paper is organised as follows. 
In section~\ref{sec:heuristics} we give heuristics for our procedures. 
In particular we demonstrate that the overall task of identifying a probbase using unlabelled questionnaires is impossible without a simplifying assumption on the joint distribution of answers and causes of death, though certain posterior specifications combinations can still be ruled out as `inviable' with a given probbase without such assumptions, in that they do not correspond to any possible true distribution of $(A,D)$. 
We then propose an assumption that answers may be partitioned into blocks such that answers to questions in different blocks are conditionally independent given cause of death, and show that (in general) this assumption is sufficient to guarantee that only one possible set of VA probabilities and corresponding probbase can correspond to an observed set of answer frequencies in the infinite data limit. 
We then introduce notation and, in section~\ref{sec:review} review recent work in this area and related methods, with particular focus on Bayesian methods to learn the joint distribution of interview answers and causes of death. 
In section~\ref{sec:methods} we then specify our methods formally, and establish theoretical properties, principally identifability and convergence of plug-in estimators.  
We then investigate robustness to violations of assumptions in section~\ref{sec:robustness}. 
All methods comprise continuous functions of probbases and/or VA values which are minimised when these inputs are correct, and which are estimable from observed data. 
In section~\ref{sec:evaluation} we develop a simulation strategy for verbal autopsy answers, and empirically test properties of our method. 
Our code is publicly available at~\texttt{https://github.com/jamesliley/VA_imputation}. 
Finally, we conclude with a brief discussion including several open questions in section~\ref{sec:discussion}.



\subsection{General heuristics and main assumption}
\label{sec:heuristics}

We describe the general ideas behind our approach and the necessity of our main assumption. A formalisation of the statements in this section is given in Appendix~\ref{apx:proofs}. 
We discuss robustness to mild violation of the assumption in section~\ref{sec:robustness}. 

We are concerned with the joint distribution of questionnaire answers (which we will informally describe by a random variable $A$, which can take $3^s$ values with $s$ possible questions and 3 possible answers) and causes of death (modelled by random variable $D$, which can take $r$ values), so that $(A,D)$ takes finitely many ($3^s \times r$) values. Our VA algorithm is taken as a set of estimates of values $\cprob{D}{A}$, where the estimate takes a candidate probbase table as a parameter. 

We do not have labelled samples from $(A,D)$. However, we presume that we have access to a number of completed questionnaires, giving us (effectively) knowledge of the marginal distribution $\prob{A}$ of $A$, and that we have a good understanding of overall cause of death frequency, giving us knowledge of the marginal distribution $\prob{D}$ of $D$. If we arrange the possible values of $\prob{A=a,D=d}$ in a matrix with values of $a$ occupying columns and values of $d$ occupying rows, then our knowledge comprises the row-sums and column-sums of the table. 

Combined with a set of estimates of $\cprob{D}{A}$ from a VA algorithm, this completely determines a distribution of $(A,D)$ (since we can write $\prob{A,D}=\cprob{D}{A}\prob{A}$), and hence determines a probbase. We can thus quickly assess whether a candidate probbase and corresponding set of VA estimates are `viable' by assessing whether they both correspond to a real distribution of $(A,D)$ with the correct marginals $\prob{A}$ and $\prob{D}$. However, this is not generally enough to guarantee that the probbase and VA estimates are correct: in fact, for any probbase (correct or not), we can find a (reasonable) set of VA probabilities such that both are consistent with the observed data, and vice versa. 

Without making use of the VA algorithm, we cannot generally identify the distribution of $(A,D)$ - there are $ \approx 3^s \times r$ unknowns, and only $\approx 3^s + r$ constraints. We also cannot identify the probbase table. 
Informally, we have:
\begin{theorem*}[Non-identifiability; informal]
\label{thm:nonidentifiability}
The probbase table is not identifiable from the marginal frequencies of answer sets and of causes of death.
\end{theorem*}

In order to do any better, we therefore require \emph{some} assumption on the joint distribution of $A$ and $D$. We presume the following:
\begin{assumption*}[Conditional independence, informal]
We may partition answers into blocks such that for any two questions in separate blocks, the answers to those questions are independent given a cause of death. 
\end{assumption*}

This assumption, while appearing mild, greatly simplifies the joint distribution of $(A,D)$. If blocks have sizes $|B_1|,|B_2|,\dots,|B_b|$ with $\sum_{\ell} |B_{\ell}|=s$, then we now only have $\approx d \sum_{\ell} 2^{|B_{\ell}|}$ unknowns, which will generally be much fewer than the number of constraints, which is $\approx 2^s + d$. Indeed, since the entire distribution can generally be inferred in this case, we could conceivably learn an entire set of probabilities $\cprob{D}{A}$ from scratch, and we describe a method for this, though in practice we will not generally have enough data to do this completely. 

We also describe and analyse two methods to recover the probbase table from observed samples from $A$, which work under this assumption. One method uses both a VA algorithm and a probbase to assess how well individual answers can be imputed when missing, and the other uses only the probbase to assess whether pairs or triples of answers are both answered together at the right frequency.

\subsection{Notation}

Throughout, we will use the notation $[n]$ to denote the set $\{1,2,\dots n\}$. Capitals denote random variables and lowercase letters denote observations. We will use indices $i$ for samples (e.g. interviews), $j$ for causes of death, $k$ for questions on the interview, and $\ell$ for blocks. We denote the r-simplex by $\Delta_r$, the discrete uniform distribution over the set $S$ by $U(S)$, and the indicator function for set $S$ by $\ind{S}$. We will use the convention that for a set $X$ usually containing elements indexed by a set $S$, the symbols $X_S$ and $X_{-S'}$ for some $S'\subseteq S$ mean the tuple of $X$ containing only those elements with indices in $S'$ or $S \setminus S'$.

Suppose we are considering $r$ causes of death $d_1,d_2,\dots, d_r$. We take the cause of death as a random variable $D$, and denote $\pi_j=\prob{D=d_j}$ for $j \in [r]$, assembling these values into a column vector $\pi$ or diagonal matrix $I(\pi)$ as needed. 
We presume that a VA interview contains $s$ questions in total, each of which has an answer in $\{1,0,NA\}$, where the interpretation of `1' `0', `NA' is contextual, though intend `NA' to generally indicate not-at-random missing answers.
We denote the full set of answers as a random variable $A=\{A_1,A_2,\dots,A_s\}$, where $A_k$ is the answer to question $k \in [s]$ and $A_S$, $A_{-S}$ are answers to subsets of questions as above. 
We denote the set of all sequences $\{0,1,NA\}^s$ as $\mathbb{A}$, with $\mathbb{A}_S$, $\mathbb{A}_{-S}$ as sets of all possible answers to subsets of questions.  
We will assume a known partition of of $[s]$ into $b$ blocks $\{B_{\ell}\}_{\ell \in [b]}$, so $(B_{\ell_1} \cap B_{\ell_2})=\emptyset$ for $\ell_1 \neq \ell_2$ and $\cup_{\ell} B_{\ell}=[s]$. We denote by $B(k)$ the block containing $k \in [s]$. 

An important quantity is the probability of answering `1' to a given question given a particular cause of death. We denote:
\begin{equation*}
q_{kj}=\cprob{A_k=1}{D=d_j}
\end{equation*}
and $q=\{q_{kj}\}_{k \in [s],j \in [r]}$ (or $\hat{q}$, $\tilde{q}$), interpreted as a $s \times r$ probbase matrix with entries in $(0,1)$. 
We deliberately distinguish this from $\cprob{A_k=1}{D=d_j,A_k\in \{0,1\}}$, which we do not attempt to estimate in general. We will be dealing with imperfect estimates of $q$, which we will call $\hat{q}=\{\hat{q}_{jk}\}$. 

We consider verbal autopsy algorithms $V(a_S,\hat{q})$ generically as functions which take a potentially incomplete set of answers $a$ and a candidate probbase $\hat{q}$,  and return a vector of length $r$ representing posterior distribution over causes of death $d_j$; that is, we take $V:\{0,1,NA,\text{missing}\}^s \times (0,1)^{rs} \to \Delta_r$, interpreting $V(a_S,\hat{q})$ as treating the values $a_{-S}$ as `missing'. 
By `a VA algorithm' we mean a function in this class. 
We differentiate missingness in input to the VA algorithm (`missing'), which we will generally consider as missingness at random, from the `NA' answer in question sets, which we allow to be treated as missing-not-at-random.

We take our data as a set of $n>1$ completed interviews $\mathcal{A}=\{\alpha^1, \alpha^2, \dots \alpha^n\}$, where $\alpha^i\in \{0,1,NA\}^s$. 
Where necessary, we will assume $\alpha^i$ are independent and all follow the distribution of $A$. 
For a subset $S \subseteq [s]$ we denote:
\begin{equation*}
p(a_S)=\prob{A_S=a_S}, \hspace{20pt} \eprob{a_S}=\frac{|\{i':\alpha^{i'}_S=\alpha_S\}|}{n}
\end{equation*}
that is, $\eprob{a_S}$ is the empirical frequency of observing the set of answers $a_S$.  

Our methods all comprise `objective functions', denoted by capital letters; e.g. $N(\hat{q},\hat{V})$. We denote estimators of the same as e.g. $\hat{N}(\hat{q},\hat{V})$, which we refer to as `estimated objective functions'. 

\section{Related work}
\label{sec:review}

Early verbal autopsy approaches used physician-coded analysis of questionnaires; that is, a physician reviewed the questionnaire and assigned a set of likely causes of death. This was recognised as being susceptible to inter-physician variability, and probabilistic methods were developed~\citep{byass03}.

The first statistical VA algorithm aiming to establish a Bayesian posterior distribution over cause of death (CoD) given a set of symptoms was developed in~\citep{byass03} and extended in~\citep{byass19}. The development of automated VA algorithms was motivated by a need to reduce workload on physicians and standardise data assessment. Given a questionnaire $a \in \{0,1,NA\}^s$ for which $a_k=1$ for $k \in \mathcal{K}$, the InterVA4 algorithm assigns a verbal autopsy score $V_j(a,q)$ as
\begin{equation}
\cprob{D=d_j}{A=a} \approx V_j(a,q) = \frac{\pi_j \prod_{k \in \mathcal{K}} q_{jk}}{\sum_r \pi_r \prod_{k \in \mathcal{K}} q_{rk} } \label{eq:interva4}
\end{equation}


These methods (InterVA4 and InterVA5) are notable in that the posterior distribution is not directly estimated, in that absent symptoms (those $k$ for which $a_k \in \{0,NA\}$) are not considered in calculations. An extension in which absent symptoms are included in the posterior calculation was developed in~\citep{miasnikof15}: if we define $\mathcal{K}=\{k: a_k \in \{0,1\}\}$ then the associated VA score $V_j(a,q)$ is defined as:
\begin{equation}
\cprob{D=d_j}{A=a} \approx V_j(a,q) = \frac{\pi_j \prod_{k \in \mathcal{K}} q_{jk}^{a_k}(1-q_{jk})^{1-a_k}}{\sum_r \pi_r \prod_{k \in \mathcal{K}} q_{rk}^{a_k}(1-q_{rk})^{1-a_k} } \label{eq:va_nb}
\end{equation}
This essentially uses a `Naive Bayes classifier'~\citep{james13}: if, for any $k_1,k_2,j$ we have that $\cprob{A_{k_1}=a_{k_1},A_{k_2}=a_{k_2}}{D=d_j}=\cprob{A_{k_1}=a_{k_1}}{D=d_j}\cprob{A_{k_2}=a_{k_2}}{D=d_j}$ then we have:
\begin{align*}
V_j(a,q) &= \frac{\prob{D=d_j} \prod_{k \in \mathcal{K}} \cprob{A_k=1}{D=d_j}^{a_k}\cprob{A_k = 0}{D=d_j}^{1-a_k}}{\sum_r \prob{D=d_r} \prod_{k \in \mathcal{K}} \cprob{A_k=1}{D=d_r}^{a_k}\cprob{A_k = 0}{D=d_r}^{1-a_k} } \\
&= \frac{\prob{D=d_j} \cprob{A_{\mathcal{K}}=a_{\mathcal{K}}}{D=d_j}}{\sum_r \prob{D=d_r} \cprob{A_{\mathcal{K}}=a_{\mathcal{K}}}{D=d_r}} \\
&= \cprob{D=d_j}{A_{\mathcal{K}}=a_{\mathcal{K}}} \nonumber 
\end{align*}
The mechanism for developing a probbase matrix in InterVA4 and InterVA5 is to consult a panel of expert physicians to estimate each probability, which is encoded qualitatively as one of $\{A,A+,\dots,F\}$. In InterVA4 and InterVA5, these are replaced with numeric values according to a fixed lookup table~\citep{byass03,byass19}. It is also possible to learn a probbase directly from labelled training data~\citep{miasnikof15}.

A method called InSilicoVA~\citep{mccormick16} proposes a hierarchical Bayesian model which includes an explicit recalibration of the probbase matrix, in which a mapping is learned from 
letter codes to probabilities. 
The specification enables incorporation of additional information into the VA method, such as physician-assigned causes of death. 
A major aim of the InSilicoVA algorithm is to learn about the cause-specific mortality fraction at the population level, and the COD distributions at the individual level~\citep{mccormick16}. 

There has been significant attention on comparison of verbal autopsy algorithms, and the degree to which parameters (prior, probbase) matter to CoD assignments~\citep{clark18}. 
Further work on verbal autopsy analysis has extended the general Bayesian model~\citep{li19,zhu24,zhu25}, explored of sensitivity to the prior~\citep{fottrell11}, and expanded the scope of the method to include circumstances of mortality~\citep{ambruoso22}. 
Other verbal autopsy methods include the `Tariff' method~\citep{serina15} and a recently-developed method using large language models~\citep{chu25}.

A general Bayesian model is proposed in~\citep{li19} in which the joint distribution of causes of death and symptoms is modelled allowing for dependence between symptoms, using a spike-and-slab prior on the covariance to incentivise sparsity. 
Posterior estimates of cause of death can then be attained by sampling from the posterior on distributional parameters. 
This entails learning a probbase in a sense; or rather, starting with an estimate and modifying it according to the data in order to make a cause-of-death assignment. 

\subsection{Relation to our methods}

Our methods are similar in spirit to~\citep{li19} in that we use the joint distribution of answers to infer the conditional distribution of causes of death. Indeed, \cite{li19} is adaptable to learning distributional parameters - including the probbase - from a marginal distribution of only $A$. However, a Bayesian model for the distribution of $A,D$ essentially requires a specification of the form of the likelihood $\cprob{A=a}{D=d_j}$. 
Our methods technically aim to evaluate the accuracy of a VA algorithm using only unlabelled data without making any specifications on the form of the likelihood function. 


\section{Theoretical analysis}
\label{sec:methods}

\subsection{Viability}
\label{sec:viability}

Suppose that, given the correct probbase $q$, our VA algorithm returns the correct posterior probabilities; that is:

\begin{assumption}
\label{asm:va_calibration}
For a VA algorithm $V(\cdot)$, we assume that if $\hat{q}=q$ then:
\begin{equation}
V_0(a_{-B_{\ell}},\hat{q})_j=\cprob{D=d_j}{A_{-B_{\ell}}=\alpha_{-B_{\ell}}} \text{ for all }a \in \{0,1,NA\}^s:\prob{A_{-B_{\ell}}=a_{-B_{\ell}}}>0 \nonumber 
\end{equation}
\end{assumption}

Under this assumption, we have for all $k \in [r], j \in [s]$, we have:
\begin{align}
q_{kj} \pi_j &= \cprob{A_k=1}{D=d_j}\prob{D=d_j} = \prob{A_k=1,D=d_j} \nonumber \\
&= \sum_{a:a_k=1} \prob{A=a,D=d_j} = \sum_{a:a_k=1} \cprob{D=d_j}{A=a}\prob{A=a} \nonumber \\
&= \sum_{a:a_k=1} V_j(a) p(a) \label{eq:vcond}
\end{align}
If we denote our first objective function as
\begin{equation*}
N(\hat{q},\hat{V})= \sum_{k \in [r],j \in [s]} \left[\hat{q}_{kj}\pi_j - \sum_{a:a_k=1} \hat{V}_j(a,\hat{q}) p(a) \right]^2
\end{equation*}
then we immediately have:

\begin{restatable}{thm}{viabilityUse}
\label{thm:viabilityUse}

If the VA algorithm $V$ in use satisfies assumption~\ref{asm:va_calibration} for the true probbase $q$, then for any $\hat{q} \in (0,1)^{s \times r}$ and VA algorithm $\hat{V}$, we have:
\begin{equation*}
N(\hat{q},\hat{V}) \geq N(q,V)=0
\end{equation*}

\end{restatable}

Given data $\mathcal{A}=\{\alpha^1, \alpha^2, \dots \alpha^n\}$, we may use the simple plug-in estimator for our estimated objective function:
\begin{equation}
\hat{N}(\hat{q},\hat{V})= \sum_{k \in [r],j \in [s]} \left[\hat{q}_{kj}\pi_j - \sum_{a:a_k=1} \hat{V}_j(a,\hat{q}) \eprob{a} \right]^2 \label{eq:viability_estimator}
\end{equation}
noting that terms in the sum will be 0 if $\{i:\alpha^i=a\}=\emptyset$. We now establish a convergence rate for $\hat{N}(\hat{q},\hat{V})$. Here and hereafter we keep constants explicit in asymptotic rates, to emphasise where they depend on the potentially very large value $3^s$; we also note that while the number $3^s$ of potential questionnaire answers is very large, the entropy of the marginal distribution of $A$ will tend to be fairly low (most answer sets are extremely improbable) so we establish convergence both in terms of the domain and entropy of the distribution of $A$. Both are established by probabilistically bounding the quantity
\begin{equation*}
\sum_{a} |p(a)-\eprob{a}|
\end{equation*}
given that either the domain or the entropy of $A$ is bounded. 

\begin{restatable}{thm}{viabilityConvergence}
\label{thm:viabilityConvergence}

Suppose the distribution of $A$ has entropy $h$. With probability at least $1-\delta$, we have for sufficiently large $n$:
\begin{equation*}
|\hat{N}(\hat{q},\hat{V}) - N(\hat{q},\hat{V})| \leq r s \sqrt{\frac{2}{n}\left(3^s \log(2) + \log\left(\frac{1}{\delta}\right)\right)}
\end{equation*}
and
\begin{equation*}
|\hat{N}(\hat{q},\hat{V}) - N(\hat{q},\hat{V})| \leq r s \frac{h+1}{\log(n)} + 2 r s \sqrt{\frac{h\log(2)}{\log(n)} + \frac{1}{n}\log\left(\frac{2}{\delta}\right)}
\end{equation*}

\end{restatable}
proved in appendix~\ref{apx:proofs}. This suggests a straightforward means to check whether a candidate probbase $\hat{q}$ and corresponding VA estimates $\hat{V}(\cdot,\hat{q})$ are `viable', in that they both correspond to a distribution of $(A,D)$ with the correct marginals. 

However, any joint distribution of $(A,D)$ with marginals $\prob{A=a}=p(a)$ and $\prob{D=d_j}=\pi_j$ corresponds to a viable probbase-VA algorithm pair, and (as per the simple argument in Section~\ref{sec:heuristics}) there is a large space of such distributions. Unless we are lucky enough that the pair $(\hat{q},\hat{V}(\cdot,\hat{q}))$ is viable \emph{only} when $\hat{q}=q$, we cannot rely on this alone to identify $q$.

\subsection{Coherence}

Henceforth, we presume that $A_{k_1},A_{k_2}$ are conditionally independent given $D$ if $k_1 \subseteq B_{\ell_1}$, $k_2 \subseteq B_{\ell_2}$ with $\ell_1 \neq \ell_2$. That is, we presume that, for $a_1,a_2 \in \{0,1,NA\}$:
\begin{assumption}
\label{asm:cond_indep}
If $k_1 \subseteq B_{\ell_1}$, $k_2 \subseteq B_{\ell_2}$ with $\ell_1 \neq \ell_2$, then for any $j \in [r]$:
\begin{equation*}
\cprob{A_{k_1}=a_{k_1},A_{k_2}=a_{k_2}}{D=d_j} = \cprob{A_{k_1}=a_{k_1}}{D=d_j} \times 
\cprob{A_{k_2}=a_{k_2}}{D=d_j}
\end{equation*}
\end{assumption}

In this case, under assumption~\ref{asm:va_calibration} we have:
\begin{align}
p(a) &= \prob{A_{B_1}=a_{B_1},A_{B_2}=a_{B_2},\dots,A_{B_b}=a_{B_b}} \nonumber \\
&=\sum_{j=1}^r \pi_j \prod_{\ell}\cprob{A_{B_{\ell}}=a_{B_{\ell}}}{D=d_j} =\sum_{j=1}^r \pi_j^{1-b} \prod_{\ell}[V_j(a_{B_{\ell}},q) p(a_{B_{\ell}})] \label{eq:coherence} 
\end{align}
We note that if this holds for $b$ blocks $B_{\ell}$, then it also holds for fewer than $b$ blocks, since we may combine blocks and maintain the conditional independence property. We will suppose we have three blocks $B_1, B_2, B_3$ of sufficient size. Then we may show that~\eqref{eq:coherence} only holds when the VA probability assignments are correct; that is:

\begin{restatable}{thm}{coherenceIdentifiability}
\label{thm:coherenceIdentifiability}
Suppose that the true distribution of $(A,D)$ satisfies assumption~\ref{asm:cond_indep} with three blocks $B_1,B_2,B_3$ such that $3^{|B_{\ell}|} \geq r>1$. Denote $V_j(a_{B_{\ell}})=\cprob{D=d_j}{A_{B_{\ell}}=a_{B_{\ell}}}$, and let $\hat{V}$ be some estimate of this quantity attained with a VA algorithm. For almost all such distribtions of $(A,D)$, the identity:
\begin{equation*}
p(a) =\sum_{j=1}^r \pi_j^{-2} \hat{V}_j(a_{B_{1}})\hat{V}_j(a_{B_{2}})\hat{V}_j(a_{B_{3}}) p(a_{B_{1}})p(a_{B_{2}})p(a_{B_{3}})
\end{equation*}
holds for all $a$ if and only if $\hat{V}(a_{B_{\ell}})=V(a_{B_{\ell}})$ for all $a_{B_{\ell}}$. 

\end{restatable}

If only two blocks $B_1, B_2$ are available, then the VA function is not generally identifiable: if the values $p(a)=p(A_{B_1},A_{B_2})$ are arranged in a $|B_1| \times |B_2|$ matrix $T$, then we may write:
\begin{equation*}
T=\beta_1 I(\pi) \beta_2^T
\end{equation*}
where $\beta_{\ell}$ is a $(3^{|B_{\ell}|} \times r)$ matrix with entries $\pi_j^{1/2}\hat{V}_j(a_{B_{\ell}})p(a_{B_{\ell}})$, and $I(\pi)$ is a diagnoal matrix with entries $\pi_j$. But for any orthogonal $r \times r$ matrix $R$, setting $\gamma_{\ell}=\gamma_{\ell} R$, we have
\begin{equation*}
\gamma_1 I(\pi) \gamma_2=\beta_1 R I(\pi) R^T \beta_2^T = \beta_1 I(\pi) R R^T \beta_2^T = \beta_1 I(\pi) \beta_2^T=T
\end{equation*}
hence $\beta_1$ and $\beta_2$ are not identifiable (that is, they cannot be uniquely identified on the basis of knowing only $T$), unless $\approx r^2$ values are already known to be correct. 

We may estimate values $p(a)$ from an observed dataset of observations $\alpha^i$ of $A$, and compute an analogous quantity to~\eqref{eq:viability_estimator} by summing squared differences:
\begin{equation}
\hat{C}(\hat{V}) = \sum_{a \in \mathbb{A}} \left[\eprob{a} - \sum_{j=1}^r \pi_j^{-2} \hat{V}_j(a_{B_1})\hat{V}_j(a_{B_2})\hat{V}_j(a_{B_3})\eprob{a_{B_1}}\eprob{a_{B_2}}\eprob{a_{B_3}}\right]^2
\end{equation}
where the sum is over all possible answer sets $a$. As for~\eqref{eq:viability_estimator}, since terms in the sum are only non-negative for those $a$ appearing somewhere amongst the values $\{\alpha^i\}_{i \in [n]}$, the sum has at most $n$ terms and can be tractably computed. 

Denoting $C(\hat{V})$ as the corresponding quantity with estimated probabilities $\hat{\mathbb{P}}()$ replaced with true probabilities $p()$ under the distribution $(A,D)$, we now have a less-than-reassuring consistency result:
\begin{restatable}{thm}{coherenceConvergence}
\label{thm:coherenceConvergence}
Suppose the entropy of $A$ is $h$. For sufficiently large $n$, we have, with probability at least $1-\delta$
\begin{equation}
|\hat{C}(\hat{V}) - C(\hat{V})| \leq 6  \sqrt{2} \cdot 3^s  \sqrt{\frac{1}{n} \log\left(\frac{4}{\delta}\right) + \frac{3^s}{n} \log(2)} \label{eq:coherence_convergence}
\end{equation}
and
\begin{equation*}
|\hat{C}(\hat{V}) - C(\hat{V})| \leq \frac{(2m+1)(h+1)}{\log(n)} +  \frac{24 m h}{\log(n)}\sqrt{\log\left(\frac{1}{\delta}\right) + 3\log(2)}
\end{equation*}
where $m=\sum_j \pi_j^{-2}$.

\end{restatable}


We consider that this is not a practically tractable way to learn VA probabilities, though it may be usable to tune individual cases, or if a VA algorithm is known up to a probbase. 


\subsection{Imputation error}

We define the `Imputation error' associated with a VA algorithm $V$ and a probbase $\hat{q}$ as:
\begin{equation}
I(\hat{q},\hat{V}) = \EE{K \sim U([s]),A,D}{\loss{F_K\left[\hat{V}\left(A_{-B(K)},\hat{q}\right),\hat{q}\right]}{\ind{A_K=1}}}
\end{equation}
where
\begin{equation*}
F_k(v,\hat{q}) = \sum_{j=1}^r v_j \hat{q}_{kj} \label{eq:fk_expansion}
\end{equation*}
and
\begin{equation*}
\loss{\hat{Y}}{Y}=-\left(Y \log(\hat{Y}) + (1-Y) \log(1-\hat{Y})\right)
\end{equation*}
%

%
%

That is, the imputation error is the expected inaccuracy when imputing an answer to a randomly-chosen question on the basis of other answers, using the candidate probbase to define the imputation rule. Supposing question $a_k$ is in block $B_{\ell}$ we can write:
\begin{align}
\overbrace{\cprob{A_k=1}{A_{-B_{\ell}}=a_{-B_{\ell}}}}^{\begin{smallmatrix} \text{Imputed probability of} \\ \text{answering `1' to question} \\ \text{$k$ given other answers} \end{smallmatrix}} 
&= \sum_{j=1}^r \cprob{A_k=1}{A_{-B_{\ell}}=a_{-B_{\ell}},D=d_j} \times \cprob{D=d_j}{A_{-B_{\ell}}=a_{-B_{\ell}}} \nonumber \\
&= \sum_{j=1}^r  \underbrace{\cprob{A_k=1}{D=d_j}}_{\text{Probbase}}  \underbrace{\cprob{D=d_j}{A_{-B_{\ell}}=a_{-B_{\ell}}}}_{\text{Standard VA}} \label{eq:main_expansion}
\end{align}
applying assumption~\ref{asm:cond_indep} in the second step. This justifies the use of $F_k(v,\hat{q})$ as an estimator of $\cprob{A_k=1}{A_{-B_{\ell}}=a_{-B_{\ell}}}$. We consider an candidate probbase $\tilde{q}$ and VA algorithm $\tilde{V}$ such that:
\begin{equation}
(\tilde{V},\tilde{q}) = \argmin_{\hat{V},\hat{q}} I(\hat{q},\hat{V}) \label{eq:imin}
\end{equation}

We firstly aim to establish that condition~\eqref{eq:imin} holds if and only if $\tilde{q}=q$. Since the unknown value $\hat{q}$ appears both in the sum $F_k[\cdot]$ and in the function $\hat{V}(\cdot)$, we must again use assumption~\ref{asm:va_calibration} to say something about how the VA algorithm depends on $q$. 

This allows us to establish the result in the `only if' direction: if a VA algorithm is `correct', given a correct probbase, then $I(\hat{q},\hat{V})$ is minimised (possibly non-uniquely) when the probbase is correct. 

\begin{restatable}{thm}{minimalCorrect}
\label{thm:minimalCorrect}

Suppose that the joint distribution of $(A,D)$ satisfies assumption~\ref{asm:cond_indep} and that a VA algorithm $V$ satisfies assumption~\ref{asm:va_calibration}. Then for any VA algorithm $\hat{V}$ and probbase $\hat{q}$ we have:
\begin{equation*}
I(\hat{V},\hat{q}) \geq I(V,q)
\end{equation*}
\end{restatable}
While fairly simple this observation enables some practical use of $I(\cdot,\cdot)$ immediately: given VA algorithm and probbase pairs $(\hat{V}_0,\hat{q}_0)$ and $(\hat{V}_1,\hat{q}_1)$ such that $I(\hat{q}_1,\hat{V}_1) > I(\hat{q}_0,\hat{V}_0)$ (in practice, with high probability), we can conclude that either $\hat{V}_1$ or $\hat{q}_1$ is incorrect. 

The `if' direction - that property~\eqref{eq:imin} under assumption~\ref{asm:va_calibration} implies that $\hat{q}=q$ - cannot yet be concluded. Indeed, this is not the case in full generality: assumption~\ref{asm:va_calibration} determines the behaviour of the function $V$ only at the point of the true probbase $q$, but in order for $I(\hat{q},\hat{V})$ to be minimised, we essentially need 
\begin{equation*}
F_k\left[\hat{V}(a_{-B(k)},\hat{q}),\hat{q}\right] = \sum_{j \in [r]} \hat{V}_{j}(a_{-B(k)},\hat{q})\hat{q}_{kj} = \cprob{A_k=1}{A_{-B(k)}=a_{-B(k)}}
\end{equation*}
(see proof of Theorem~\ref{thm:minimalCorrect}) which is a system of $\sum_{\ell}3^{s-|B_{\ell}|}|B_{\ell}|$ constraints. There are $\sum_{\ell}3^{s-|B_{\ell}|}r$ values $\hat{V}_j(a_{-B(k)},\hat{q})$, so if (in general) $r> |B_{\ell}|$, given a matrix $\hat{q}$ we will generally be able to find a corresponding set of values $\hat{V}_j(a_{-B(k)},\hat{q})$ so as to minimise $I(\hat{q},\hat{V})$. 

We give a partial result toward the identifiability of $\hat{q}$ from condition~\ref{eq:imin} given assumption~\ref{asm:va_calibration}. We must first meaningfully constrict the possibilities for the function $V$. To do this, we will introduce a notion of `partial calibration':
\begin{definition} 
\label{def:partialCalibration}
Given block structure $\{B_{\ell}\}_{\ell \in [b]}$, a fixed marginal distribution $\pi$ and fixed marginal distributions $p(a_{B_{\ell}})$, 
we say a VA algorithm $\hat{V}$ is `partly calibrated to $\pi$, $\{p(a_{B_{\ell}})\}_{\ell \in [b]}$ and $\{B_{\ell}\}_{\ell \in [b]}$ at $\hat{q}$' if there exists a distribution $d$ of $(A,D)$ satisfying Assumption~\ref{asm:cond_indep} with block structure $\{B_{\ell}\}_{\ell \in [b]}$ such that $\pprob{d}{a_{B_{\ell}}}=p(a_{B_{\ell}})$, $\pprob{d}{D=d_j}=\pi_j$, $\cpprob{d}{D=d_j}{A_{S}=a_{S}}=\hat{V}(a_S,\hat{q})$, and $\cpprob{d}{A_k=1}{D=j}=\hat{q}_{kj}$.
\end{definition}
This notion essentially describes whether a candidate VA function $V$ and probbase $q$ both correspond to a plausible distribution of $(A,D)$ (which is essentially our previous condition of `viability') with the requirement that the marginal probabilities $\pprob{d}{A_{B_{\ell}}=a_{B_{\ell}}}$ also match the true probabilities.

In order to establish our result we consider a modified version of $I$:
\begin{equation*}
I^{\ell_1,\ell_2}(\hat{q},\hat{V}) = \EE{K \sim U(B_{\ell_1}\cup B_{\ell_2}),A,D}{\loss{F_K\left[\hat{V}\left(A_{B_{\ell_1}\cup B_{\ell_2}\setminus B(K)},\hat{q}\right),\hat{q}\right]}{\ind{A_K=1}}}
\end{equation*}
that is, equivalent to $I(\hat{q},\hat{V})$ if questions outside $B_{\ell_1}\cup B_{\ell_2}$ are ignored. With that, we have:

\begin{restatable}{thm}{imputationIdentifiability}
\label{thm:imputationIdentifiability} 

Consider block structures with three blocks $B_1,B_2,B_3$ and a given VA algorithm $V$ satisfying assumption~\ref{asm:va_calibration}. Given values of $\pi$ and $\{p(a_{B_{\ell}})\}_{\ell \in [3]}$ and block structure $\{B_{\ell}\}_{\ell \in [3]}$ let $\mathcal{Q}$ be the set of $\hat{q}$ such that $V$ is partly calibrated to $\pi$, $\{p(a_{B_{\ell}})\}_{\ell \in [b]}$ and $\{B_{\ell}\}_{\ell \in [b]}$ at $\hat{q}$, and $\hat{q}$ and $V^{\ell}(\hat{q})$ have rank at least $r$. Then for almost any true distribution of $(A,D)$, we have:
\begin{equation*}
\tilde{q}=\argmin_{\hat{q} \in \mathcal{Q}} \left[I^{1,2}(\hat{q},V) + I^{1,3}(\hat{q},V) + I^{2,3}(\hat{q},V)\right] \Leftrightarrow \tilde{q}=q
\end{equation*}
\end{restatable}
We believe that stronger results are true; indeed it seems likely that 
\begin{equation*}
\tilde{q}=\argmin_{\hat{q} \in \mathcal{Q}} I(\hat{q},V)\Leftrightarrow \tilde{q}=q
\end{equation*}
but as yet this result is open. We estimate $I(\hat{q},\hat{V})$ using our set $\mathcal{A}$ as: 
\begin{equation*}
\hat{I}(\hat{q},\hat{V})= \frac{1}{ns} \sum_{i=1}^n \sum_{k=1}^s \loss{F_k\left[\hat{V}\left(\alpha^i_{-B(k)},\hat{q}\right),\hat{q}\right]}{\ind{\alpha^i_k=1}}
\end{equation*}
which has the advantage of being unbiased. We also have the following convergence property:

\begin{restatable}{thm}{imputationConvergence}
\label{thm:imputationConvergence} 

Suppose that $\hat{q}_{jk} \geq m_q > 0$ and that the distribution of $A$ has entropy $h$. 
Then the estimator $\hat{I}(\hat{q},\hat{V})$ of $I(\hat{q},\hat{V})$ is unbiased and consistent, and for sufficiently large $n$ we have
\begin{equation*}
|\hat{N}(\hat{q},\hat{V}) - N(\hat{q},\hat{V})| \leq \log\left(\frac{1}{m_q}\right) \sqrt{\frac{2}{n}\left(3^s \log(2) + \log\left(\frac{1}{\delta}\right)\right)}
\end{equation*}
and
\begin{equation*}
|\hat{N}(\hat{q},\hat{V}) - N(\hat{q},\hat{V})| \leq \log\left(\frac{1}{m_q}\right) \frac{h+1}{\log(n)} + 2 \log\left(\frac{1}{m_q}\right) \sqrt{\frac{h\log(2)}{\log(n)} + \frac{1}{n}\log\left(\frac{2}{\delta}\right)}
\end{equation*}

\end{restatable}

\subsection{Two-way and three-way comparison}

Our final method to establish internal consistency of a VA algorithm and probbase aims to directly identify the probbase matrix from observed data, circumventing any evaluation of VA probabilities. We define the two-way and three-way agreement associated with a probbase $\hat{q}$ respectively as:
\begin{align}
R_2(\hat{q}) &:= \sum_{\substack{k_1,k_2 \in [s] \\ B(k_1) \neq B(k_2)}} \left[\prob{A_{k_1}=1,A_{k_2}=1} - \sum_j \hat{q}_{k_1 j}\hat{q}_{k_2 j} \pi_j \right]^2 \label{eq:pairwise_consistency} \\
R_3(\hat{q}) &:= \sum_{\substack{k_1,k_2,k_3 \in [s] \\ B(k_1), B(k_2),B(k_3)\text{ distinct}}} \left[\prob{A_{k_1}=1,A_{k_2}=1,A_{k_3}=1} - \sum_j \hat{q}_{k_1 j}\hat{q}_{k_2 j}\hat{q}_{k_3 j} \pi_j \right]^2 \label{eq:threeway_consistency}
\end{align}
which effectively compares the probability of seeing answers `1' to two or three questions in different blocks with what we would expect to see given the probbase. This is motivated by the expansion, for $k_1,k_2 \in [s]$ with $B(k_1) \neq B(k_2)$:
\begin{align*}
\prob{A_{k_1}=1,A_{k_2}=1} &= \sum_j \cprob{A_{k_1}=1,A_{k_2}=1}{D=d_j}\prob{D=d_j} \\
&= \cprob{A_{k_1}=1}{D=d_j}\cprob{A_{k_2}=1}{D=d_j}\pi_j \\
&= q_{k_1 j}q_{k_2 j}\pi_j
\end{align*}
and a similar derivation for three indices. 

Defining the matrix $\{P_2\}_{k_1 k_2}=\prob{A_{k_1}=1,A_{k_2}=1}$, we have that $P_2$ agrees with $qI(\pi)q^T$ except on block-diagonal elements, where $q$ is the correct probbase. Likewise, defining the tensor $\{P_3\}_{k_1 k_2 k_2}=\prob{A_{k_1}=1,A_{k_2}=1,A_{k_3}=1}$, the value of $P_3$ matches the value of $\sum_j q_{k_1 j}q_{k_2 j}q_{k_3 j} \pi_j$ for indices $k_1,k_2,k_3$ in different blocks. We then have:

\begin{restatable}{thm}{agreementIdentifiability}
\label{thm:agreementIdentifiability}

Consider distributions of $(A,D)$ satisfying assumption~\ref{asm:cond_indep} with three blocks $B_1,B_2,B_3$ where $\min_{\ell} |B_{\ell}| \geq r \geq 2$. For almost all such distributions, we have:%
\begin{equation*}
\tilde{q}=\argmin_{\hat{q} \in Q} R_2(\hat{q}) \Leftrightarrow \hat{q}= q S \text{    and    }
\tilde{q}=\argmin_{\hat{q} \in Q} R_3(\hat{q}) \Leftrightarrow \hat{q}= q
\end{equation*}
where $Q = \{\hat{q}\in (0,1)^{s \times r}:\hat{q}\pi=q\pi\}$ and $S$ is an arbitrary $r \times r$ orthogonal matrix such that $qS \in Q$.
\end{restatable}

The matrix $q$ is consequently not identifiable from $R_2$ alone, but if it is partly known (as may reasonably often be the case) then $R_2$ may be used to fill in remaining entries.

Analogous to other estimators, we use a set of completed interviews $\mathcal{A}=\{\alpha^1,\alpha^2,\dots, \alpha^n\}$ to define:
\begin{align*}
\hat{R_2}(\hat{q}) &= \sum_{\substack{k_1,k_2 \in [s] \\ B(k_1) \neq B(k_2)}} \left[\frac{\left|\{i:\alpha^i_{k_1}=\alpha^i_{k_2}=1\}\right|}{n} - \sum_j \hat{q}_{k_1 j}\hat{q}_{k_2 j} \pi_j  \right]^2 \\
\hat{R_3}(\hat{q}) &= \sum_{\substack{k_1,k_2,k_3 \in [s] \\ B(k_1), B(k_2),B(k_3) \text{ distinct}}} \left[\frac{\left|\{i:\alpha^i_{k_1}=\alpha^i_{k_2}=\alpha^i_{k_3}=1\}\right|}{n} - \sum_j \hat{q}_{k_1 j}\hat{q}_{k_2 j}\hat{q}_{k_3 j} \pi_j  \right]^2
\end{align*}
The convergence of $\hat{R_2}$ and $\hat{R_3}$.  is essentially governed by the convergence of the empirical estimators for $\prob{A_{k_1}=1,A_{k_2}=1}$. We have the following result:

\begin{restatable}{thm}{agreementConvergence}
\label{thm:agreementConvergence}

Suppose that $(A,D)$ follow a distribution satisfying assumption~\ref{asm:cond_indep} and that for any $k_1,k_2$ with $B(k_1) \neq B(k_2)$ we have $\prob{A_{k_1}=1,A_{k_2}=1} \in (0,1)$. Then with probability at least $1-\delta$ we have:
\begin{align*}
\left|\hat{R_2}(\hat{q})-R_2(\hat{q})\right| &\leq 2\sqrt{2}s^2\sqrt{\frac{1}{n}\log\left(\frac{2 s^2}{\delta}\right)} 
\intertext{    and    } 
\left|\hat{R_3}(\hat{q})-R_3(\hat{q})\right| &\leq 2\sqrt{2}s^3\sqrt{\frac{1}{n}\log\left(\frac{2 s^2}{\delta}\right)} 
\end{align*}

\end{restatable}

Both rates of convergence are reasonably fast, especially compared with the previously-described estimators. The rates are not dependent on the large number ($3^s$) of possible answer sets.

\subsection{Robustness to conditional independence and other assumptions}
\label{sec:robustness}


The usefulness of imputation error and two/three way agreement (theorems~\ref{thm:minimalCorrect}, \ref{thm:imputationIdentifiability} and~\ref{thm:agreementIdentifiability}) rely on a fairly strong assumption (assumption~\ref{asm:va_calibration}) that with the correct probbase $q$, the function $V(a_S,q)$ perfectly encodes posterior probabilities $\cprob{D=d_j}{A_{-B_{\ell}}=a_{-B_{\ell}}}$. While the VA algorithm is designed with the intent of doing exactly this, the complete satisfaction of this assumption is unlikely. Indeed, several VA algorithms are too simple to realistically achieve this; for instance, the naive-Bayes methods in~\cite{byass19}. Implicit in this assumption is the ability of VA algorithms to manage question answers which are missing-at-random: that is, $V(a_{-S},\pi)$ is taken to mean a posterior probability when the answers in $S$ are missing at random. The VA algorithm may be able to incorporate missing-not-at-random questions too (which are taken as $NA$ answers amongst the values in $a_{-S}$). We consider that this is somewhat reasonable given that VA algorithms are designed reflect medical opinion on spectra of symptoms, and in this sense have the potential to be well-calibrated, given a reasonable probbase, in that medical practitioners frequently manage missing-at-random information and incorporate missing-not-at-random information. 

Although difficult to formally establish robustness to mild violations of assumption~\ref{asm:va_calibration}, we may reasonably hope that minimisation of $I(\hat{q},\hat{V})$ finds \emph{both} a reasonably good VA algorithm and a corresponding probbase, since the two must correspond in a sense. It is possible, in the absence of assumption~\ref{asm:va_calibration}, that a VA algorithm using a somewhat incorrect probbase generally returns more accurate posterior probabilities than does one using a completely correct probbase. By finding corresponding VA and probbase pairs which fit observed data well, we may be able to find a better VA algorithm than by tuning the probbase alone.  

%

Assumption~\ref{asm:cond_indep} merits further scrutiny, since, as described in section~\ref{sec:heuristics}, we are contingent on this assumption for the problem to be tractable at all. Nonetheless, we may show that our imputation-based and two or three-way agreement approaches are somewhat robust to moderate violations of this assumption, in that the functions $I(\hat{q},\hat{V})$, $R_2(\hat{q})$, and $R_3(\hat{q})$ change only slighly when deviations from conditionl independence are small. 
We show in section~\ref{sec:evaluation} that, in practice, mild violations of this assumption have a negligible effect on the usability of our objective functions. 


We show that imputation error is somewhat robust in the event that most of the probbase is already correct, and that the VA algorithm depends on the probbase only `directly', in that the posterior probability $V_j(a_S,\hat{q}) \approx \cprob{D=d_j}{A_S=a_s}$ depends only the values $\hat{q}_{kj} \approx \cprob{A_k=a_k}{D=d_j}$ with $k\in S$. 
Conditional dependence essentially affects imputation error through changing values of the form:
\begin{align*}
&\left[\sum_j \cprob{D=d_j}{A_{-b(k)}=a_{-b(k)}}\cprob{A_k=1}{D=d_j} - 
\sum_j V_j(a_{-b(k)},Q) Q_{jk}\right]^2 
\intertext{to}
&\left[\sum_j \cprob{D=d_j}{A_{-b(k)}=a_{-b(k)}}\cprob{A_k=1}{A_{-b(k)}=a_{-b(k)},D=d_j} - 
\sum_j V_j(a_{-b(k)},Q) Q_{jk}\right]^2.
\end{align*}
Heuristically, we expect the values $\cprob{A_k=1}{A_{-b(k)}=a_{-b(k)},D=d_j}$ to lie somewhat symmetrically around the values $\cprob{A_k=1}{D=d_j}$, so in general the weighted sums comprising the first terms in the above expressions should have similar values. In general, we have:

\begin{restatable}{thm}{imputationRobustness}
\label{thm:imputationRobustness}

Given assumption~\ref{asm:va_calibration}, suppose also that, for any $j \in [r]$, block $b$, value $k \notin b$, and value of $a_{-b}$, we have:
\begin{equation}
|\cprob{A_k=1}{D=d_j}-\cprob{A_k=1}{D=d_j,A_{-b}=a_{-b}}| \leq \epsilon \label{asm:assumption_epsilon1}
\end{equation}
and that $V_j(a_S,\hat{q})$ depends only on values $Q_{kj}$ with $k \in S$. For some block $b$ we consider the set $\mathcal{Q}=\{Q_{kj}:Q_{kj}=q_{kj} \text{ if } k \notin b, \sum_j Q_{kj}=1\text{ for all } k\}$. Then for $Q \in \mathcal{Q}$ we have:
\begin{equation*}
I(\hat{q},\hat{V})=I_2(\hat{q},\hat{V}) + E + e 
\end{equation*}
where $I_2(\hat{q},\hat{V})$ is minimised when $\hat{q}=q$,
$e \leq s \frac{\epsilon}{2}$, and:
\begin{equation*}
E = O\left(\frac{1}{s} \sum_{k=1}^s \sum_{a_{-b(k)}} \prob{A_{-b(k)}=a_{-b(k)}}\left[\cprob{A_k=1}{A_{-b(k)}=a_{-b(k)}} - F_k\left[V(a_{-b(k)},Q),Q\right]\right]^3 \right)
\end{equation*}

\end{restatable}

Theorem~\ref{thm:agreementIdentifiability} for pairwise consistency is independent of the VA algorithm in use, but remains contingent on assumption~\ref{asm:cond_indep}. However, as for imputation error, the function $C(\hat{q})$ is reasonably robust to mild violations of the assumption. We have:
\begin{restatable}{thm}{agreementRobustness}
\label{thm:agreementRobustness}

Suppose that, for any $j \in [r]$, block $b$, value $k \notin b$, and value of $a_{-b}$, we have:
\begin{equation}
|\cprob{A_k=1}{D=j}-\cprob{A_k=1}{D=j,A_{-b}=a_{-b}}| \leq \epsilon \label{asm:assumption_epsilon2}
\end{equation}
%
Then for $\hat{q} \in (0,1)^{s \times r}$ we have:
\begin{align*}
|R_2(\hat{q})-R_2^A(\hat{q})| &\leq 2 \epsilon s^2 \text{ and }\\
|R_3(\hat{q})-R_3^A(\hat{q})| &\leq 4 \epsilon s^3
\end{align*}
where $R_2^A(\hat{q})$ and $R_3^A(\hat{q})$ are minimised when $\hat{q}=q$.
\end{restatable}

\subsection{Management of demographic questions}

An important mode of violation of this assumption is for questions which are expected to be correlated with all others; for instance, demographic questions on age and sex. 

If the number of such questions is reasonably small, and the relationship between demographics and causes of death is reasonably well known, this may be managed straightforwardly by stratification of samples with separate estimation of stratum-specific probbases, with and subsequent recombination. In particular, if we suppose demographics are encoded in a set of questions $A_{B_0}$, and the joint distribution $\prob{A_{B_0}=a_{B_0},D=d_j}$ is known, then we may replace $\pi=\prob{D=d_j}$ with $\cprob{D=d_j}{A_0=a_0}$, and use only samples $\alpha^i$ with $\alpha^i_{B_0}=a_{b_0}$ in the estimation  of $I^V(\hat{q})$ or $C(\hat{q})$ to attain an estimate of $\tilde{q}(a_{B_0}):= \cprob{A_k=1}{D=d_j,A_{B_0}=a_{B_0}}$ which has properties analogous to those of $\tilde{q}$ estimated using $I^V$ or $C$ in the absence of such demographic questions.  We then have the identity:
\begin{align*}
q_{kj}&=\cprob{A_k=1}{D=d_j} \\
&=\sum_{a_{B_0}} \cprob{A_k=1}{D=d_j,A_{B_0}=a_{B_0}}\cprob{A_{B_0}=a_{B_0}}{D=d_j} \\
&= \sum_{a_{B_0}} \tilde{q}(a_{B_0})\cprob{A_{B_0}=a_{B_0}}{D=d_j}
\end{align*}
allowing the recovery of the overall probbase $q$.




\section{Simulation}
\label{sec:evaluation}

We evaluated our estimated objective unctions using a simulated dataset conforming to a known probabilistic model. We considered a real dataset: the \texttt{RandomPhysician} dataset from the \texttt{openVA} package~\citep{openVA1,openVA2}, which we will refer to as $A^{\text{sim}}$. 
This dataset consists of a set of 1,000 VA questionnaires coarsely annotated with causes of death. 
This dataset is in the WHO2012 format~\citep{who12}, which we used throughout our simulations, largely due to data availability and correspondence to the available probbase matrix, which we called $q^{\text{sim}}$. We estimated population-wide cause of death frequencies $\pi_{\text{sim}}$ to be consistent with $q^{\text{sim}}$ and $A^{\text{sim}}$. We then sampled latent CoDs independently from a multinomial distribution parametrised by $\pi^{\text{sim}}$. 

We considered only non-obstetric and non-neonatal CoDs, and excluded questions which directly identify an external CoD, leaving 133 questions and 32 CoDs. We used the annotated CoDs to define three approximately conditionally independent question blocks (as per assumption~\ref{asm:cond_indep}) of size at least 32, and to define per-CoD correlation matrices between questions within these blocks. Given our latent CoDs, we then simulated answers independently for each block according to these correlation matrices and the answer frequencies in $q^{\text{sim}}$. To evaluate robustness to assumption~\ref{asm:cond_indep}, we also simulated datasets for which the conditional independence assumption did not hold. We simulated multiple independent datasets of answers. 

We consider the position of a researcher in posession of a `candidate probbase', which is only partially correct. We considered two VA algorithms: firstly, the \texttt{interVA} algorithm~\citep{li14}, which uses an approximation to a Naive Bayes assumption, and satisfies assumption~\ref{asm:va_calibration} only approximately. We also considered an `oracle' algorithm which was aware of the correlation matrices used for simulation, and returned posterior CoD probabilities according to the correct correlation matrices and the candidate probbase. 

We evaluated the suitability of our methods to the following tasks:
\begin{enumerate}
\item To differentiate an incorrect candidate probbase from a correct probbase,
\item To recover unknown true values of a small number of probbase entries, given an otherwise correct candidate probbase
\item Given a candidate probbase with various errors (as compared to values in $q^{\text{sim}}$) identify values with large errors. 
\end{enumerate}
In all cases, we presumed that the candidate probbase was roughly correct. We briefly considered the potential of recovering the entire unknonw probbase, but found that this was essentially untenable given reasonable dataset sizes ($\leq 10^5$ answer sets). 

We give full specifications for our simulations in the Supplementary Material. Code for our simulation is available in the 
GitHub repository at~\texttt{https://github.com/jamesliley/VA_imputation}

\subsection{Differentiation of correct and incorrect probbases}
\label{sec:sim_pertubation}

We repeatedly simulated datasets of various sizes. We considered several degrees of pertubation of probbases, and considered the proportion of perturbed probbases for which the estimated objective function function (out of viability, coherence, imputation error and two- or three- way agreement) returned a `better' (e.g., lower) value when evaluated on the perturbed probbase than when on the true probbase. 

We first considered heavily perturbed probbases: either random  that is, $\hat{q}_{kj} \iidsim U(0,1)$, or with all elements perturbed about their true values by about 1\%. 
In all cases, for all dataset sizes, for both VA methods, no test function returned a lower score for a random probbase than for the real probbase, with the exception of coherence, for which a lower score was returned in fewer than 1 in 1000 cases. For all test functions except coherence, essentially no random probbase had a lower score than the real probbase, with some exceptions at dataset sizes $\leq 1000$ (see Supplementary Figure~\ref{supp_fig:pertubation_moderate}).

\begin{figure}[H]
\centering
    \begin{subfigure}[t]{0.45\textwidth}
    \includegraphics[width=\textwidth]{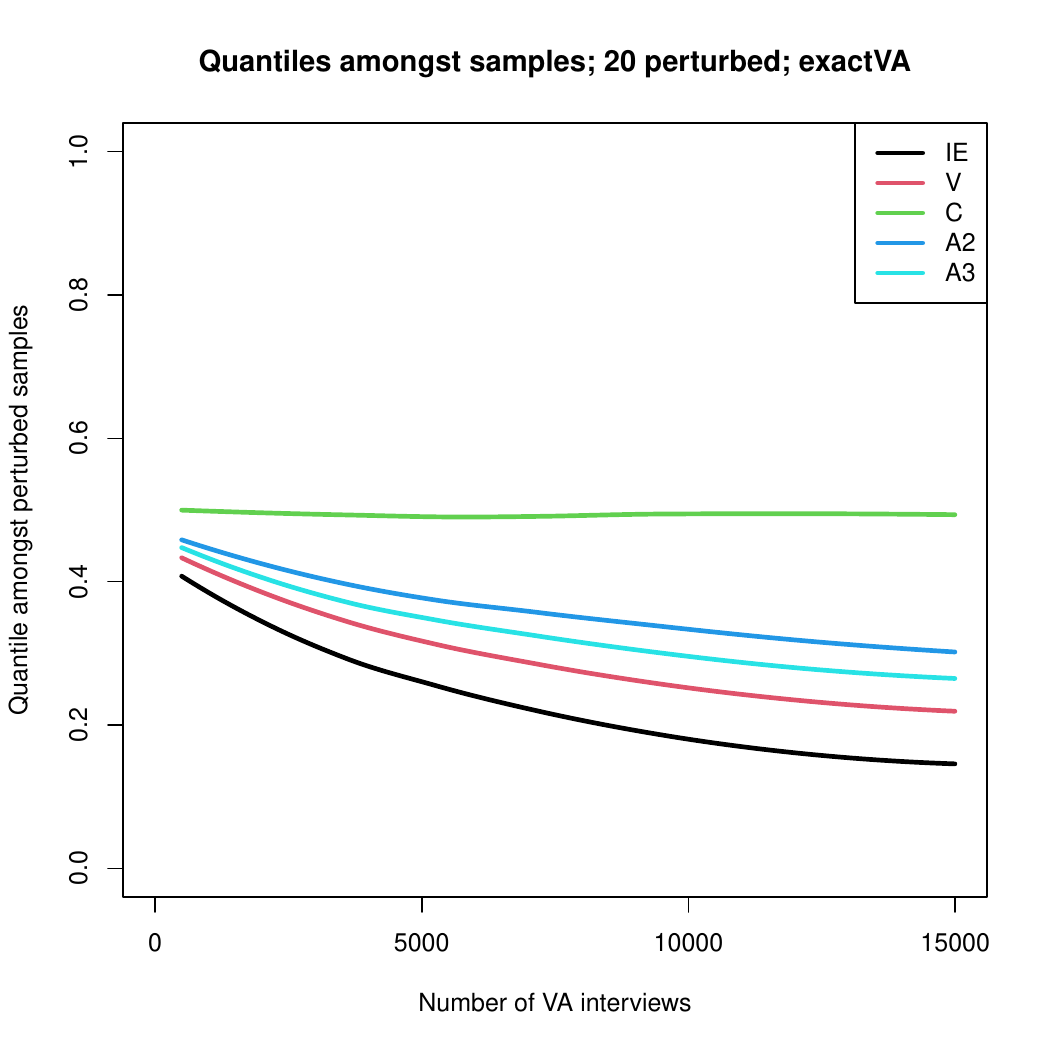}
    \end{subfigure}
    \begin{subfigure}[t]{0.45\textwidth}
    \includegraphics[width=\textwidth]{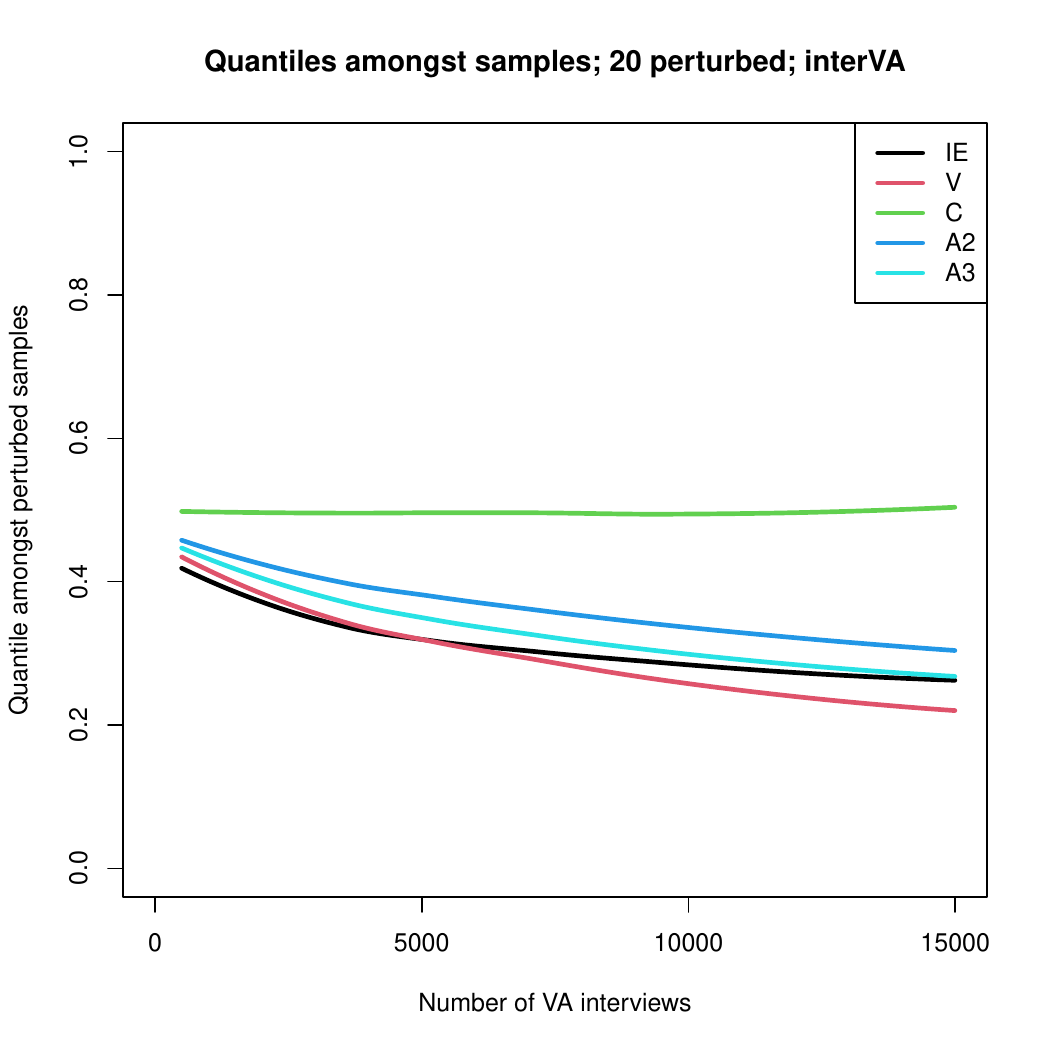}
\end{subfigure}
\caption{Differentiation between true probbase and slightly perturbed probbases ($N(0,(1/100)^2)$ noise added to 20 uniformly-randomly-chosen elements) using estimated objective functions (Imputation error, IE: $\hat{I}$, viability, V: $\hat{N}$, coherence, C: $\hat{C}$, two-way agreement, A2: $\hat{R}_2$ and three-way agreement, A3: $\hat{R}_3$). Leftmost plot shows computations for oracle VA algorithm; rightmost plot for InterVA algorithm. Lines show estimated mean (LOESS) quantile of test function evaluated on true probbase amongst test function evaluated on perturbed probbases, when using a database with total number of samples given by the value on the x-axis. Pointwise standard errors are less than the widths of the lines.}
\label{fig:pertubation}
\end{figure}

We then considered `slightly perturbed' probbases, for which we perturbed 20 uniformly-randomly chosen probbase values by about 1\%. Test function values evaluated on the true probbase were between the first and fifth decile of those evaluated on slightly perturbed probbases. The empirically best-performing test function was imputation error when we used an exact VA algorithm, and either viability or imputation error when we used the InterVA algorithm (Figure~\ref{fig:pertubation}). Performance of both test functions was better than two- or three- way agreement, and coherence was essentially unable to distinguish the true probbase from perturbed probbases at this level of pertubation. 

Encouragingly, performance was not empirically highly contingent on assumption~\ref{asm:va_calibration}, with essentially similar results for the oracle VA algorithm (which satisfies assumption~\ref{asm:va_calibration}) and the InterVA algorithm (which does not). Findings were essentially identical when using simulations for which assumption~\ref{asm:cond_indep} was violated (Supplementary Figure~\ref{supp_fig:pertubation_noci}).

\subsection{Recovery of unknown probbase entries}
\label{sec:sim_reconstruction}

We next assessed whether minimisation of estimated objective functions could be used to fix incorrect probbase elements. For each simulation, we chose between two and 20 elements of the probbase uniformly at random, considered their values unknown, and found the values for these elements in $[0,1]$ which minimised each estimated objective function. We then compared these discovered values with the true values of the probbase, and considered the mean absolute error. 

\begin{figure}[H]
\centering
    \begin{subfigure}[t]{0.45\textwidth}
    \includegraphics[width=\textwidth]{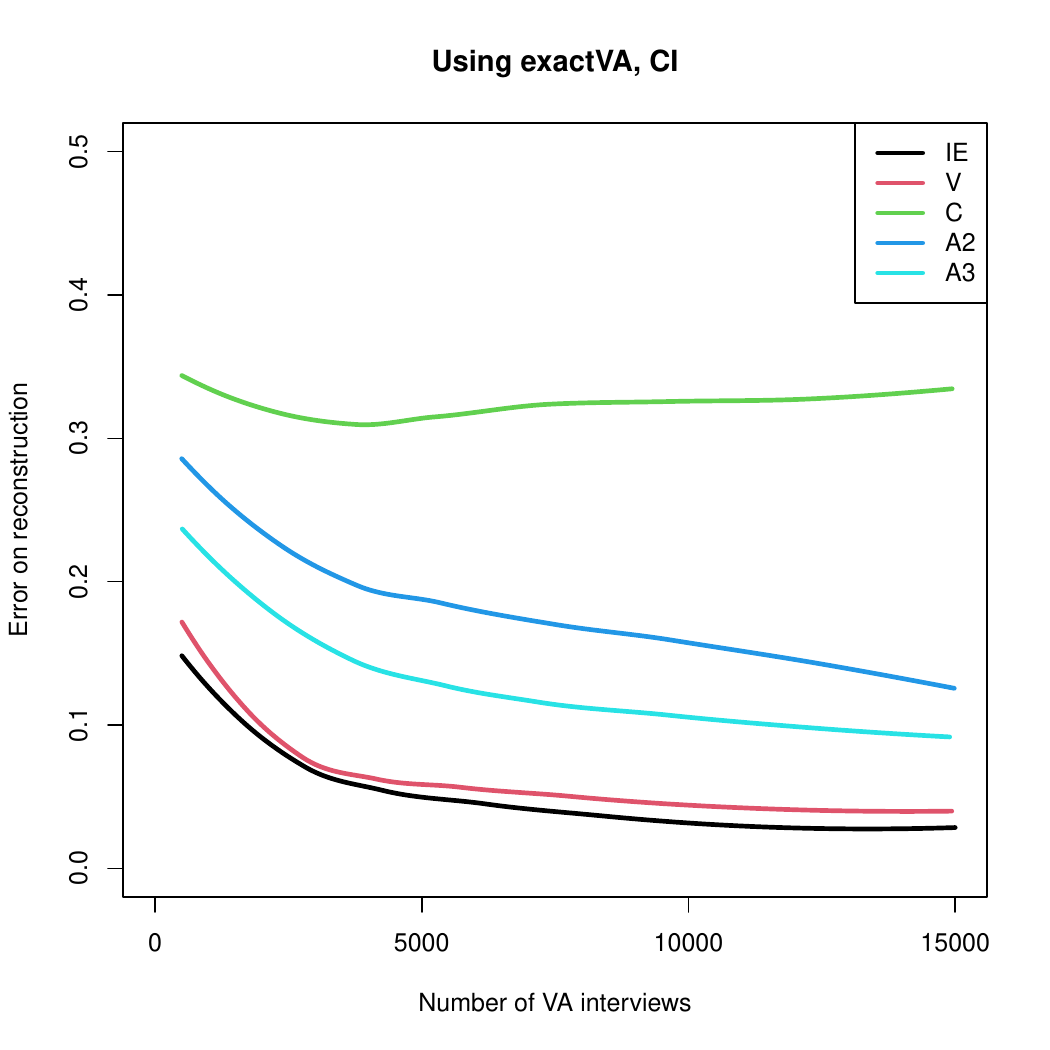}
    \end{subfigure}
    \begin{subfigure}[t]{0.45\textwidth}
    \includegraphics[width=\textwidth]{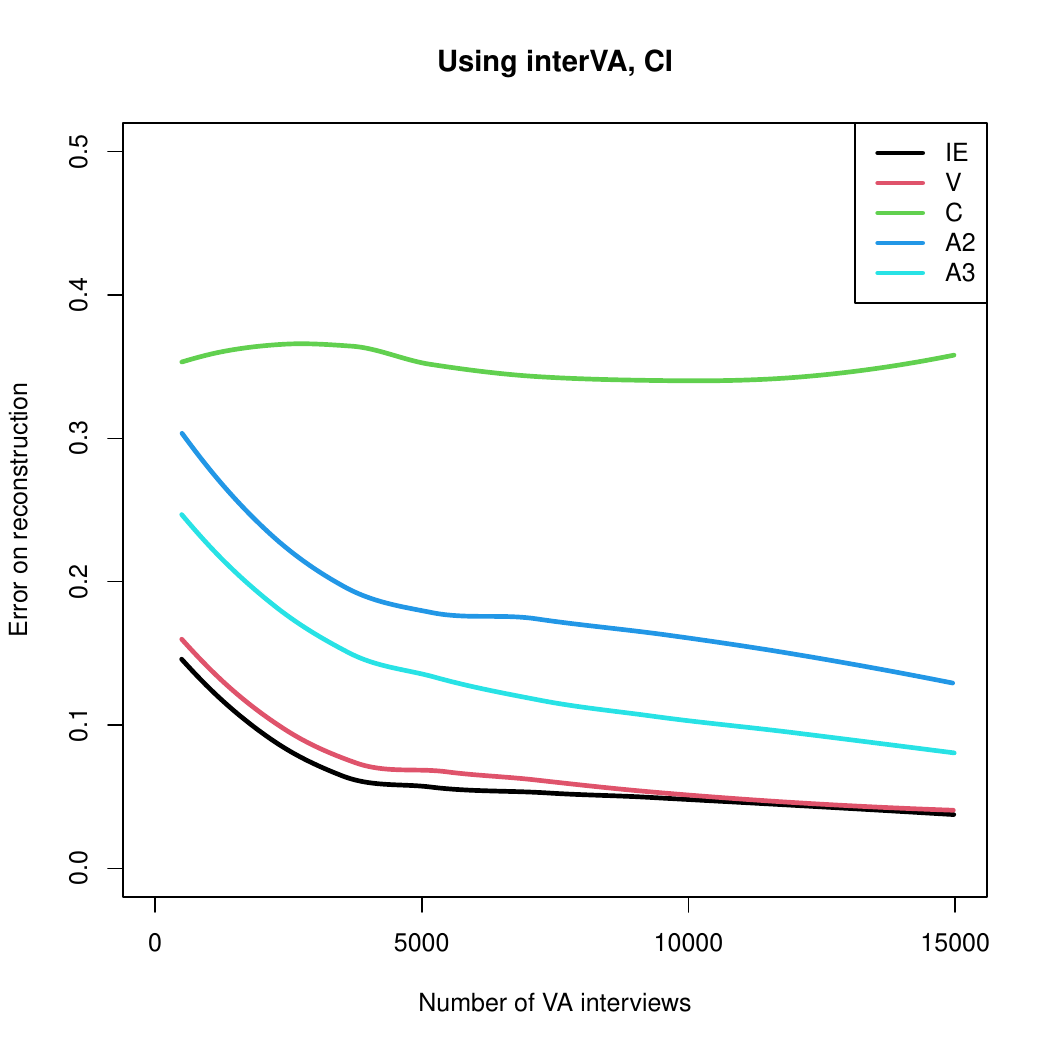}
\end{subfigure}
\caption{Reconstruction of 2-20 missing probbase values by minimisation of estimated objective functions (Imputation error, IE: $\hat{I}$, viability, V: $\hat{N}$, coherence, C: $\hat{C}$, two-way agreement, A2: $\hat{R}_2$ and three-way agreement, A3: $\hat{R}_3$). The leftmost plot shows computations for an oracle VA algorithm; the rightmost plot for the InterVA algorithm. Lines show mean absolute difference between recovered values and true values when using a database with total number of samples given by the value on the x-axis. Pointwise standard errors are less than the widths of the lines.}
\label{fig:reconstruction}
\end{figure}

When at least 2000 samples were available, it was possible to reconstruct missing values with absolute error $<10\%$ using viability or imputation error, when using either the oracle VA or InterVA. Other methods were variably successful, but we could not generally reconstruct missing values using coherence (Figure~\ref{fig:reconstruction}). Results were unchanged when using simulations for which assumption~\ref{asm:cond_indep} was violated (Supplementary Figures~\ref{supp_fig:reconstruction_noci}). 

\subsection{Identification of perturbed probbase entries}
\label{sec:sim_search}

Finally, we evaluated whether incorrect probbase entries could be discovered. We randomly chose one-fifth of probbase entries and perturbed them by about 10\%. For each test function, we attempted to detect the perturbed elements by finding the alternative value of each probbase element which minimised the value of the estimated objective function, and considering the distance between this alternative value and the given probbase value. 

\begin{figure}[H]
\centering
    \begin{subfigure}[t]{0.45\textwidth}
    \includegraphics[width=\textwidth]{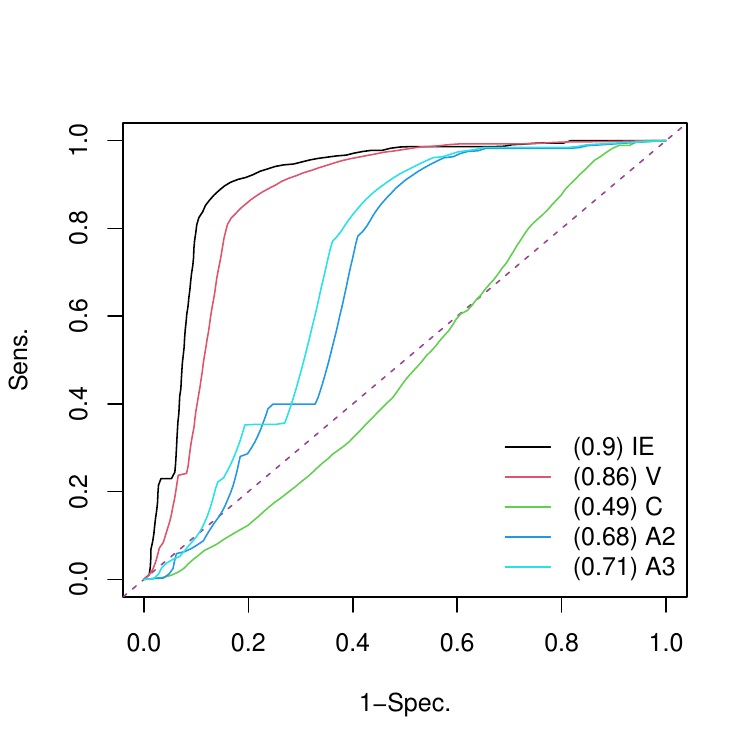}
    \end{subfigure}
    \begin{subfigure}[t]{0.45\textwidth}
    \includegraphics[width=\textwidth]{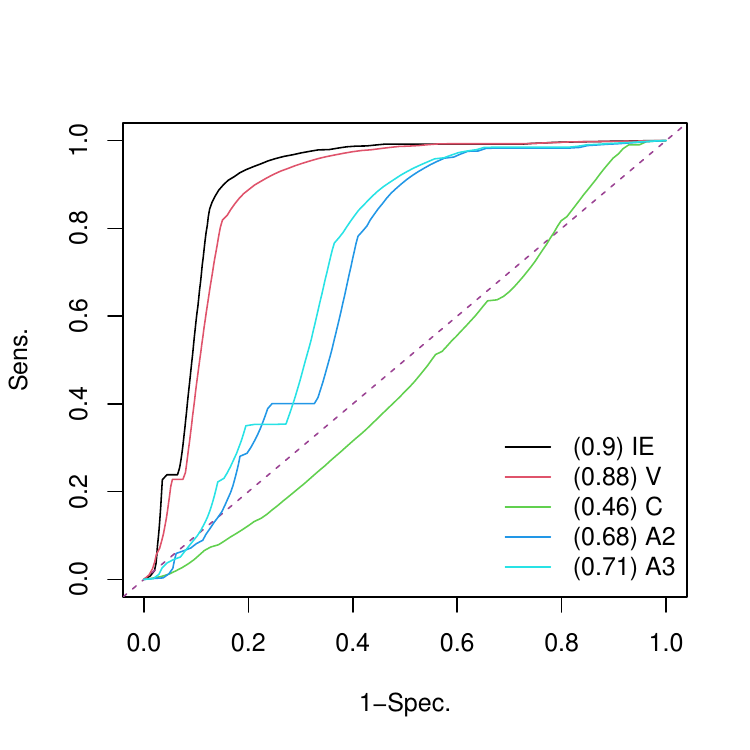}
\end{subfigure}
\caption{Search for probbase elements perturbed by at least 10\% (a randomly-chosen 20\% of elements). Curves show ROC plots using test functions (Imputation error, IE: $\hat{I}$, viability, V: $\hat{N}$, coherence, C: $\hat{C}$, two-way agreement, A2: $\hat{R}_2$ and three-way agreement, A3: $\hat{R}_3$). The leftmost plot shows computations for an oracle VA algorithm; the rightmost plot for the InterVA algorithm. Bracked values in legend show AUROC. }
\label{fig:search}
\end{figure}

We evaluated the predictive ability to detect a pertubation of at least 10\%, evaluating ability using ROC curves. We found that imputation error and viability could detect perturbed elements accurately, with AUROC$\approx 0.9$ for both the oracle VA and InterVA; two- or three- way agreement moderately accurately, with AUROC$\approx 0.7$ for both oracle VA and InterVA; and coherence could not detect perturbed elements better-than-randomly, with AUROC$\approx 0.5$ (Figure~\ref{fig:search}). Results were the same when using simulations for which assumption~\ref{asm:cond_indep} was violated (Supplementary Figures~\ref{supp_fig:search_noci})

\section{Discussion}
\label{sec:discussion}


\subsection{Summary of findings}

We propose a suite of methods to learn posterior probability assignments for verbal autopsy, or `VA algorithm', taken as parametrised functions of ternary strings. 
In general, we focus on finding `probbase' values, which is generally what is needed to calibrate VA algorithms. 
Our `coherence' method relies only on the VA algorithm without any assumptions on its parametrisation, and could, with enough data, be used to calibrate a VA algorithm agnostic of the probbase values. 
Our `two-way/three-way agreement' methods, by contrast, concern only the probbase and make no use of the VA algorithm. 
Our `viability' and `imputation error' methods use both the VA algorithm and the probbase.

We generally assume that questions may be partitioned into blocks which are conditionally independent given cause of death. Typically, we need at least three blocks for the probbase or VA assignments to be identifiable for our methods, and each block must be of size at least the number of causes of death. Two blocks may be sufficient if the estimation of the probbase is otherwise constrained; for instance, if the probbase is already known to be partially correct; but in general adversarial examples can be constructed in which the objective function is minimised for an incorrect probbase. Our `viability' method does not require an assumption of conditional independence blocks, but the probbase is not identifiable from minimisation of this objective function. 

We roughly summarise our methods in table~\ref{tab:summary}. In practice, for evaluation of a probbase or VA algorithm, a combination of all methods could be used, since they have a range of advantages and disadvantages.

\begin{table}
\begin{tabular}{l|llllllll}
Obj.  & Asm.~\ref{asm:cond_indep} & Min. $q$ & Min. $V$ &  Iden. $q$ & Iden. $V$ & Prac. \\ \hline
$N(\hat{q},\hat{V})$ (Via.) & No & Yes & Yes & No & No & High \\ 
$C(\hat{V})$ (Coh.) & Yes &  No & Yes$^1$ & No & Yes & Low \\ 
$I(\hat{q},\hat{V})$ (Imp. er.) & Yes & Yes & Yes & Yes$^2$ & Yes$^2$ & High \\ 
$R_2(\hat{q})$ (2-way agr.) & Yes & Yes &  No & No & No & Med. \\ 
$R_3(\hat{q})$ (3-way agr.) & Yes & Yes & No & Yes & No & Med. \\ 
\end{tabular}
\caption{Performance of objective functions (viability, coherence, imputation error, two- and three- way agreement). Column `Asm.~\ref{asm:cond_indep}' indicates whether assumption~\ref{asm:cond_indep} is required for use.  Column `Min. $q$' indicates whether the objective function is minimised when $\hat{q}=q$. Column `Min. $V$ indicates whether the objective function is minimised when assumption~\ref{asm:va_calibration} holds. Columns `Iden. $q$' and `Iden. $V$' indicate whether the objective is minimised \emph{only} if $\hat{q}=q$ or $\hat{V}=V$ respectively. Column `Prac.' is an overview of practical usefulness (high, medium, or low), which largely depends on convergence rate of the estimator. $^1$: in a stronger sense; see Theorem~\ref{thm:coherenceIdentifiability}. $^2$: in a weak sense; see Theorem~\ref{thm:imputationIdentifiability}. }
\label{tab:summary}
\end{table}[H]

In terms of practical performance, the best of our objective functions were imputation error and viability. Whilst both viability and two-way agreement lack identifiability properties, this did not have severe effect in practice. Coherence was largely impractical, except on very perturbed probbases, which was likely due to the very slow rate of convergence of the estimated objective function. The coherence estimator compares observed and expected distributions of length-$s$ strings, and for realistically large $s$, the observed distribution is too coarse to be useful, in that observed strings are typically unique and have very low marginal probability. 

Our assumption~\ref{asm:va_calibration} does not hold for the InterVA method and would not be expected to hold for practical VA methods in general. However, our methods seem reasonably practically robust to this assumption, in that performance in all three practical tasks in section~\ref{sec:evaluation} was essentially the same between InterVA and our oracle algorithm. 

Our assumption~\ref{asm:cond_indep} of a block structure to the covariance of answer probabilities conditional on causes of death is also tenuous. This assumption may not hold in practice, in particular because some questions (in particular, demographic questions such as age and sex) violate this assumption by being essentially correlated with all answers. However, our viability method does not depend on this assumption, and our imputation error method is somewhat robust to violations of it (as per Theorem~\ref{thm:imputationRobustness}). 

Our objective functions are differentiable in their arguments, and if VA algorithms are also differentiable in their parameters, gradient-based methods can be used for optimisation. 
Given the dependence on large datasets of (assumed) independent and identically-distributed samples from $(A,D)$, there is scope for stochastic gradient methods to be used. 
We have not explored in detail whether entire probbases could be reconstructed given sufficient data, although initial attempts were not promising given $\approx 10,000$ samples, although a comprehensive multidimensional optimisation was beyond our scope and may have some success if paralellised. 
In general, however, there is little reason to need to do this, as we usually expect the probabse to be roughly correct, and methods such as ours to be useful for fine-tuning. 









\subsection{Practicalities}

Learning probbase matrices is difficult, but is a reasonably important task for VA operation~\cite{clark18}. The two dominant non-Bayesian methods - asking physicians to manually generate a probbase, or to assign CoDs to a training corpus of VA samples - are both susceptible to inter-clinician variation, or systematic errors in medical practice. Moreover, answer probabilities change across populations~\cite{clark18}, potentially substantially, meaning the process of training a probbase must either be repeated often, or a degree of miscalibration accepted. Even within a population, probbases may change with circumstances of mortality~\citep{ambruoso22}, with time, or across demographics. Our method may give a quick option for doing this, or at least for abbreviating the process, by prioritising elements of probbases to review (as in section~\ref{sec:sim_search}) and by suggesting corrected values (as in section~\ref{sec:sim_reconstruction}.

%

We consider that our method may find application for unusual, hard-to-reach populations, for which existing probbases are likely to be miscalibrated and for which CoD- labelled data is likely to be unavailable. Our method is adaptable to a range of VA algorithms and models. We consider that the assessment of VA algorithms and probbases in the absence of labelled data is an important problem, and our approach contributes to a set of methods for doing this.

\subsection{Wider scope}

Our method is somewhat similar to a maximum-likelihood estimator, in that we consider a continuous objective depending on our object of inference (the probbase). We cannot directly use a true maximum-likelihood estimator, however, as the likelihood function - the probability of observing a set of answers given a cause of death - is not directly tractable in general, particularly if the verbal autopsy algorithm is complex (e.g., as in~\citep{chu25}, which makes use of large language models). We are in an unusual setting in which the  likelihood is unspecified, but we have a given form for the posterior (the probability of a given cause of death given a set of answers), given by the VA algorithm in question. 

An important philosophy of our approach is that in this circumstance, the posterior distribution of cuases of death is much easier to work with than the likelihood. In general, medical practitioners are well accustomed to estimating posterior distributions over potential diagnoses given symptoms, and will be able to make use of probbase-like information in doing so, but are less able to meaningfully estimate the probability of answering questions in a particular way, particular more than one question. Clinical practice is reflected in design of VA algorithms: early VA algorithms, including InterVA, were developed in partnernship with clinicians, and modern algorithms, particularly when making use of large language models~\cite{chu25} reflect medical reasoning standards directly.




\section{Acknowledgements}

We thank Edward Fottrell and Lucia D'Ambruoso for their suggestions and help in motivating and preparing this manuscript. 
We also thank the community of researchers working on verbal autopsies, and all patients, families and health workers worldwide who have been part of the verbal autopsy process.

This work has made use of the Hamilton HPC Service of Durham University.

\setcounter{section}{0}
\renewcommand\thesection{S\arabic{section}}

\setcounter{figure}{0}
\renewcommand\thefigure{S\arabic{figure}}

\setcounter{thm}{0}
\setcounter{lem}{0}
\setcounter{lemma}{1}

\setcounter{section}{0}
\renewcommand\thesection{A\arabic{section}}

\section{Proofs}
\label{apx:proofs}

\subsection{Relating to section~\ref{sec:heuristics}}

We firstly restate and prove the first theorem:

\begin{theorem}
Consider a distribution of a random variable $(A,D)$, where $A$ takes values in $\{0,1,NA\}^s$ and $D$ takes values in $[r]$, taken as a point in $\Delta_{r \times 3^s}$, with $r > 1$ and $s \geq 1$. Then the conditional probabilities $\cprob{A_k=1}{D=j}$ are almost surely not identifiable from the marginal probabilities $\prob{A}$ and $\prob{D}$. 
\end{theorem}

\begin{proof}
Every such distribution can be written as a $r \times 3^s$ matrix with $(j,k)$th entry $\prob{A=c_k,D=j}$, where $c_k$ is a list of $\{0,1,NA\}$ sequences. In this case, we have knowledge of the row and column sums of the matrix, which are $\prob{D=j}$ and $\prob{A=c_k}$ respectively.

If we order $c_k$ such that all the sequences starting with 
`1' occur first, then the value $\cprob{A_1=1}{D=1}$ is the sum of the first $3^{s-1}$ elements of the first row of the matrix. 

Choose four element of the matrix in a grid, only one of which is amongst the $3^{s-1}$ elements of the first row. Then we may freely modify these four elements whilst preserving their row and column sums; if the elements are $u_{jk}$ then we can change
\begin{equation*}
\left(\begin{matrix}
u_{11} & u_{12} \\
u_{21} & u_{22} \\
\end{matrix} \right) \to 
\left(\begin{matrix}
u_{11} + x & u_{12} -  x \\
u_{21} - x & u_{22} + x \\
\end{matrix}\right)
\end{equation*}
for any $x \in \mathbb{R}$. There are almost surely uncountably many such $x$ giving rise to matrix entries in $(0,1)$. By substituting these in the matrix above, we attain new matrices with the same row and column sums as the original matrix, but different values of $\cprob{A_1=1}{D=1}$. We thus cannot identify the value $\cprob{A_1=1}{D=1}$ uniquely from only $\prob{A}$ and $\prob{D}$.

\end{proof}

We note that this result is not contingent on values of $A$ taking the value $NA$; we can restrict to $A$ only taking values in $\{0,1\}$, but we do require the additional condition that $2^s r \geq 6$ in this case. 

Suppose, as per the assumption, that we may divide $[s]$ into mutually exclusive blocks $B_1, B_2 \dots B_b$. Then, since we have:
\begin{equation*}
\prob{A=a,D=d}=\cprob{A=a}{D=d}\prob{D=d}=\prob{D=d}\prod_{ell} \cprob{A_{B_{\ell}}=a_{B_{\ell}}}{D=d}
\end{equation*}
the full distribution of $(A,D)$ is determined by the values $\cprob{A_{B_{\ell}}=a_{B_{\ell}}}{D=d}$. There are $d\left(\sum_{\ell}3^{|B_{\ell}|}-1\right)$ free parameters amongst these values, and $3^s + 3^d - 2$ constraints from knowledge of $\prob{A}$ and $\prob{D}$, so in general there will be at most finitely many distributions of $(A,D)$ with the given block structure and marginals $\prob{A}$ and $\prob{D}$.

\subsection{Relating to identifiability}

\viabilityUse

\begin{proof}
See main text. 
\end{proof}

\coherenceIdentifiability

\begin{proof}
This essentially follows directly from~\cite[Theorem 4a]{kruskal77}, which asserts that a product representation of this type is unique up to scaling and order permutation if matrices are of a sufficient rank. Assemble the values $\prob{A_{B_1}=a_{B_1},A_{B_2}=a_{B_2},A_{B_3}=a_{B_3}}$ into a rank-3 tensor $T$ of dimension $3^{|B_1|} \times 3^{|B_2|} \times 3^{|B_3|}$. Assemble the values $\pi_j^{-\frac{2}{3}}V_j(a_{B_{\ell}})p(a_{B_{\ell}})$ into $3^{|B_{\ell}|} \times r$ matrices $v^{\ell}$, and the $\pi_j^{-\frac{2}{3}}\hat{V}_j(a_{B_{\ell}})p(a_{B_{\ell}})$ into $3^{|B_{\ell}|} \times r$ matrices $\hat{v}^{\ell}$. Then we have:
\begin{equation*}
\sum_j v^1_{k_1j}v^2_{k_2j}v^3_{k_3j}=T_{k_1k_2k_3}=\sum_j \hat{v}^1_{k_1j}\hat{v}^2_{k_2j}\hat{v}^3_{k_3j}
\end{equation*}
Since we have written $T_{k_1k_2k_3}$ as a product of three terms in this way, its rank $R$ (in the sense of~\cite{kruskal77}) is at most $r$. For almost all distributions of $(A,D)$ admitting the block structure $B_1,B_2,B_3$, the matrices $v^{\ell}$ are full-rank; that is, have rank $r$. We then have
\begin{equation*}
2R+2 \leq 2r + 2 \leq 3r = \text{rank}(v^1) + \text{rank}(v^2) + \text{rank}(v^3)
\end{equation*}
so by~\cite[Theorem 4a]{kruskal77} we attain that $\hat{v}^{\ell}=v^{\ell}I_{\ell}$, for diagonal matrices $I_{\ell}$ with $I_1I_2I_3=I$ (disregarding permutation of $\pi_j$ values). But from the constraints that $\sum_j V_j(a_{B_{\ell}})=\sum_j \hat{V}_j(a_{B_{\ell}})=1$, we must have $\hat{v}^{\ell}=v^{\ell}$, and hence $\hat{V}(a_{B_{\ell}})=V(a_{B_{\ell}})$ for all $a_{B_{\ell}}$. 

\end{proof}

\minimalCorrect

\begin{proof}

We firstly establish that, if some VA algorithm $\tilde{V}$ and probbase $\tilde{q}$ is such that for all $k \in [s]$ and $a \in \{0,1,NA\}^s$ with $p(a_{-B_{\ell}})>0$, we have
\begin{equation*}
F_k\left[\tilde{V}(\alpha_{-B(k)},\tilde{q}),\tilde{q}\right] = \cprob{A_k=1}{A_{-B(k)}=\alpha_{-B(k)}}
\end{equation*}
then we also have 
\begin{equation}
(\tilde{V},\tilde{q}) \in \arg \min_{\hat{V},\hat{q}} I(\hat{q},\hat{V}) \nonumber
\end{equation}
Indeed, we have
\begin{align*}
I(\hat{q}) &= \EE{K \sim U([s]),A}{\loss{F_K\left[V\left(A_{-B(k)},\hat{q}\right),\hat{q}\right]}{\ind{A_K=1}}} \\
&= \frac{1}{s}\sum_{a \in \mathbb{A}} \sum_{k=1}^s \prob{A=a} \loss{F_k\left[V\left(a_{-B(k)},\hat{q}\right),\hat{q}\right]}{\ind{a_k=1}} \\
&= \frac{1}{s}\sum_{\ell=1}^b \sum_{k \in B_{\ell}} \sum_{a \in \mathbb{A}} \prob{A_{B_{\ell}}=a_{B_{\ell}},A_{-B_{\ell}}=a_{-B_{\ell}}} \loss{F_k\left[V\left(a_{-B_{\ell}},\hat{q}\right),\hat{q}\right]}{\ind{a_k=1}} \\
&= -\frac{1}{s}\sum_{\ell=1}^b \sum_{k \in B_{\ell}} \sum_{a \in \mathbb{A}} \prob{A_{B_{\ell}}=a_{B_{\ell}},A_{-B_{\ell}}=a_{-B_{\ell}}} \left[ \ind{a_k=1} \log \left( F_k\left[V\left(a_{-B_{\ell}},\hat{q}\right),\hat{q}\right]\right)  + \right. \\
&\phantom{= \frac{1}{s} \sum \sum \sum \prob{A_{\ell}=\alpha_{\ell},A_{-\ell}=\alpha_{-\ell}}} \left.\ind{a_k \neq 1} \log \left( 1- F_k\left[V\left(a_{-B_{\ell}},\hat{q}\right),\hat{q}\right]\right)  \right] \\
&= -\frac{1}{s}\sum_{\ell=1}^b \sum_{k \in B_{\ell}} \sum_{a_{-B_{\ell}} \in \mathbb{A}_{-B_{\ell}}} \left[\prob{A_{-B_{\ell}}=a_{B_{\ell}},A_{k}=1} \log \left( F_k\left[V\left(a_{-B_{\ell}},\hat{q}\right),\hat{q}\right] \right)  + \right. \\
&\phantom{= \frac{1}{s} \sum \sum \sum } \left. \prob{A_{-B_{\ell}}=a_{-B_{\ell}},A_{k} \neq 1}  \log\left(1-F_k\left[V\left(a_{-B_{\ell}},\hat{q}\right),\hat{q}\right]\right) \right] 
\end{align*}
where in the final step we sum over all sequences of values in block ${\ell}$ except $k$, recognising that $\loss{F_k\left[V\left(a_{-B_{\ell}},\hat{q}\right),\hat{q}\right]}{\ind{a_k=1}}$ does not depend on these values. We have:
\begin{align*}
\prob{A_{-B_{\ell}}=a_{-B_{\ell}},A_k=1} &\log\left[F_k\left[V\left(a_{-B_{\ell}},\hat{q}\right),\hat{q}\right]\right]  + \prob{A_{-B_{\ell}}=a_{-B_{\ell}},A_k \neq 1} \log\left[1-F_k\left[V\left(a_{-B_{\ell}},\hat{q}\right),\hat{q}\right]\right] \\
&= p(a_{-B_{\ell}})\left( \cprob{A_k=1}{A_{-B_{\ell}}=a_{-B_{\ell}}} \log\left[F_k\left[V\left(a_{-B_{\ell}},\hat{q}\right),\hat{q}\right]\right]   \right. \\
&\phantom{= p(a_{-B_{\ell}})(}\left. + \left(1-\cprob{A_k = 1}{A_{-B_{\ell}}=a_{-B_{\ell}}}\right) \log\left[1-F_k\left[V\left(a_{-B_{\ell}},\hat{q}\right),\hat{q}\right]\right]\right) \\
&:= p(a_{-B_{\ell}})\left( p_k \log(r_k) + (1-p_k) \log(1-r_k) \right) 
\end{align*}
where $p_k:=\cprob{A_k=1}{A_{-B_{\ell}}=a_{-B_{\ell}}}$ and $r_k=F_k\left[V\left(A_{-B_{\ell}},\hat{q}\right),\hat{q}\right]$. This is is maximised when $p_k=r_k$, so each term in the expansion of $I(\hat{q})$ is individually maximised when $F_k\left[V\left(A_{-B_{\ell}},\hat{q}\right)\right]=\cprob{A_k=1}{A_{-B_{\ell}}=a_{-B_{\ell}}}$ for all $a_{-B_{\ell}}$ for which $p(a_{-B_{\ell}})>0$.

Given this we now need only show that under the conditions of the theorem, the minimum is achieved. If $\hat{q}_{kj}=q_{kj}=\cprob{A_k=1}{D=d_j}$ then, as in expansion~\eqref{eq:main_expansion}:
\begin{align*}
F_k\left[V\left(a_{-B_{\ell}},\hat{q}\right),\hat{q}\right] &= \sum_{j \in [r]}V\left(a_{-B_{\ell}},\hat{q}\right)_j\hat{q}_{kj} \\
&= \sum_{j \in [r]} \cprob{D=d_j}{A_{-B_{\ell}}=a_{-B_{\ell}}}\cprob{A_k=1}{D=d_j} &&\text{(assump.~\ref{asm:va_calibration})} \\
&= \sum_{j \in [r]} \cprob{D=d_j}{A_{-B_{\ell}}=a_{-B_{\ell}}}\cprob{A_k=1}{D=d_j,A_{-B_{\ell}}=a_{-B_{\ell}}} &&\text{(assump.~\ref{asm:cond_indep})}\\
&= \sum_{j \in [r]}\cprob{A_k=1,D=d_j}{A_{-B_{\ell}}=a_{-B_{\ell}}} \\
&= \prob{A_k=1}{A_{-B_{\ell}}=a_{-B_{\ell}}} 
\end{align*}
so this minimum is achieved when $\hat{q}=q$.  

\end{proof}

\imputationIdentifiability

\begin{proof}

It is helpful to define matrices depending on a particular distribution $\delta$ of $(A,D)$. For $\ell \in \{1,2,3\}$, let $a_{\ell}^1, a_{\ell}^2, \dots a_{\ell}^{3^{|B_{\ell}|}}$ denote an enumeration of $\{0,1,NA\}$ sequences of length $|B_{\ell}|$. We define:
\begin{itemize}
\item $q_{\delta} \in \mathbb{R}^{s \times r}$ as the matrix with $(k,j)$th entry $\cpprob{\delta}{A_k=1}{D=d_j}$.
\item $V_{\delta}^{\ell} \in \mathbb{R}^{3^{|B_{\ell}|} \times r}$ as the matrix with $(k,j)th$ entry $\cpprob{\delta}{D=j}{A_{B_{\ell}} = a_{\ell}^k}$.
\item $M_{\ell}^{\delta} \in \mathbb{R}^{r \times 3^{B_{\ell}}}$ as a matrix with $(j,k)$th entry $\pprob{\delta}{A_{B_{\ell}}=a_{\ell}^k,D=j}$,
\item $R_{\ell}^{\delta} \in \mathbb{R}^{r \times |B_{\ell}|}$ as a matrix with $(j,k)$th entry $\pprob{\delta}{A_{k}=1,D=j}$
\item $q_{\ell}^{\delta} \in \mathbb{R}^{r \times |B_{\ell}|}$ as a matrix with $(j,k)$th entry $\cpprob{\delta}{A_{k}=1}{D=j}$
\end{itemize}
When the distribution $\delta$ is the true distribution of $(A,D)$, we omit the superscript. 
We note that $\hat{q}=q$ implies that $I^{1,2}(\hat{q},V)$, $I^{1,3}(\hat{q},V)$, and $I^{2,3}(\hat{q},V)$ simultaneously achieve their minimum by theorem~\ref{thm:minimalCorrect}. We firstly consider only two blocks $B_1$ and $B_2$. 
The condition
\begin{equation*}
\tilde{q}=\argmin_{\hat{q} \in \mathcal{Q}} I^{1,2}(\hat{q},V) 
\end{equation*}
holds if and only if
\begin{align}
\cprob{A_k=1}{A_{-B_2}=a_{-B_2}} &= \sum_j V\left(a_{-B_2},\tilde{q}\right) \tilde{q}_{kj} &&\text{for $k \in B_1$} \nonumber \\
\cprob{A_k=1}{A_{-B_1}=a_{-B_1}} &= \sum_j V\left(a_{-B_1},\tilde{q}\right) \tilde{q}_{kj} &&\text{for $k \in B_2$} \label{eq:opt_condition}
\end{align}
for any $a$ such that $p(a_{-B_1})>0$ or $p(a_{-B_2})>0$, which is almost surely all $a$. Define 
\begin{itemize}
\item $Y_{12} \in \mathbb{R}^{3^{|B_2|}\times |B_1|}$ with $(k_2,k_1)$th entry $\cprob{A_{B_1[k_1]}=1}{A_{-B_2}=\alpha_{2}^{k_2}}$
\item $Y_{21} \in \mathbb{R}^{3^{|B_1|}\times |B_2|}$ with $(k_1,k_2)$th entry $\cprob{A_{B_2[k_2]}=1}{A_{-B_1}=\alpha_{1}^{k_1}}$
\end{itemize}
Now condition~\eqref{eq:opt_condition} is equivalent to the pair of conditions
\begin{align*}
V_2^{\delta} q_1^{\delta} &= Y_{12} = V_2 q_1 \\
V_1^{\delta} q_2^{\delta} &= Y_{21} = V_2 q_1
\end{align*}
holding for some $\delta$ such that there exists a $\tilde{q}\in \mathcal{Q}$ for which $\tilde{q}$ and $V(\cdot,\tilde{q})$ correspond to the distribution $\delta$, in the sense of definition~\ref{def:partialCalibration}. Indeed, $q_1^{\delta}$ and $q_2^{\delta}$ are submatrices of $\tilde{q}$. 

Since $|B_{\ell}|> r$, the matrices $V_1,V_2,q_1,q_2$, $V_1^{\delta},V_2,q_1,q_2$ all have rank $r$ by assumption. Since the column space of $Y_{12}$ is spanned by both $V_2^{\delta}$ and $V_2$, we must have 
\begin{align*}
V_2^{\delta}=V_2 A &\hspace{30pt} q_1^{\delta}=A^{-1} q_1 \\
V_1^{\delta}=V_1 B &\hspace{30pt} q_2^{\delta}=B^{-1} q_2 
\end{align*}
for some invertible matrices $A,B \in \mathbb{R}^{r \times r}$. Since the row sums of both $q_1, q_2, q_1^{\delta}, q_2^{\delta}$ must all be 1, we must also have $A\mathbf{1}=\mathbf{1}$ and $B\mathbf{1}=\mathbf{1}$.

We can now write $R_1^{\delta}$ in two ways. Firstly, we may attain $R_1^{\delta}$ from $q_1^{\delta}$ by accounting for the probabilities $\pi$; secondly, we may attain $R_1^{\delta}$ from $M_1^{\delta}$ as $R_1^{\delta}=M_1^{\delta}G_1$. Moreover, since $M_{\ell}^{\delta} = (V_{\ell}^{\delta})^T I_{\ell}$, we have $M_{1}^{\delta} = B^T M_1$ and $M_2^{\delta}=A^T M_2$. Thus:
\begin{align*}
R_1^{\delta} &= I(\pi) q_1^{\delta} = I(\pi) A^{-1} q_1 = I(\pi) A^{-1} I(\pi)^{-1} I(\pi) q_1 = I(\pi) A^{-1} I(\pi)^{-1} R_1 \\
R_1^{\delta} &= M_1^{\delta} G_1 = B^T M_1 G_1 = B^T R_1
\end{align*}
and since $R_1$ is of rank $r$, we have: $B=\left(I(\pi) A^{-1} I(\pi)^{-1}\right)^T = I(\pi)^{-1} [A^{-1}]^T I(\pi)$. In order for $B\mathbf{1}=\mathbf{1}$, we require
\begin{align*}
I(\pi)^{-1} (A^{-1})^T I(\pi) &= 1 \\
\Leftrightarrow \hspace{10pt} (A^{-1})^T \bm{\pi} &= \bm{\pi}
\end{align*}
where $\bm{\pi}$ is the vector of values $\pi_j$. Since $A\mathbf{1}=\mathbf{1}$, we must thus also have $A\bm{\pi}=\bm{\pi}$. We attain that
\begin{equation*}
\left(\begin{matrix} q_1^{\delta} \\ q_2^{\delta} \end{matrix} \right) = \argmin_{\hat{q} \in \mathcal{Q}} I^{1,2}(\hat{q},V) \Leftrightarrow \left(\begin{matrix} q_1^{\delta} \\ q_2^{\delta} \end{matrix} \right)=\left(\begin{matrix} A \\ B \end{matrix}\right) \left(\begin{matrix} q_1 \\ q_2 \end{matrix} \right) 
\end{equation*}
for some invertible matrix $A$ with eigenvectors $\pi$ and $\mathbf{1}$ corresponding to unit eigenvalues, and $B=I(\pi)^{-1} [A^{-1}]^T I(\pi)$.

We now move to the three-block case. Applying the two-block case to the three pairs of blocks $(B_1,B_2)$, $(B_2,B_3)$ and $(B_1,B_3)$, we attain that 
\begin{equation*}
\tilde{q}= \left(\begin{matrix} q_1^{\delta} \\ q_2^{\delta} \\ q_3^{\delta} \end{matrix} \right) = \argmin_{\hat{q} \in \mathcal{Q}} \left[I^{1,2}(\hat{q},V)+I^{2,3}(\hat{q},V)+I^{1,3}(\hat{q},V)\right] \Leftrightarrow \tilde{q}=\left(\begin{matrix} A \\ B \\C \end{matrix}\right) \left(\begin{matrix} q_1 \\ q_2 \\ q_3 \end{matrix} \right) = \left(\begin{matrix} A \\ B \\C \end{matrix}\right) q
\end{equation*}
where $A,B,C$ are invertible, and $B=I(\pi)^{-1} [A^{-1}]^T I(\pi)$, $C=I(\pi)^{-1} [B^{-1}]^T I(\pi)$, and $A=I(\pi)^{-1} [C^{-1}]^T I(\pi)$. Substituting the first identity in the second gives:
\begin{equation*}
C=I(\pi)^{-1} [B^{-1}]^T I(\pi) = I(\pi)^{-1} [[I(\pi)^{-1} [A^{-1}]^T I(\pi)]^{-1}]^T I(\pi) = A
\end{equation*}
and since now we have $I(\pi)A = A^{-1} I(\pi)$, we must have that $A$, $B$, and $C$ are all identity matrices, and hence $\tilde{q}=q$. 

 \end{proof}

\agreementIdentifiability

\begin{proof}
For $R_3$, we again appeal to~\cite[Theorem 4a]{kruskal77}, which asserts that a product representation of this type is unique up to scaling and order permutation if matrices are of a sufficient rank. 

As above, denote $P_3$ as the tensor with $(k_1,k_2,k_3)$th element $\prob{A_{k_1}=k_1,A_{k_2}=k_2,A_{k_3}=k_3}$. Given blocks $\ell_1,\ell_2,\ell_3$, let $P_3^{123}$ be the subtensor of $P_3$ containing elements $\{P_3\}_{k_1k_2k_3}$ with $k_1 \in B_1,k_2 \in B_2$, and $k_3 \in B_3$. Then
\begin{equation*}
(P_3^{123})_{k_1k_2k_3} = \sum_j q'_{k_1j}q'_{k_2j}q'_{k_2j}=\sum_j \tilde{q}'_{k_1j}\tilde{q}'_{k_2j}\tilde{q}'_{k_2j} 
\end{equation*}
where $q'=q I(\pi^{-\frac{1}{3}})$; that is, a diagonal matrix with entries $\{\pi_j^{-\frac{1}{3}}\}_{j \in [r]}$. For almost any distribution of $(A,D)$, the submatrices $q_{k_{\ell} \cdot}:k_{\ell} \in B_{\ell}$ are all of full Kruskal rank; that is, every matrix attained from a subset of rows of size $r$ is full rank. Since $P_3^{123}$ can be written as a product as above, its rank $R$ in the sense of~\cite{kruskal77} is at most $r$, and since:
\begin{equation*}
2R+2 \leq 2r + 2 \leq 3r = \text{rank}(q_{k_{1} \cdot}:k_{1} \in B_{1}) + \text{rank}(q_{k_{2} \cdot}:k_{2} \in B_{2}) + \text{rank}(q_{k_{3} \cdot}:k_{3} \in B_{3})
\end{equation*}
we may apply Theorem 4a to attain that:
\begin{equation*}
\tilde{q}'= q' \Lambda P
\end{equation*}
for some $r \times r$ diagonal matrix $\Lambda$ and permutation matrix $P$. But since we must have 
\begin{equation*}
q'\Lambda P I(\pi^{\frac{4}{3}}) = \tilde{q}'I(\pi^{\frac{4}{3}}) = \tilde{q} I(\pi) = q I(\pi)=q'I(\pi^{\frac{4}{3}})
\end{equation*}
then we must have $\Lambda P =I_r$, and hence $\tilde{q}=q$.

For $R_2$, as previously, let $I(\pi)$ be a diagonal matrix with entries $\pi_j=\prob{D=d_j}$. Consider the matrix $q\, I(\pi) q^T$. Now $[q\, I(\pi) q^T]_{k_1k_2}$ agrees with $\prob{A_{k_1}=1,A_{k_2}=1}$ if $B(k_1) \neq B(k_2)$. 

Since $R_2(q)=0$ and $R_2$ is nonnegative, we have $R_2(\tilde{q})=0$. Without loss of generality, we assume that elements of $B_1$, $B_2$ and $B_3$ are consecutive. We decompose $\tilde{q}$ and $q$ as 
\begin{align*}
q\,I(\pi)^{\frac{1}{2}}=\left(\begin{matrix} q_1 \\ q_2 \\ q_3 \end{matrix}\right), &\hspace{15pt} \tilde{q} \,I(\pi)^{\frac{1}{2}}=\left(\begin{matrix} \tilde{q}_1 \\ \tilde{q}_2 \\ \tilde{q}_3 \end{matrix}\right), \\
\text{  so  } \left(\begin{matrix} q_1 \\ q_2 \\ q_3 \end{matrix}\right)\left(\begin{matrix} q_1^T & q_2^T & q_3^T \end{matrix}\right)=q\, I(\pi) q^T,  &\hspace{15pt} \left(\begin{matrix} \tilde{q}_1 \\ \tilde{q}_2 \\ \tilde{q}_3 \end{matrix}\right)\left(\begin{matrix} \tilde{q}_1^T & \tilde{q}_2^T & \tilde{q}_3^T \end{matrix}\right)=\tilde{q}\, I(\pi) \tilde{q}^T
\end{align*}
where $q_{\ell}$ and $\tilde{q}_{\ell}$ correspond to the elements of $B_{\ell}$. The agreement of $\prob{A_{k_1}=1,A_{k_2}=1}$ and $\sum_j \tilde{q}_{k_1 j} \tilde{q}_{k_2 j} \pi_j$, and hence $[\tilde{q} I(\pi) \tilde{q}^T]_{k_1k_2}$ and $[q I(\pi) q^T]_{k_1k_2}$, when $B(k_1) \neq B(k_2)$, is equivalent to the agreement of submatrices:
\begin{equation}
\tilde{q}_1 \tilde{q}_2^T = q_1 q_2^T \hspace{15pt} \tilde{q}_2 \tilde{q}_3^T = q_2 q_3^T \hspace{15pt} \tilde{q}_3 \tilde{q}_1^T = q_3 q_1^T \label{eq:qcond}
\end{equation}
Since matrix $q_{\ell}$ almost surely has rank $|B_{\ell}| \geq r$, we must have $q_{\ell} = \tilde{q}_{\ell}$. To see this, let $p_{\ell}$, $\tilde{p}_{\ell}$ denote some choice of $r$ rows of $q_{\ell}$, $\tilde{q}_{\ell}$ respectively (so $p_{\ell}$ and $\tilde{p}_{\ell}$ are square). Assume hereafter all square matrices are full rank, which is almost surely the case. Then the identities above for $q_{\ell}$, $\tilde{q}_{\ell}$ also hold for $p_{\ell}$, $\tilde{p}_{\ell}$ and we have:
\begin{equation*}
\tilde{p}_1 = p_1 p_2^T (\tilde{p}_2^T)^{-1} =(\tilde{p}_3^{-1}p_3 p_1^T)^T = p_1 p_3^T (\tilde{p}_3^T)^{-1} \\
\end{equation*}
so
\begin{equation*}
\tilde{p}_3^T = \left[(p_1 p_3^T)^{-1} p_1 p_2^T (\tilde{p}_2^T)^{-1}\right]^{-1} = \tilde{p}_2^T (p_2^T)^{-1} p_1^{-1}p_1 p_3^T = \tilde{p}_2^T (p_2^T)^{-1} p_3^T  \\
\end{equation*}
and
\begin{equation*}
p_2 p_3^T = \tilde{p}_2 \tilde{p}_3^T = \tilde{p}_2 \tilde{p}_2^T (p_2^T)^{-1} p_3^T \Rightarrow \tilde{p}_2 \tilde{p}_2^T = p_2 p_3^T (p_3^T)^{-1} p_2^T = p_2 p_2^T\\
\end{equation*}
and similarly for other $\tilde{p}_{\ell}$. So we must have $\tilde{p}_{\ell} = p_{\ell} U_{\ell} $, where $U_{\ell}$ is an orthogonal matrix. 

The space of $r \times r$ orthogonal matrices has dimension $\frac{r(r-1)}{2}$. If $\tilde{p}_1$ and $\tilde{p}_1'$ are $r-$ row submatrices of $q_1$ (with corresponding matrices $p_1$, $p_1'$), where the rows chosen are identical but for one, then $\tilde{p}_1=U_1 p_1$ and $\tilde{p}_1'=U_1' p_1'$. But $U_1$ is fully defined by the $r-1$ rows common to $\tilde{p}_1$ and $\tilde{p}_1'$, and hence $U_1=U_1'$.  By adjusting which $r$ rows of $q_{\ell}$, $\tilde{q}_{\ell}$ we choose, we can thus assert that $\tilde{q}_{\ell} = U_{\ell} q_{\ell}$. Moreover, from the identities in~\eqref{eq:qcond}, we have:
\begin{equation*}
q_1 q_2^T = \tilde{q}_1 \tilde{q}_2^T = q_1 U_1 U_2^T q_2^T
\end{equation*}
and similarly for $q_2 q_3^T$ and $q_3 q_1^T$, which, given that $q_1,q_2$ are full rank, implies $U_1=U_2=U_3=U$ for some orthogonal matrix $U$. Given any such matrix $U$, the matrix $\tilde{q}=q U$ satisfies $\tilde{q} \tilde{q}^T = q U U^T q^T=q q^T$ and hence is readily seen to satisfy:
\begin{equation*}
R_2(\tilde{q})=0=R_2(q)=\min_{\hat{q} \in Q} R_2(\hat{q})
\end{equation*}

\end{proof}

\subsection{Relating to consistency, unbiasedness, and convergence rate}

For brevity in establishing concentration results we first state two general lemmas:
\begin{lemma}
\label{lem:sumabsdiff}
If we have $\{a_i,b_i,\hat{a}_i,\hat{b}_i\}_{i \in [n]} \in [a,b]$, then:
\begin{equation*}
\left|\sum_i (\hat{a}_i-\hat{b}_i)^2 - \sum_i (a_i-b_i)^2\right| \leq 2(b-a) \left[\sum_i |\hat{a}_i - a_i| + \sum_i |\hat{b}_i - b_i|\right]
\end{equation*}
\end{lemma}

\begin{proof}
We have:
\begin{align*}
\left|\sum_i (\hat{a}_i-\hat{b}_i)^2 - \sum_i (a_i-b_i)^2\right| &= \left|\sum_i \left[(\hat{a}_i-\hat{b}_i)^2 - (a_i-b_i)^2\right]\right| \\
&\leq \sum_i \left|(\hat{a}_i-\hat{b}_i)^2 - (a_i-b_i)^2\right| \\
&= \sum_i \left|\hat{a}_i-\hat{b}_i + a_i-b_i\right| \left|(\hat{a}_i-\hat{b}_i) - (a_i-b_i)\right| \\
&\leq \max_i \left|\hat{a}_i-\hat{b}_i + a_i-b_i\right| \sum_i \left|(\hat{a}_i-a_i) - (\hat{b}_i-b_i)\right| \\
&\leq 2(b-a) \left[\sum_i \left|\hat{a}_i-a_i\right|  + \sum_i \left|\hat{b}_i-b_i\right|\right] 
\end{align*}
\end{proof}

\begin{lemma}
\label{lem:multinomial}
Suppose $A$ follows a multinomial distribution on $m$ outcomes $\mathbb{A}$ with entropy $h$. Given $n$ independent and identically distributed samples from $A$, let $\hat{p}(a)$ be the empirical frequency of outcome $A=a$, and $p(a)=\prob{A=a}$. Then we have, for $\epsilon>0$
\begin{enumerate}
\item $\prob{\sum_{a \in \mathbb{A}} |\hat{p}(a)-p(a)| \geq \epsilon} \leq 2^m e^{-n \frac{\epsilon^2}{2}}$,
\item $\prob{\sum_{a \in \mathbb{A}} |\hat{p}(a)-p(a)| \geq \frac{h+1}{\log(n)} + \epsilon} \leq 2^{h\frac{n}{\log(n)}} e^{-\frac{1}{2} \epsilon^2\left(n - \frac{(h+1)n}{\log(n)}\right)} + e^{-2\frac{n}{\log(n)^2}} $.
\end{enumerate}
For $\delta \in (0,1)$, with probability at least $1-\delta$ and $n>0$ we have:
\begin{equation*}
\sum_{a \in \mathbb{A}} |\hat{p}(a)-p(a)| \leq \sqrt{\frac{2}{n}\left(m \log(2) + \log\left(\frac{1}{\delta}\right)\right)}
\end{equation*}
For sufficiently large $n$ not depending on $|\mathbb{A}|$ (strictly $n$ such that $e^{-2\frac{n}{\log(n)^2}}<\frac{\delta}{2}$ and $\frac{h+1}{\log(n)} < \frac{1}{2}$) with probability at least $1-\delta$ we have
\begin{equation*}
\sum_{a \in \mathbb{A}} |\hat{p}(a)-p(a)| \leq\frac{h+1}{\log(n)} + 2\sqrt{\frac{h\log(2)}{\log(n)} + \frac{1}{n}\log\left(\frac{2}{\delta}\right)}
\end{equation*}

\end{lemma}

\begin{proof}
Statement (1) is the Bretagnolle-Huber-Carol inequality~\citep{wellner13}. 

For statement (2), choose $\eta$ with $e^{-1}>\eta>0$, and denote
\begin{equation*}
|\{a:p(a) \geq \eta\}|=m_{\eta}
\end{equation*}
Then 
\begin{equation*}
h=H(A)=-\sum_{a} p(a) \log[p(a)] \geq -\sum_{a:p(a) \geq \eta} p(a) \log[p(a)] \geq m_{\eta} \eta \log\left(\frac{1}{\eta}\right)
\end{equation*}
hence $m_{\eta} \leq \frac{h}{\eta \log\left(\frac{1}{\eta}\right)}$. Also:
\begin{equation*}
h=-\sum_{a} p(a) \log[p(a)] \geq -\sum_{a:p(a) < \eta} p(a) \log[p(a)] \geq \sum_{a:p(a) < \eta} p(a) \log\left(\frac{1}{\eta}\right)
\end{equation*}
so $p_{\eta}:=\sum_{a:p(a) < \eta} p(a) \leq \frac{h}{\log\left(\frac{1}{\eta}\right)}$. Now:
\begin{align*}
\sum_{a \in \mathbb{A}} |\hat{p}(a)-p(a)| &= \sum_{a:p(a) < \eta} |\hat{p}(a)-p(a)| + \sum_{a:p(a) \geq \eta} |\hat{p}(a)-p(a)| \\
&:= T_1 + T_2
\end{align*}
Suppose we observe $n_{\eta}$ samples amongst the values $\{a:p(a)< \eta\}$. Each such sample contributes at most $\frac{1}{n}$ to $T_1$, so $T_1 \leq \frac{n_{\eta}}{n}$. Given that there are $n-n_{\eta}$ samples amongst the values $\{a:p(a) \geq \eta\}$, the probability of a sample taking the value $a$ is $\frac{1}{1-p_{\eta}}p(a)$, so $T_2$ is the value of $\sum_{a \in \mathbb{A}'} |\hat{p}'(a)-p'(a)|$ where $\mathbb{A}'$, $\hat{p}'()$, $p'()$ are the corresponding values amongst a multinomial distribution over $n-n_{\eta}$ options $\{a:p(a)\geq \eta\}$. Hence, for any $k \in [n], \eta \in (0,1)$, and $\epsilon_1$ such that $\epsilon_1>\frac{k}{n}$
we have: 
\begin{align*}
\prob{\sum_{a \in \mathbb{A}} |\hat{p}(a)-p(a)| \geq \epsilon_1} &= \prob{T_1 + T_2 \geq \epsilon_1} \\
&\leq  \prob{T_1 + T_2 \geq \epsilon_1} \\
&\leq \cprob{T_1 + T_2 \geq \epsilon_1}{n_{\eta} \leq k}\prob{n_{\eta} \leq k} + \cprob{T_1 + T_2 \geq \epsilon_1}{n_{\eta}>k}\prob{n_{\eta}>k} \\
&\leq \cprob{T_2 \geq \epsilon_1 - \frac{k}{n}}{n_{\eta}\leq k}\prob{n_{\eta}\leq k} + \prob{n_{\eta}>k} \\
&\leq \cprob{T_2 \geq \epsilon_1 - \frac{k}{n}}{n_{\eta} = k} + \prob{n_{\eta}>k} 
\end{align*}
If $\frac{k}{n} \leq p_{\eta}$, then $\prob{n_{\eta}>k}\geq \frac{1}{2}$, so we consider $k$ with $\frac{k}{n} > p_{\eta}$. Denoting $\epsilon=\epsilon_1 - \frac{k}{n}>0$ and applying the Bretagnolle-Huber-Carol inequality to the first term and a Hoeffding bound to the second:
\begin{align*}
\prob{\sum_{a \in \mathbb{A}} |\hat{p}(a)-p(a)| \geq \epsilon + \frac{k}{n}} &\leq 2^{m_{\eta}}e^{-\frac{1}{2}(n-k)\epsilon^2}  + e^{-2n\left(\frac{k}{n} - p_{\eta}\right)^2} \\
&\leq 2^{\frac{h}{\eta \log\left(\frac{1}{\eta}\right)}}e^{-\frac{1}{2}(n-k)\epsilon^2} + e^{-2n\left(\frac{h}{\log\left(\frac{1}{\eta}\right)}-\frac{k}{n}\right)^2} 
\end{align*}
We now choose rates on $k$ and $\eta$ with $n$. For the bound on the left-hand-side to be meaningful, we require $\frac{k}{n} \to 0$ and hence $p_{\eta}\to 0$, so we require $\eta \to 0$ (and as fast as possible). In order for the right hand side to have limit 0, we require (at least) $\eta \log\left(\frac{1}{\eta}\right) = o(n)$. 
We set $\eta = \frac{1}{n}$ and $k=\frac{(h+1)n}{\log(n)}$ (so that $\frac{k}{n} = \frac{h+1}{\log(n)}> \frac{h}{\log(n)} \geq p_{\eta}$) to give, for any $\epsilon>0$:
\begin{align*}
\prob{\sum_{a \in \mathbb{A}} |\hat{p}(a)-p(a)| \geq \frac{h+1}{\log(n)} + \epsilon} \leq 2^{h\frac{n}{\log(n)}} e^{-\frac{1}{2} n \epsilon^2\left(1 - \frac{h+1}{\log(n)}\right)} + e^{-2\frac{n}{\log(n)^2}} 
\end{align*}
Given $\delta \in (0,1)$, for $n$ large enough that $\frac{h+1}{\log(n)}<\frac{1}{2}$, and $e^{-2\frac{n}{\log(n)^2}} \leq \frac{\delta}{2}$, then with probability at least $1-\delta$, we have: 

\begin{align*}
\sum_{a \in \mathbb{A}} |\hat{p}(a)-p(a)| &\leq \frac{h+1}{\log(n)} + \sqrt{\frac{2}{n\left(1-\frac{h+1}{\log(n)}\right)}\left(h\log(2)\frac{n}{\log(n)} + \log\left(\frac{1}{\delta - e^{-2\frac{n}{\log(n)}}}\right)\right)} \\
&\leq \frac{h+1}{\log(n)} + 2\sqrt{\frac{h\log(2)}{\log(n)} + \frac{1}{n}\log\left(\frac{2}{\delta}\right)}
\end{align*}

We may alternatively set $\eta = \frac{\log(n)}{n}$ and $k=\frac{2n(h+1)}{\log(n)}$ (so that for $n>e^1$, we have $\frac{k}{n} = \frac{h+1}{\frac{1}{2}\log(n)}> \frac{h}{\frac{1}{2}\log(n)} \geq \frac{h}{\log(n)-\log(\log(n))} = \frac{h}{\log\left(\frac{1}{\eta}\right)} \geq p_{\eta}$) to give, for any $\epsilon>0$:
\begin{align*}
\prob{\sum_{a \in \mathbb{A}} |\hat{p}(a)-p(a)| \geq \frac{2(h+1)}{\log(n)} + \epsilon} &\leq 2^{h\frac{n}{\log(n)^2 - \log(n) \log(\log(n))}} e^{-\frac{1}{2} n \epsilon^2\left(1 - \frac{2(h+1)}{\log(n)}\right)} + e^{-8\frac{n}{\log(n)^2}} \\
&\leq 2^{h\frac{2n}{\log(n)^2}} e^{-\frac{1}{2} n \epsilon^2\left(1 - \frac{2(h+1)}{\log(n)}\right)} + e^{-8\frac{n}{\log(n)^2}}
\end{align*}
Given $\delta \in (0,1)$, for $n$ large enough that $\frac{2(h+1)}{\log(n)}<\frac{1}{2}$, and $e^{-8\frac{n}{\log(n)^2}} \leq \frac{\delta}{2}$, then with probability at least $1-\delta$, we have: 
\begin{align*}
\sum_{a \in \mathbb{A}} |\hat{p}(a)-p(a)| &\leq \frac{2(h+1)}{\log(n)} + \sqrt{\frac{2}{n\left(1-\frac{2(h+1)}{\log(n)}\right)}\left(2h\log(2)\frac{n}{(\log(n))^2} + \log\left(\frac{1}{\delta - e^{-8\frac{n}{(\log(n))^2}}}\right)\right)} \\
&\leq \frac{2(h+1)}{\log(n)} + \frac{2}{\log(n)}\sqrt{2 h\log(2) + \frac{(\log(n))^2}{n}\log\left(\frac{2}{\delta}\right)}
\end{align*}

\end{proof}

\viabilityConvergence

\begin{proof}
By lemma~\ref{lem:sumabsdiff} we have:
\begin{align*}
|\hat{N}(\hat{q},\hat{V}) - N(\hat{q},\hat{V})| &\leq \sum_{k \in [r],j \in [s]} \left|\sum_{a:a_k=1} \hat{V}_j(a,\hat{q})p(a) - \sum_{a:a_k=1} \hat{V}_j(a,\hat{q})\eprob{a}\right| \\
&= \sum_{k \in [r],j \in [s]} \left|\sum_{a:a_k=1} \hat{V}_j(a,\hat{q})\left[p(a) - \eprob{a}\right]\right| \\
&\leq \sum_{k \in [r],j \in [s]} \sum_{a:a_k=1} \hat{V}_j(a,\hat{q})\left|p(a) - \eprob{a}\right| \\
&\leq r s \sum_{a \in \mathbb{A}} \left|p(a) - \eprob{a}\right| 
\end{align*}
since $V_j(\cdot,\cdot) \leq 1$. Applying Lemma~\ref{lem:multinomial} we firstly have
\begin{equation*}
\prob{\sum_{a \in \mathbb{A}} \left|p(a) - \eprob{a}\right| \geq \epsilon} \leq 2^{3^s} e^{-n \frac{\epsilon^2}{2}}
\end{equation*}
hence, setting 
\begin{equation*}
\epsilon=r s \sqrt{\frac{2}{n}\left(3^s \log(2) + \log\left(\frac{1}{\delta}\right)\right)}
\end{equation*}
we have:
\begin{equation*}
\prob{|\hat{N}(\hat{q},\hat{V}) - N(\hat{q},\hat{V})| \geq \epsilon} \leq \delta
\end{equation*}
Moreover, from the second part of Lemma~\ref{lem:multinomial}, for $\delta < 2e^{-1}$ and appropriately large $n$, we have with probability at least $1-\delta$:
\begin{align*}
|\hat{N}(\hat{q},\hat{V}) - N(\hat{q},\hat{V})| &\leq r s \sum_{a \in \mathbb{A}} \left|p(a) - \eprob{a}\right| \\
&\leq r s \left[\frac{h+1}{\log(n)} + 2\sqrt{\frac{h\log(2)}{\log(n)} + \frac{1}{n}\log\left(\frac{2}{\delta}\right)}\right]
\end{align*}

\end{proof}

\coherenceConvergence

\begin{proof}
We apply lemma~\ref{lem:sumabsdiff} with $a=0,b=1$ to attain:
\begin{align*}
|\hat{C}(\hat{V}) - C(\hat{V})| &\leq \sum_{a \in \mathbb{A}} |\eprob{a}-p(a)| + \sum_{a \in \mathbb{A}} c_a |\eprob{a_{B_1}}\eprob{a_{B_2}}\eprob{a_{B_3}} - p(a_{B_1})p(a_{B_2})p(a_{B_3})| \\
&:= T_1 + T_2
\end{align*}
denoting $c_a = \sum_j \pi_j^{-2} \hat{V}_j(a_{B_1})\hat{V}_j(a_{B_2})\hat{V}_j(a_{B_3}) \leq \sum_j \pi_j^{-2}:=m$. Now:
\begin{equation*}
\prob{T_1 + T_2 \geq \epsilon} \leq \prob{T_1 \geq \frac{\epsilon}{2}} + \prob{T_2 \geq \frac{\epsilon}{2}}
\end{equation*}
By the first part of Lemma~\ref{lem:multinomial}:
\begin{equation}
\prob{T_1 \geq \frac{\epsilon}{2}} \leq 2^{3^s}e^{-n\frac{\epsilon^2}{8}} \label{eq:T1bound}
\end{equation}
For $T_2$, we note that
\begin{align*}
|\eprob{a_{B_1}}\eprob{a_{B_2}}\eprob{a_{B_3}} - &p(a_{B_1})p(a_{B_2})p(a_{B_3})| \\
&= |\eprob{a_{B_1}}\eprob{a_{B_2}}\eprob{a_{B_3}} - p(a_{B_1})\eprob{a_{B_2}}\eprob{a_{B_3}} + \\
&\phantom{=|} p(a_{B_1})\eprob{a_{B_2}}\eprob{a_{B_3}}  - p(a_{B_1})p(a_{B_2})\eprob{a_{B_3}}  + \\
&\phantom{=|} p(a_{B_1})p(a_{B_2})\eprob{a_{B_3}} - p(a_{B_1})p(a_{B_2})p(a_{B_3})| \\
&\leq |\eprob{a_{B_1}} - p(a_{B_1})|\eprob{a_{B_2}}\eprob{a_{B_3}} + \\
&\phantom{=} |\eprob{a_{B_2}} - p(a_{B_2})|p(a_{B_1})\eprob{a_{B_3}}  + \\
&\phantom{=} |\eprob{a_{B_3}} - p(a_{B_3})|p(a_{B_1})p(a_{B_2}) \\
&\leq |\eprob{a_{B_1}}-p(a_{B_1})| + |\eprob{a_{B_2}}-p(a_{B_2})| + |\eprob{a_{B_1}}-p(a_{B_1})|
\end{align*}
since $p(\cdot), \hat{p}(\cdot) \leq 1$, and hence, for sufficiently large $n$: 
\begin{align}
\prob{T_2 \geq \frac{\epsilon}{2}} &=\prob{\sum_{a \in \mathbb{A}} c_a |\eprob{a_{B_1}}\eprob{a_{B_2}}\eprob{a_{B_3}} - p(a_{B_1})p(a_{B_2})p(a_{B_3})| \geq \frac{\epsilon}{2}} \label{eq:T2bound1} \\
&\leq \prob{m\sum_{a \in \mathbb{A}} \sum_{\ell} |\eprob{a_{B_{\ell}}} - p(a_{B_{\ell}})| \geq \frac{\epsilon}{2}} \nonumber \\
&\leq \sum_{\ell=1}^3 \prob{\sum_{a \in \mathbb{A}} |\eprob{a_{B_{\ell}}} - p(a_{B_{\ell}})| \geq \frac{\epsilon}{6 m}} \nonumber \\
&= \sum_{\ell=1}^3 \prob{3^{s-|B_{\ell}|}\sum_{a_{B_{\ell}}} |\eprob{a_{B_{\ell}}} - p(a_{B_{\ell}})| \geq \frac{\epsilon}{6 m} }  \nonumber \\
&\leq \sum_{\ell=1}^3 2^{3^{|B_{\ell}|}} \exp\left[-\frac{n}{72 m^2}3^{2|B_{\ell}|-2s}\epsilon^2\right] &&\text{(Lemma~\ref{lem:multinomial})} \nonumber \\
&= \sum_{\ell=1}^3  \exp\left[-3^{|B_{\ell}|}\left(\frac{n}{72 m^2}3^{|B_{\ell}|-2s}\epsilon^2 - \log(2)\right)\right] \nonumber 
\intertext{Denote by $N_{\ell}$ the value of $3^{|B_{\ell}|}$ corresponding to the largest term in the sum. Now:}
&\leq 3\exp\left[-N_{\ell}\left(\frac{n}{72 m^2 3^{2s}}N_{\ell}\epsilon^2 - \log(2)\right)\right] \label{eq:T2bound2}
\end{align}
We note that this bound is meaningful only for very large $n$: for the exponent in the penultimate line to be negative, we need $n> 72 \log(2) m^2 3^{2s}N_{\ell}^{-1}$. 
Since $m \geq 1$ and $N_{\ell} < 3^s$, bound~\eqref{eq:T2bound2} dominates bound~\eqref{eq:T1bound}; hence for large enough $n$:
\begin{equation*}
\prob{T_1+T_2 \geq \epsilon} \leq 4\exp\left[-N_{\ell}\left(\frac{n}{72 m^2 3^{2s}}N_{\ell}\epsilon^2 - \log(2)\right)\right] 
\end{equation*}
hence, choosing $\delta$ equal to the right hand side, we attain that 
\begin{equation*}
|\hat{C}(\hat{V}) - C(\hat{V})| \leq  6 \sqrt{2}\frac{3^s}{N_{\ell}} \sqrt{\frac{1}{n} \log\left(\frac{4}{\delta}\right) + \frac{N_{\ell}}{n} \log(2)} \leq 6  \sqrt{2} \cdot 3^s  \sqrt{\frac{1}{n} \log\left(\frac{4}{\delta}\right) + \frac{3^s}{n} \log(2)}
\end{equation*}
with probability at least $1-\delta$. 

Supposing now that $A$ has fixed entropy $h$, we may proceed by noting that for $\epsilon>0$
\begin{equation*}
\prob{T_1 + T_2 \geq \frac{(2m+1)(h+1)}{\log(n)} + 2 \epsilon} \leq \prob{T_1  \geq \frac{h+1}{\log(n)} + \epsilon} + \prob{T_2 \geq \frac{2m(h+1)}{\log(n)} +  \epsilon}
\end{equation*}
We may bound the first term by Lemma~\ref{lem:multinomial} directly; for appropriate $\delta$ and large enough $n$ we have: 
\begin{equation*}
\prob{T_1  \geq \frac{h+1}{\log(n)} + \epsilon} \leq 2^{h\frac{n}{\log(n)}} e^{-\frac{1}{2} \epsilon^2\left(n - \frac{(h+1)n}{\log(n)}\right)} + e^{-2\frac{n}{\log(n)^2}} \label{eq:T1boundcoherence}
\end{equation*}
To bound the second term, we take a similar approach to the proof of Lemma~\ref{lem:multinomial}. Denote
\begin{equation*}
\mathbb{A}_{\eta}:= \{a:p(a) \geq \eta\}
\end{equation*}
and $m_{\eta}:=|A_{\eta}|$. Then, as in the proof of Lemma~\ref{lem:multinomial}, we have $m_{\eta} \leq \frac{h}{\eta \log\left(\frac{1}{\eta}\right)}$ and  $p_{\eta}:=\sum_{a:p(a) < \eta} p(a) \leq \frac{h}{\log\left(\frac{1}{\eta}\right)}$. We now write:
\begin{align*}
\sum_{a \in \mathbb{A}} c_a &|\eprob{a_{B_1}}\eprob{a_{B_2}}\eprob{a_{B_3}} - p(a_{B_1})p(a_{B_2})p(a_{B_3})| = \\
&\sum_{\mathbb{A}_{\eta}} c_a |\eprob{a_{B_1}}\eprob{a_{B_2}}\eprob{a_{B_3}} - p(a_{B_1})p(a_{B_2})p(a_{B_3})| + \\
&\sum_{\mathbb{A} \setminus \mathbb{A}_{\eta}} c_a |\eprob{a_{B_1}}\eprob{a_{B_2}}\eprob{a_{B_3}} - p(a_{B_1})p(a_{B_2})p(a_{B_3})| \\
&:= T_2^A + T_2^B
\end{align*}
Suppose we observe $n_{\eta}$ samples amongst the values $\{a:p(a)< \eta\}$. Each such sample contributes at most $\frac{m}{n}$ to $T_2^A$, so $T_2^A \leq \frac{n_{\eta} m}{n}$. Given that there are $n-n_{\eta}$ samples amongst the values $\{a:p(a) \geq \eta\}$, the probability of any such sample taking the value $a \in \mathbb{A}_{\eta}$ is $\frac{1}{1-p_{\eta}}p(a)$, so $T_2^B$ is the value of 
\begin{equation*}
\sum_{\mathbb{A}_{\eta}} c_a |\hat{p}'(a_{B_1})\hat{p}'(a_{B_2})\hat{p}'(a_{B_3})-p'(a_{B_1})p'(a_{B_2})p'(a_{B_3})|
\end{equation*}
where $\hat{p}'(a)=\hat{p}(a)$ and $p'(a)=\frac{1}{1-p_{\eta}}p(a)$. We can consider this as corresponding to quantity~\eqref{eq:T2bound1} for $n-n_{\eta}$ samples of a multinomial distribution over elements in $\{a:p(a)\geq \eta\}$ with true and estimated probabilities $p'(a), \hat{p}'(a)$. Hence, for any $k, \eta$, and $\epsilon_1$ such that $\epsilon_1>\frac{k}{n}$ and $\frac{k}{n}>p_{\eta}$ we have:
\begin{align*}
\prob{T_2 \geq \epsilon_1} &= \prob{T_2^A + T_2^B \geq \epsilon_1} \\
&\leq \cprob{T_2^A + T_2^B \geq \epsilon_1}{n_{\eta} \leq k}\prob{n_{\eta} \leq k} + \cprob{T_1 + T_2 \geq \epsilon_1}{n_{\eta}>k}\prob{n_{\eta}>k} \\
&\leq \cprob{T_2^B \geq \epsilon_1 - m\frac{k}{n}}{n_{\eta}\leq k}\prob{n_{\eta}\leq k} + \prob{n_{\eta}>k} \\
&\leq \cprob{T_2^B \geq \epsilon_1 - m\frac{k}{n}}{n_{\eta} = k} + e^{-2n\left(\frac{h}{\log\left(\frac{1}{\eta}\right)}-\frac{k}{n}\right)^2} 
\end{align*}
Applying the analogous argument to that in~\eqref{eq:T2bound2} with $\epsilon_1 - m\frac{k}{n}$ substituted for $\frac{\epsilon}{2}$, $n-k$ substituted for $n$, and $m_{\eta}$ substituted for $3^s$, we attain:
\begin{align*}
\prob{T_2 \geq \epsilon_1} &\leq  \sum_{\ell=1}^3  \exp\left[-3^{|B_{\ell} \cap \mathbb{A}_{\eta}|}\left(\frac{n-k}{18 m^2 m_{\eta}^2}3^{|B_{\ell} \cap \mathbb{A}_{\eta}|}\left(\epsilon_1-m\frac{k}{n}\right)^2 - \log(2)\right)\right] + e^{-2n\left(\frac{h}{\log\left(\frac{1}{\eta}\right)}-\frac{k}{n}\right)^2}  
\intertext{and denoting by $N_{\eta}$ the value of $3^{|B_{\ell} \cap A_{\eta}|}$ which corresponds to the largest value of the summand:}
&\leq 3\exp\left[-N_{\eta}\left(\frac{n-k}{72 m^2 m_{\eta}^2}N_{\eta}\left(\epsilon_1-m\frac{k}{n}\right)^2- \log(2)\right)\right] + e^{-2n\left(\frac{h}{\log\left(\frac{1}{\eta}\right)}-\frac{k}{n}\right)^2}  \\
&\leq  3 \exp\left[-N_{\eta}\left(\frac{\left(\eta \log\left(\frac{1}{\eta}\right)\right)^2}{72 m^2 h^2}(n-k)N_{\eta}\left(\epsilon_1-m\frac{k}{n}\right)^2 - \log(2) \right)\right] + e^{-2n\left(\frac{h}{\log\left(\frac{1}{\eta}\right)}-\frac{k}{n}\right)^2}  
\intertext{Given the term in the first exponent, we can no longer use $\eta=\frac{1}{n}$; so substituting  $\eta = n^{-\frac{1}{2}}$ and $k=\frac{2n(h+1)}{\log(n)}$ (so that $\frac{k}{n} = \frac{h+1}{\log\left(n^{\frac{1}{2}}\right)} > \frac{h}{\log\left(\frac{1}{\eta}\right)} > p_{\eta}$) gives}
&\leq  3 \exp\left[-N_{\eta}\left(\frac{N_{\eta}\log(n)^2}{288 m^2 h^2}\left(1-\frac{2(h+1)}{\log(n)}\right)\left(\epsilon_1-\frac{2m(h+1)}{\log(n)}\right)^2 - \log(2)\right)\right] + e^{-8\frac{n}{\log(n)^2}}  \\
\intertext{
If we set $\epsilon = \epsilon_1-\frac{2m(h+1)}{\log(n)}$ then we have}
\prob{T_2 \geq \frac{2m(h+1)}{\log(n)}+ \epsilon} &\leq  3 \exp\left[-N_{\eta}\left(\frac{N_{\eta}\log(n)^2}{288 m^2 h^2}\left(1-\frac{2(h+1)}{\log(n)}\right)\epsilon^2 - \log(2)\right)\right] + e^{-8\frac{n}{\log(n)^2}}
\end{align*}
For fixed $\epsilon$, the first term in the bound above dominates the second term and both terms in bound~\ref{eq:T1boundcoherence}, so for large enough $n$ (depending on $\epsilon$) we have:
\begin{align*}
&\prob{|\hat{C}(\hat{V}) - C(\hat{V})| \geq \frac{(2m+1)(h+1)}{\log(n)} + 2 \epsilon} \leq \\
&\hspace{20pt} 4 \exp\left[-N_{\eta}\left(\frac{N_{\eta}\log(n)^2}{288 m^2 h^2}\left(1-\frac{2(h+1)}{\log(n)}\right)\epsilon^2 - \log(2)\right)\right]
\end{align*}
and with probability at least $1-\delta$ we have:
\begin{align*}
|\hat{C}(\hat{V}) - C(\hat{V})| &\leq \frac{(2m+1)(h+1)}{\log(n)} +  \frac{12\sqrt{2} m h}{\log(n)}\sqrt{\frac{\frac{1}{N_{\eta}}\log\left(\frac{4}{\delta}\right) + \log(2)}{N_{\eta}\left(1-\frac{2(h+1)}{\log(n)}\right)}}
\intertext{For $n> e^{4(h+1)}$:}
&\leq \frac{(2m+1)(h+1)}{\log(n)} +  \frac{24 m h}{\log(n)}\sqrt{\frac{1}{N_{\eta}^2}\log\left(\frac{4}{\delta}\right) + \frac{\log(2)}{N_{\eta}}} \\
&\leq \frac{(2m+1)(h+1)}{\log(n)} +  \frac{24 m h}{\log(n)}\sqrt{\log\left(\frac{1}{\delta}\right) + 3\log(2)} \\
\end{align*}
as required. 

\end{proof}

\imputationConvergence

\begin{proof}

We have:
\begin{align*}
\hat{I}(\hat{q},\hat{V})&=\frac{1}{ns}\sum_{i=1}^n \sum_{\ell=1}^b \sum_{k=1}^{|B_{\ell}|} \mathcal{L}\left\{F_k\left[\hat{V}(\alpha^i_{-B_{\ell}},\hat{q}),\hat{q}\right],\ind{\alpha^i_k=1}\right\} \\
&=\frac{1}{ns}\sum_{i=1}^n \sum_{k=1}^{s} \mathcal{L}\left\{F_k\left[\hat{V}(\alpha^i_{-B(k)},\hat{q}),\hat{q}\right],\ind{\alpha^i_k=1}\right\} \\
&= \frac{1}{ns}\sum_{a \in \mathbb{A}} \left|\{i:\alpha^i=a\}\right| \times \left(\sum_{k=1}^s \mathcal{L}\left\{F_k\left[\hat{V}(a_{-B(k)},\hat{q}),\hat{q}\right],\ind{a_k=1}\right\}\right) \\
&:= \frac{1}{s} \sum_{a \in \mathbb{A}} \hat{p}(a)  i_{a}(\hat{q})
\end{align*}
defining 
$i_{\alpha}(\hat{q})$ as the term in brackets. Likewise, we have:
\begin{align*}
I(\hat{q}) &= \EE{K \sim U([s]),A}{\loss{F_K\left[\hat{V}\left(A_{-B(K)},\hat{q}\right),\hat{q}\right]}{\ind{A_K=1}}} \\
&= \sum_{k=1}^s \sum_{a \in \mathbb{A}} P(A=a,K=k) \loss{F_k\left[\hat{V}\left(a_{-B(k)},\hat{q}\right),\hat{q}\right]}{\ind{a_k=1}} \\
&= \frac{1}{s} \sum_{a \in \mathbb{A}} P(A=a) \times \left(\sum_{k=1}^s \mathcal{L}\left\{F_k\left[\hat{V}(a_{-B(k)},\hat{q}),\hat{q}\right],\ind{a_k=1}\right\}\right) \\
&:= \frac{1}{s} \sum_{a \in \mathbb{A}} p(a) i_{a}(\hat{q})
\end{align*}
Since $F_k[\cdot] \in (0,1)$ the value $i_{a}(\hat{q})$ is finite for all $a \in \mathbb{A}$. The only random term in the expression for $\hat{I}(\hat{q},\hat{V})$ is $\hat{p}(a)$, so we have:
\begin{equation}
\E{\hat{I}(\hat{q},\hat{V})} = \frac{1}{s} \sum_{a \in \mathbb{A}} \E{\hat{p}(a)} i_{a}(\hat{q}) = \frac{1}{s} \sum_{a \in \mathbb{A}} p(a) i_{a}(\hat{q}) = I(\hat{q},\hat{V}) \nonumber 
\end{equation}
so $\hat{I}(\hat{q},\hat{V})$ is an unbiased estimator of $I(\hat{q},\hat{V})$. Moreover, we have:
\begin{align*}
\left|\hat{I}(\hat{q},\hat{V})-I(\hat{q},\hat{V})\right| &= \frac{1}{s}\left|\sum_{ a \in \mathbb{A}} \left(\hat{p}(a) - p(a)\right) i_{a}(\hat{q})\right| \\
&\leq \left(\max |i_a(\hat{q})|\right) \sum_{ a \in \mathbb{A}} \left|\hat{p}(a) - p(a)\right| \\
\end{align*}
We have 
\begin{align*}
\max |i_a(\hat{q})| &= \max \left| \sum_{k=1}^s \mathcal{L}\left\{\sum_j \hat{V}_j(a_{-B(k)},\hat{q})\hat{q}_{kj},\ind{a_k=1}\right\}\right| \\
&\leq \max \left| \log \left\{\sum_j \hat{V}_j(a_{-B(k)},\hat{q})\hat{q}_{kj}\right\}\right| \\
&\leq \max \left| \log \left\{\min \hat{q}_{kj}\sum_j \hat{V}_j(a_{-B(k)},\hat{q})\right\}\right| \\
&\leq -\log\left(\min_{j,k} \hat{q}_{kj}\right)
\end{align*}
The result then follows as a direct application of Lemma~\ref{lem:multinomial}.

\end{proof}

\agreementConvergence

\begin{proof}
Let $p_{k_1k_2}=\prob{A_{k_1}=1,A_{k_2}=1}$ and $\hat{p}_{k_1k_2}=\frac{\{i:\alpha^i_{k_1}=\alpha^i_{k_2}=1\}}{n}$.  We have as an initial step:
\begin{align*}
\left[\hat{R_2}(\hat{q})-R_2(\hat{q})\right]^2 &= \left[\sum_{\substack{k_1,k_2 \in [s] \\ B(k_1) \neq B(k_2)}} \left(  \left[\hat{p}_{k_1k_2} - \sum_j \hat{q}_{k_1 j}\hat{q}_{k_2 j} \pi_j  \right]^2 -   \left[p_{k_1k_2} - \sum_j \hat{q}_{k_1 j}\hat{q}_{k_2 j} \pi_j  \right]^2 \right)\right]^2 \\ 
&\leq s^2 \sum_{\substack{k_1,k_2 \in [s] \\ B(k_1) \neq B(k_2)}}  \left[ \left[\hat{p}_{k_1k_2} - \sum_j \hat{q}_{k_1 j}\hat{q}_{k_2 j} \pi_j  \right]^2 - \left[p_{k_1k_2} - \sum_j \hat{q}_{k_1 j}\hat{q}_{k_2 j} \pi_j  \right]^2 \right]^2 \\
&= s^2 \sum_{\substack{k_1,k_2 \in [s] \\ B(k_1) \neq B(k_2)}}  \left[\hat{p}_{k_1k_2} + p_{k_1k_2} - 2\sum_j \hat{q}_{k_1 j}\hat{q}_{k_2 j} \pi_j  \right]^2 \left[\hat{p}_{k_1k_2} - p_{k_1k_2}\right]^2 \\
&\leq  4s^2 \sum_{\substack{k_1,k_2 \in [s] \\ B(k_1) \neq B(k_2)}}  \left[\hat{p}_{k_1k_2} - p_{k_1k_2}\right]^2 \\
&\leq  4s^2 \sum_{k_1,k_2 \in [s]}  \left[\hat{p}_{k_1k_2} - p_{k_1k_2}\right]^2 
\end{align*}
%
%
We can write:
\begin{align*}
\hat{p}_{k_1k_2}-p_{k_1k_2} &= \sum_{i=1}^n \frac{\ind{\alpha_{k_1}^i=\alpha_{k_2}^i=1}}{n} - \prob{A_{k_1}=A_{k_2}=1} \\
&= \frac{1}{n}\sum_{i=1}^n \left[\ind{\alpha_{k_1}^i=1}\ind{\alpha_{k_2}^i=1} - \prob{A_{k_1}=A_{k_2}=1}\right] 
\end{align*}
where each entry in the sum is independent, identically distributed, and bounded in $[-1,1]$. By a Hoeffding bound on the sum, for $\epsilon>0$
\begin{equation*}
\prob{\left|\hat{p}_{k_1k_2}-p_{k_1k_2} \right| \geq \epsilon} \leq 2 e^{-n\frac{\epsilon^2}{2}}
\end{equation*}
Applying a union bound across $k_1,k_2$ gives, for $\epsilon>0$
\begin{align*}
\prob{\sum_{k_1,k_2 \in [s]}  \left[\hat{p}_{k_1k_2} - p_{k_1k_2}\right]^2 \geq \epsilon} &\leq s^2\prob{ \left[\hat{p}_{k_1k_2} - p_{k_1k_2}\right]^2 \geq \frac{\epsilon}{s^2}} \\
&\leq 2 s^2 e^{-n\frac{\epsilon}{2s^2}}
\end{align*}
With $\delta = 2 s^2 e^{-n\frac{\epsilon}{2s^2}}$ we have:
\begin{equation*}
\prob{\left[\hat{R_2}(\hat{q})-R_2(\hat{q})\right]^2 \leq \frac{8s^4}{n}\log\left(\frac{2 s^2}{\delta}\right)} \leq \prob{\sum_{k_1,k_2 \in [s]}  \left[\hat{p}_{k_1k_2} - p_{k_1k_2}\right]^2 \geq \frac{2s^2}{n}\log\left(\frac{2 s^2}{\delta}\right)} \leq \delta
\end{equation*}
as required. We use essentially the same approach for $R_3$: denoting $p_{k_1k_2k_2}=\prob{A_{k_1}=1,A_{k_2}=1,A_{k_3}=1}$ and $\hat{p}_{k_1k_2k_3}=\frac{\{i:\alpha^i_{k_1}=\alpha^i_{k_2}=\alpha^i_{k_3}=1\}}{n}$, we have:
\begin{align*}
\left[\hat{R_3}(\hat{q})-R_3(\hat{q})\right]^2 &\leq  4s^3 \sum_{k_1,k_2,k_3 \in [s]}  \left[\hat{p}_{k_1k_2k_3} - p_{k_1k_2k_3}\right]^2 
\end{align*}
and writing
\begin{align*}
\hat{p}_{k_1k_2k_3}-p_{k_1k_2k_3} &= \frac{1}{n}\sum_{i=1}^n \left[\ind{\alpha_{k_1}^i=1}\ind{\alpha_{k_2}^i=1}\ind{\alpha_{k_3}^i=1} - \prob{A_{k_1}=A_{k_2}=A_{k_3}=1}\right] 
\end{align*}
we again have, by a Hoeffding bound on the sum, for $\epsilon>0$
\begin{equation*}
\prob{\left|\hat{p}_{k_1k_2k_3}-p_{k_1k_2k_3} \right| \geq \epsilon} \leq 2 e^{-n\frac{\epsilon^2}{2}}
\end{equation*}
so, for $\epsilon>0$
\begin{align*}
\prob{\sum_{k_1,k_2,k_3 \in [s]}  \left[\hat{p}_{k_1k_2k_3} - p_{k_1k_2k_3}\right]^2 \geq \epsilon} &\leq s^3\prob{ \left[\hat{p}_{k_1k_2k_3} - p_{k_1k_2k_3}\right]^2 \geq \frac{\epsilon}{s^2}} \\
&\leq 2 s^3 e^{-n\frac{\epsilon}{2s^2}}
\end{align*}
With $\delta = 2 s^3 e^{-n\frac{\epsilon}{2s^2}}$ we have:
\begin{equation*}
\prob{\left[\hat{R_3}(\hat{q})-R_3(\hat{q})\right]^2 \leq \frac{8s^6}{n}\log\left(\frac{2 s^2}{\delta}\right)} \leq \prob{\sum_{k_1,k_2 \in [s]}  \left[\hat{p}_{k_1k_2} - p_{k_1k_2}\right]^2 \geq \frac{2s^3}{n}\log\left(\frac{2 s^2}{\delta}\right)} \leq \delta
\end{equation*}
as required.

\end{proof}

\subsection{Relating to robustness}

\imputationRobustness

\begin{proof}

As per assumptions in the theorem statement, suppose that $\hat{q}_{jk}=q_{jk}$ for $k \in ([s]\setminus b)$, for some block $b$. Then we have $V_j(A_{-b},\hat{q})=P(D=d_j|A_{-b}=a_{-b})$, since this VA assignment should depend only on those $\hat{q}_{jk}$ with $k$, again by assumption. 

Firstly, by straightforward expansion, we have:
\begin{align*}
\cprob{A_k=1}{D=d_j} &= \frac{1}{\prob{D=d_j}}\sum_{a_{-b(k)}} \prob{A_k=1,D=d_j,A_{-b(k)}=a_{-b(k)}} \\
&= \frac{1}{\prob{D=d_j}}\sum_{a_{-b(k)}} \cprob{A_k=1}{D=d_j,A_{-b(k)}=a_{-b(k)}}\prob{D=d_j,A_{-b(k)}=a_{-b(k)}} \\
&= \sum_{a_{-b(k)}} \cprob{A_k=1}{D=d_j,A_{-b(k)}=a_{-b(k)}}\cprob{A_{-b(k)}=a_{-b(k)}}{D=d_j} 
\end{align*}
Now:
\begin{align*}
I(\hat{q},\hat{V}) &= \EE{K \sim U(r)}{\EE{A}{\mathcal{L}\left[A_k,F_K\left[V\left(A_{-b(K)},\hat{q}\right),\hat{q}\right]\right]}} \\
&= \frac{1}{s}\sum_{k=1}^s \sum_{a \in \mathbb{A}} \prob{A=a} \mathcal{L}\left[a_k,F_K\left[V\left(a_{-b(k)},\hat{q}\right),\hat{q}\right]\right] \\
&= \frac{1}{s}\sum_{k=1}^s \sum_{a_{-b(k)}} \sum_{a_k} \sum_{a_{b(k)\setminus k}} \underbrace{\prob{A_{-b(k)}=a_{-b(k)},A_{b(k)\setminus k}=a_{b(k) \setminus k},A_k=a_k}}_{\prob{A=a}} \underbrace{\mathcal{L}\left[a_k,F_K\left[V\left(a_{-b(k)},\hat{q}\right),\hat{q}\right]\right]}_{\text{Independent of $a_{b(k)\setminus k}$}} \\
&= \frac{1}{s}\sum_{k=1}^s \sum_{a_{-b(k)}} \sum_{a_k} \mathcal{L}\left[a_k,F_K\left[V\left(a_{-b(k)},\hat{q}\right),\hat{q}\right]\right] \sum_{a_{b(k)\setminus k}} \prob{A_{-b(k)}=a_{-b(k)},A_{b(k)\setminus k}=a_{b(k) \setminus k},A_k=a_k} \\
&= \frac{1}{s}\sum_{k=1}^s \sum_{a_{-b(k)}} \sum_{a_k} \mathcal{L}\left[a_k,F_K\left[V\left(a_{-b(k)},\hat{q}\right),\hat{q}\right]\right] \prob{A_{-b(k)}=a_{-b(k)},A_k=a_k} \\
&= \frac{1}{s}\sum_{k=1}^s \sum_{a_{-b(k)}} \left[\prob{A_k=1,A_{-b(k)}=a_{-b(k)}} \log\left[F_K\left[V\left(a_{-b(k)},\hat{q}\right),\hat{q}\right]\right] + \right. \\
&\phantom{= \sum \sum  \hspace{5pt}} \left.\prob{A_k=0,A_{-b(k)}=a_{-b(k)}} \log\left[1-F_K\left[V\left(a_{-b(k)},\hat{q}\right),\hat{q}\right]\right]\right] \\
&= \frac{1}{s}\sum_{k=1}^s \sum_{a_{-b(k)}} \prob{A_{-b(k)}=a_{-b(k)}}\left[\cprob{A_k=1}{A_{-b(k)}=a_{-b(k)}} \log\left[F_K\left[V\left(a_{-b(k)},\hat{q}\right),\hat{q}\right]\right] + \right. \\
&\phantom{= \sum \sum \prob{A_{-b(k)}=a_{-b(k)}} \hspace{5pt}} \left.\cprob{A_k=0}{A_{-b(k)}=a_{-b(k)}} \log\left[1-F_K\left[V\left(a_{-b(k)},\hat{q}\right),\hat{q}\right]\right]\right] \\
&:=\frac{1}{s} \sum w \left[\zeta \log[\eta] + (1-\zeta) \log[1-\eta]\right]
\end{align*}
noting $w$ depends on $a_{-b(k)}$ and $\zeta$ and $\eta$ depend on $k, a_{-b(k)}$. Now:
\begin{align*}
\sum w \left[\zeta \log[\eta] + (1-\zeta) \log[1-\eta]\right] &= \sum w \left[\zeta \log[\zeta] + (1-\zeta) \log[1-\zeta]\right] - \\
&\phantom{=} \sum \frac{w}{2\zeta (1-\zeta)}(\eta- \zeta)^2 + \\
&\phantom{=} O\left[\sum w (\eta - \zeta)^3\right]
\end{align*}
where the first term does not depend on $\eta$, and hence does not depend on $\hat{q}$. We may write the middle term as
\begin{align*}
\sum \frac{w}{2\zeta (1-\zeta)}(\eta- \zeta)^2 &= \frac{1}{s} \sum_{k=1}^s \sum_{a_{-b(k)}} \frac{\prob{A_{-b(k)}=a_{-b(k)}}}{\cprob{A_k=1}{A_{-b(k)}=a_{-b(k)}}\cprob{A_k=0}{A_{-b(k)}=a_{-b(k)}}} \times \\
&\phantom{=\arg \min \sum \sum\hspace{5pt}} \left[\cprob{A_k=1}{A_{-b(k)}=a_{-b(k)}} - F_K\left[V\left(a_{-b(k)},\hat{q}\right),\hat{q}\right]\right]^2 \\
&:= \frac{1}{s} \sum_{k=1}^s \sum_{a_{-b(k)}}  w_k(a_{-b(k)}) \times \\
&\phantom{=\arg \min \sum \sum\hspace{5pt}} \left[\cprob{A_k=1}{A_{-b(k)}=a_{-b(k)}} - \sum_j V_j\left(a_{-b(k)},\hat{q}\right) \hat{q}_{jk}\right]^2 \\
\intertext{and from the earlier derivation:}
&= \frac{1}{s}  \sum_{k=1}^s \sum_{a_{-b(k)}}  w_k(a_{-b(k)}) \times \\
&\phantom{= \arg \min } \left[\sum_j \cprob{D=d_j}{A_{-b(k)}=a_{-b(k)}}\cprob{A_k=1}{A_{-b(k)}=a_{-b(k)},D=d_j} - \right. \\
&\phantom{=\arg \min [}\left. \sum_j V_j(a_{-b(k)},\hat{q}) \hat{q}_{jk}\right]^2 
\intertext{From the assumptions in the theorem statement, $V_j(a_{-b},\hat{q})$ depends only on values $\hat{q}_{jk}$ with $k \notin b$, and $\hat{q}_{jk}=q_{jk}$ when $k \notin b$. Hence $V_j(a_{-b},\hat{q})=V_j(a_{-b},q)=\cprob{D=j}{A_{-b}=a_{-b}}$, so we may partition over $k \in b$ and $k \notin b$ as}
&= \frac{1}{s} \sum_{k \notin b} \sum_{a_{-b(k)}}  w_k(a_{-b(k)}) \times \\
&\phantom{=} \left[\sum_j \cprob{D=d_j}{A_{-b(k)}=a_{-b(k)}}\cprob{A_k=1}{A_{-b(k)}=a_{-b(k)},D=d_j} - \right. \\
&\phantom{=\arg \min [}\left. \sum_j V_j(a_{-b(k)},\hat{q}) \cprob{A_k=1}{D=d_j}\right]^2  + \\
&\phantom{=} \frac{1}{s} \sum_{k \in b} \sum_{a_{-b(k)}}  w_k(a_{-b(k)}) \times \\
&\phantom{=} \left[\sum_j \cprob{D=d_j}{A_{-b(k)}=a_{-b(k)}}\left[\cprob{A_k=1}{A_{-b(k)}=a_{-b(k)},D=d_j} -  \hat{q}_{jk}\right]\right]^2 \\
&:= \frac{1}{s} \sum_{k \notin b} \sum_{a_{-b(k)}} w_k(a_{-b(k)}) \left[\sum_j \left[p_{jk}(a_{-b(k)}) u_j(a_{-b(k)}) - q_{jk}V_j(a_{-b(k)},\hat{q})\right]\right]^2 + \\
&\phantom{= } \frac{1}{s} \sum_{k \in b} \sum_{a_{-b(k)}} w_k(a_{-b(k)}) \left[\sum_j u_j(a_{-b(k)})\left[p_{kj}(a_{-b(k)}) - \hat{q}_{kj}\right]\right]^2 \\
&:= T^A + T^B
\end{align*}
where we have defined the shorthands
\begin{align*}
u_j(a_S) &:= \cprob{D=d_j}{A_S=a_S} \\
p_{kj}(a_S) &:= \cprob{A_k=1}{A_S=a_S,D=d_j} \\
q_{kj} &:= \cprob{A_k=1}{D=d_j} \\
w_k(a_S) &:= \frac{\prob{A_{-b(k)}=a_{-b(k)}}}{\cprob{A_k=1}{A_{-b(k)}=a_{-b(k)}}\cprob{A_k=0}{A_{-b(k)}=a_{-b(k)}}} 
\end{align*}
%
%
We now define 
\begin{align*}
I_1(\hat{q},\hat{V}) &:= \frac{1}{s}\sum_{k \notin b} \sum_{a_{-b(k)}} w_k(a_{-b(k)}) \left[\sum_j q_{jk}\left[u_j(a_{-b(k)}) - V_j(a_{-b(k)},\hat{q})\right]\right]^2 + \\
&\phantom{= } \frac{1}{s} \sum_{k \in b} \sum_{a_{-b}} w_k(a_{-b}) \left[\sum_j u_j(a_{-b})\left[q_{kj} - \hat{q}_{kj}\right]\right]^2 \\
&:= T_1^A + T_1^B
\end{align*}
e.g., replacing $p_{kj}(a_{-b(k)})$ with $q_{jk}$ (which differ, by the assumption in the theorem statement, by at most $\epsilon$). Note that $I_1(\hat{q},\hat{V})$ achieves its minimum if $q=\hat{q}$ (and $u_j=V_j$, which by Assumption~\ref{asm:va_calibration} happens if $q=\hat{q}$).  Dealing with $T^A$ and $T^B$, we denote:
\begin{equation*}
\alpha = \sum_j u_j(a_{-b(k)})p_{kj}(a_{-b(k)}) \text{,} \hspace{10pt} \beta = \sum_j V_j(a_{-b(k)},\hat{q})q_{kj} \hspace{5pt} \text{   and   }\hspace{5pt}  \gamma = \sum_j u_j(a_{-b(k)})q_{kj}  \text{,} \hspace{10pt}  \delta = \sum_j u_j(a_{-b(k)})\hat{q}_{kj} 
\end{equation*}
noting that only $\beta$ and $\delta$ depends on $\hat{q}$, and note that the squared expression in $T^A$ satisfies:
\begin{align*}
\left[\sum_j \left[u_j(a_{-b(k)})p_{kj}(a_{-b(k)}) - V_j(a_{-b(k)})q_{kj}\right]\right]^2 &= (\alpha - \beta)^2 \\
&= (\alpha-\gamma)^2 - 2\gamma^2 + 2\alpha \gamma + (\beta -\gamma)^2 + 2\beta (\gamma-\alpha) \\
&= (\alpha-\gamma)^2 - 2\gamma^2 + 2 \alpha \gamma + \\
&\phantom{=} \left[\sum_j q_{jk}\left[u_j(a_{-b(k)}) - V_j(a_{-b(k)})\right]\right]^2 + \\
&\phantom{=} 2\left(\sum_j V_j(a_{-b(k)})q_{kj} \right)\left(\sum_j u_j(a_{-b(k)})\left[q_{kj}-p_{kj}(a_{-b(k)})\right] \right)
\end{align*}
where $(\alpha-\gamma)^2 - 2\gamma^2 + 2 \alpha \gamma$ does not depend on $\hat{q}$, and the squared term is the same as the squared term in $T_1^A$. We can thus write:
\begin{equation*}
T^A = T_1^A + \left(\text{terms not depending on $\hat{q}$}\right) + e^A
\end{equation*}
where
\begin{align*}
|e^A| &= \left|\sum_{k \in b} \sum_{a_{-b}} w_k(a_{-b}) \left(\sum_j V_j(a_{-b})q_{kj} \right)\left(\sum_j u_j(a_{-b})\left[q_{kj}-p_{kj}(a_{-b})\right] \right) \right| \\
&\leq \sum_{k \in b} \sum_{a_{-b}} w_k(a_{-b}) \left(\sum_j V_j(a_{-b})q_{kj} \right)\sum_j u_j(a_{-b})\left|q_{kj}-p_{kj}(a_{-b})\right| \\
&\leq \sum_{k \in b} \sum_{a_{-b}} w_k(a_{-b}) \left(\sum_j V_j(a_{-b})q_{kj} \right)\sum_j u_j(a_{-b})\epsilon \\
&= \epsilon \sum_{k \in b} \sum_{a_{-b}} w_k(a_{-b}) \sum_j V_j(a_{-b})q_{kj}  \\
&\leq \epsilon \sum_{k \in b} \sum_{a_{-b}} w_k(a_{-b}) \sum_j V_j(a_{-b})\\
&= \epsilon \sum_{k \in b} \sum_{a_{-b}} w_k(a_{-b}) 
&\intertext{and noting $\sum_{a_{-b}} w_k(a_{-b})\leq \frac{1}{4}\sum_{a_{-b}}\prob{A_{-b}=a_{-b}}=\frac{1}{4}$}
&\leq |b|\frac{\epsilon}{4} \ 
\end{align*}
%

For $T^B$ and $T_1^B$, we note that the squared term in $T^B$ can be written as
\begin{align*}
\left[\sum_j u_j(a_{-b})\left[p_{kj}(a_{-b(k)}) - \hat{q}_{kj}\right]\right]^2 &= (\alpha - \delta)^2 \\
&= (\alpha - \gamma)^2 + (\gamma-\delta)^2 - 2\gamma^2 + 2\alpha \gamma + 2 \delta(\gamma- \alpha) \\
&= (\alpha - \gamma)^2 - 2\gamma^2 + 2\alpha \gamma + \\
&\phantom{=}  \left[\sum_j u_j(a_{-b})\left[q_{kj} - \hat{q}_{kj}\right]\right]^2 + \\
&\phantom{=} 2 \left(\sum_j u_j(a_{-b(k)})\hat{q}_{kj} \right) \left(\sum_j u_j(a_{-b(k)})\left[q_{kj}-p_{kj}(a_{-b(k)})\right] \right)
\end{align*}
where the squared term is the squared term in $T_1^B$. We can then write 
\begin{equation*}
T^B = T_1^B + \left(\text{terms not depending on $\hat{q}$}\right) + e^B
\end{equation*}
and analogously to above
\begin{align*}
|e^B|&= \left|\sum_{k \notin b} \sum_{a_{-b(k)}} w_k(a_{-b(k)}) \left(\sum_j u_j(a_{-b(k)})\hat{q}_{kj} \right)\left(\sum_j u_j(a_{-b(k)})\left[q_{kj}-p_{kj}(a_{-b(k)})\right] \right) \right| \\
&\leq \sum_{k \notin b} \sum_{a_{-b(k)}} w_k(a_{-b(k)}) \left(\sum_j u_j(a_{-b(k)})\hat{q}_{kj} \right)\sum_j u_j(a_{-b(k)})\left|q_{kj}-p_{kj}(a_{-b(k)})\right| \\
&\leq \sum_{k \notin b} \sum_{a_{-b(k)}} w_k(a_{-b(k)}) \left(\sum_j u_j(a_{-b(k)})\hat{q}_{kj} \right)\sum_j u_j(a_{-b(k)})\epsilon \\
&= \epsilon \sum_{k \notin b} \sum_{a_{-b(k)}} w_k(a_{-b(k)}) \sum_j u_j(a_{-b(k)})\hat{q}_{kj}   \\
&\leq \epsilon \sum_{k \notin b} \sum_{a_{-b(k)}} w_k(a_{-b(k)}) \sum_j u_j(a_{-b(k)})\\
&= \epsilon \sum_{k \notin b} \sum_{a_{-b(k)}} w_k(a_{-b(k)}) \\
&\leq (s-|b|)\frac{\epsilon}{4} 
\end{align*}
and thus
\begin{equation*}
I(\hat{q},\hat{V})=I_1(\hat{q},\hat{V}) + J + e + E := I_2(\hat{q},\hat{V}) + e + E
\end{equation*}
where $I_2(\hat{q}$ attains its minimum at $\hat{q}=q$, $J$ does not depend on $\hat{q}$, $|e| \leq |e^A| + |e^B| \leq s\frac{\epsilon}{2}$, and the final term is
\begin{align*}
E &= O\left(\sum w (\eta - \zeta)^3\right) \\
&= O\left(\frac{1}{s} \sum_{k=1}^s \sum_{a_{-b(k)}} \prob{A_{-b(k)}=a_{-b(k)}}\left[\cprob{A_k=1}{A_{-b(k)}=a_{-b(k)}} - F_k\left[V(a_{-b(k)},\hat{q}),\hat{q}\right]\right]^3 \right)
\end{align*}

\end{proof}

\agreementRobustness
\begin{proof}

In the absence of assumption~\ref{asm:cond_indep}, we have, for any pairwise distinct $k_1,k_2,k_3 \in [r]$:
\begin{align*}
\prob{A_{k_1}=1,A_{k_2}=1} &= \sum_j \cprob{A_{k_1}=1}{A_{k_2}=1,D=d_j}\cprob{A_{k_2}=1}{D=d_j}\prob{D=d_j} \\
&= \sum_j \cprob{A_{k_1}=1}{A_{k_2}=1,D=d_j}q_{k_2 j} \pi_j
\intertext{and }
\prob{A_{k_1}=1,A_{k_2}=1,A_{k_3}=1} &= \sum_j \cprob{A_{k_1}=1}{A_{k_2}=1,A_{k_3}=1,D=d_j} \times \\
&\phantom{= \sum_j } \cprob{A_{k_2}=1}{A_{k_3}=1,D=d_j} \times \\
&\phantom{= \sum_j } q_{k_3 j}\pi_j
\end{align*}
Denote:
\begin{align*}
R_2^A(\hat{q}) &= \sum_{\substack{k_1,k_2 \in [r] \\ B(k_1) \neq B(k_2)}} \left[\sum_j \cprob{A_{k_1}=1}{D=d_j}\cprob{A_{k_2}=1}{D=d_j}\prob{D=d_j} - \sum_j \hat{q}_{k_1 j}\hat{q}_{k_2 j} \pi_j \right]^2 \\
&= \sum_{\substack{k_1,k_2 \in [r] \\ B(k_1) \neq B(k_2)}} \left[\sum_j q_{k_1 j} q_{k_2 j} \pi_j - \sum_j \hat{q}_{k_1 j}\hat{q}_{k_2 j} \pi_j \right]^2 \\
\intertext{and }
R_3^A(\hat{q}) &= \sum_{\substack{k_1,k_2,k_3 \in [r] \\ B(k_{\ell})\text{ distinct}}} \left[\sum_j q_{k_1 j} q_{k_2 j} q_{k_3 j} \pi_j - \sum_j \hat{q}_{k_1 j}\hat{q}_{k_2 j}\hat{q}_{k_3 j} \pi_j \right]^2 
\end{align*}
which both achieve their minima when $\hat{q}=q$. Now a direct expansion gives:
\begin{align*}
|R_2(\hat{q})-R_2^A(\hat{q})| &= \left| \sum_{\substack{k_1,k_2 \in [r] \\ B(k_1) \neq B(k_2)}} \left[\sum_j \cprob{A_{k_1}=1}{A_{k_2}=1,D=d_j}q_{k_2 j} \pi_j - \sum_j \hat{q}_{k_1 j}\hat{q}_{k_2 j} \pi_j \right]^2  - \right.\\
&\phantom{=} \left. \sum_{\substack{k_1,k_2 \in [r] \\ B(k_1) \neq B(k_2)}} \left[\sum_j q_{k_1 j} q_{k_2 j} \pi_j - \sum_j \hat{q}_{k_1 j}\hat{q}_{k_2 j} \pi_j  \right]^2  \right| \\
&\leq \sum_{\substack{k_1,k_2 \in [r] \\ B(k_1) \neq B(k_2)}} \left|  \left[\sum_j \cprob{A_{k_1}=1}{A_{k_2}=1,D=d_j}q_{k_2 j} \pi_j - \sum_j \hat{q}_{k_1 j}\hat{q}_{k_2 j} \pi_j \right]^2  - \right.\\
&\phantom{= \sum } \left. \left[\sum_j q_{k_1 j} q_{k_2 j} \pi_j - \sum_j \hat{q}_{k_1 j}\hat{q}_{k_2 j} \pi_j  \right]^2  \right| \\
&= \sum_{\substack{k_1,k_2 \in [r] \\ B(k_1) \neq B(k_2)}} \left| \sum_j \left[ \cprob{A_{k_1}=1}{A_{k_2}=1,D=d_j}q_{k_2 j} \pi_j  + q_{k_1 j} q_{k_2 j} \pi_j - 2\hat{q}_{k_1 j}\hat{q}_{k_2 j} \pi_j \right]\right| \times \\
&\phantom{= \sum } \left| \sum_j \left[ \cprob{A_{k_1}=1}{A_{k_2}=1,D=d_j}q_{k_2 j} \pi_j  - q_{k_1 j} q_{k_2 j} \pi_j\right]\right| \\
\intertext{Since $\sum_j \pi_j = 1$, we have for $p_j \in (0,1)$ that $\sum_j p_j \pi_j \leq 1$, so:}
&\leq 2\sum_{\substack{k_1,k_2 \in [r] \\ B(k_1) \neq B(k_2)}} \left| \sum_j q_{k_2 j} \pi_j\left[ \cprob{A_{k_1}=1}{A_{k_2}=1,D=d_j}  - q_{k_1 j} \right]\right| \\
&\leq 2 \epsilon \sum_{\substack{k_1,k_2 \in [r] \\ B(k_1) \neq B(k_2)}} \left| \sum_j q_{k_2 j} \pi_j \right| \\
&= 2 \epsilon \sum_{\substack{k_1,k_2 \in [r] \\ B(k_1) \neq B(k_2)}} \prob{A_{k_2}=1} \\
&\leq 2 \epsilon s^2
\end{align*}
as required. For $R_3$, we firstly for brevity, we denote:
\begin{equation*}
T_j(k_1,k_2,k_3):= \cprob{A_{k_1}=1}{A_{k_2}=1,A_{k_3}=1,D=d_j}\cprob{A_{k_2}=1}{A_{k_3}=1,D=d_j}q_{k_3 j} \pi_j
\end{equation*}
and note that, for $\epsilon<1$:
\begin{equation*}
\cprob{A_{k_1}=1}{A_{k_2}=1,A_{k_3}=1,D=d_j}\cprob{A_{k_2}=1}{A_{k_3}=1,D=d_j} - q_{k_1 j} q_{k_2 j} < 2\epsilon
\end{equation*}
so we have:
\begin{align*}
&|R_3(\hat{q})-R_3^A(\hat{q})| = \\
&\left| \sum_{\substack{k_1,k_2,k_3 \in [r] \\ B(k_{\ell})\text{ distinct}}} \left[\sum_j T_j(k_1,k_2,k_3) - \sum_j \hat{q}_{k_1 j}\hat{q}_{k_2 j}\hat{q}_{k_3 j} \pi_j \right]^2 \right.  - \\
&\phantom{=} \left. \sum_{\substack{k_1,k_2,k_3 \in [r] \\ B(k_{\ell})\text{ distinct}}} \left[\sum_j q_{k_1 j} q_{k_2 j} q_{k_3 j} \pi_j - \sum_j \hat{q}_{k_1 j}\hat{q}_{k_2 j}\hat{q}_{k_3 j} \pi_j  \right]^2  \right| \\
&\leq \sum_{\substack{k_1,k_2,k_3 \in [r] \\ B(k_{\ell})\text{ distinct}}} \left| \left[\sum_j T_j(k_1,k_2,k_3) - \sum_j \hat{q}_{k_1 j}\hat{q}_{k_2 j}\hat{q}_{k_3 j} \pi_j \right]^2 \right. - \\
&\phantom{=} \left. \sum_{\substack{k_1,k_2,k_3 \in [r] \\ B(k_{\ell})\text{ distinct}}} \left[\sum_j q_{k_1 j} q_{k_2 j} q_{k_3 j} \pi_j - \sum_j \hat{q}_{k_1 j}\hat{q}_{k_2 j}\hat{q}_{k_3 j} \pi_j  \right]^2  \right| \\
&= \sum_{\substack{k_1,k_2,k_3 \in [r] \\ B(k_{\ell})\text{ distinct}}} \left| \sum_j T_j(k_1,k_2,k_3) +  q_{k_1 j} q_{k_2 j} q_{k_3 j} \pi_j - 2\hat{q}_{k_1 j}\hat{q}_{k_2 j}\hat{q}_{k_3 j} \pi_j \right| \times \\
&\phantom{= \sum \sum} \left| \sum_j T_j(k_1,k_2,k_3) -  \hat{q}_{k_1 j}\hat{q}_{k_2 j}\hat{q}_{k_3 j} \pi_j \right| \\
&\leq 2\sum_{\substack{k_1,k_2,k_3 \in [r] \\ B(k_{\ell})\text{ distinct}}} \left| \sum_j q_{k_3 j} \pi_j\times \right. \\
&\phantom{= \sum \sum} \left[\cprob{A_{k_1}=1}{A_{k_2}=1,A_{k_3}=1,D=d_j}\cprob{A_{k_2}=1}{A_{k_3}=1,D=d_j} - q_{k_1 j} q_{k_2 j}\right]\bigg| \\
&\leq 4 \epsilon \sum_{\substack{k_1,k_2,k_3 \in [r] \\ B(k_{\ell})\text{ distinct}}} \left| \sum_j q_{k_2 j} \pi_j \right| \\
&= 4 \epsilon \sum_{\substack{k_1,k_2,k_3 \in [r] \\ B(k_{\ell})\text{ distinct}}} \prob{A_{k_2}=1} \\
&\leq 4 \epsilon s^3
\end{align*}
as required. 
\end{proof}

\section{Simulation Details}

We simulated 10,000 datasets of questionnaires, using each simulated dataset for only one of the three main methodical evaluations described in section~\ref{sec:evaluation}. 

\subsection{Data sources, setup, and exclusions}

We began with the WHO2012 verbal autopsy questionnaire, and associated probbase~\cite{who12,li14}, which we converted to numerical values according to the scheme in the \texttt{openVA} package function \texttt{InterVA}~\cite{li14,byass19}. More recent VA data formats include 2016 and 2020 updates of the WHO format, and an alternative format from the Population Health Metrics Research Consortium~\citep{murray11}. We denoted the probbase $q^{\text{sim}}$. Prior to exclusions, the probbase was of dimension $246 \times 81$. 

We also made use of the \texttt{RandomPhysician} dataset from \texttt{openVA}. This dataset contains 1000 completed VA questionnaires in the WHO2012 format, along with crude CoD categories (non-communicable disease, tuberculosis/aquired immune deficiency syndrome, communicable disease, maternal, external, and unknown). 

We removed several rows and columns of the probbase matrix for our simulation. Our choice was largely to simplify the simulation in this initial case; we removed obstetric and paediatric CoDs since they can affect only a minority of individuals in question, and we removed external CoDs since they are generally identified definitively by a single question in the VA questionnaire:
\begin{itemize}
\item Columns 1-16 do not correspond to CoDs and are used in VA processing only
\item Columns 77-81 correspond to circumstances of mortality rather than CoD
\item Columns 49-56 correspond to paediatric/neonatal CoDs
\item Columns 68-76 correspond to obstetric CoDs
\item Columns 57-67 correspond to external CoDs
\end{itemize}
We also removed the following rows:
\begin{itemize}
\item Rows 177-178, concerning stillbirths, were excluded since no samples in the \texttt{RandomPhysician} dataset answered these questions affirmatively
\item Rows 213-224 are direct questions about specific external CoDs (e.g. `Did [the deceased] die of snakebite?')
\item Rows 237-246 concern circumstances of mortality and were not answered in the \texttt{RandomPhysician} dataset. 
\end{itemize}
After exclusions, the probbase included 133 questions and 32 causes of death. 

\subsection{Data simulation method}

Denoting as usual by $\alpha^i_k$ the $k$th answer and by $\delta^i$ the latent cause of death in the $i$th simulation, we firstly simulated latent causes of death as $\delta^i = d_j \text{ where } j \sim \text{Multi}(\pi^{\text{sim}})$ where $\text{Multi}()$ denotes a multinomial distribution. After this, for each block $b$, we simulated answers as:
\begin{align*}
\{(\alpha^i_k|\delta^i=d_j)\}_{k \in b} &= \ind{Z^i_b > \Phi^{-1}(\{q^{\text{sim}}_{kj}\}_{k \in b})} \text{ where } Z^i_b \iidsim N\left(0,\Sigma_{b,j}\right)
\end{align*}
given covariance matrices $\Sigma_{b,j}$. We did not directly simulate missing answers. 

We also simulated datasets which violated assumption~\ref{asm:cond_indep}. In this case, we simply simulated 
\begin{align*}
\{(\alpha^i_k|\delta^i=d_j)\}_{k \in [s]} &= \ind{Z^i > \Phi^{-1}(q^{\text{sim}}_{\cdot j})} \text{ where } Z^i \iidsim N\left(0,\Sigma_{j}\right)
\end{align*}
for a single covariance matrix $\Sigma_j$ per CoD. The simulation parameters were then:
\begin{itemize}
\item $\hat{q}^{\text{sim}}$, the probbase 
\item $\pi_{\text{sim}}$, the prior over CoDs
\item Covariance matrices $\Sigma_{b,j}$ or $\Sigma_j$, and implicitly the block conditional independence structure
\item $n$, the number of simulations
\end{itemize}

\subsection{Simulation parameters}

We attained the probbase $q^{\text{sim}}$ directly from the \texttt{openVA} package, as above. For the prior $\pi^{\text{sim}}$, since for a true probbase $q_{jk}=\cprob{A_k=1}{D=d_j}$ and prior $\pi_j=\prob{D=d_j}$ we expect that 
\begin{equation*}
\sum_j \pi_j q_{kj} = \sum_j \pi_j \cprob{A_k=1}{D=d_j} \prob{D=d_j} = \prob{A_k=1}
\end{equation*}
we presumed that we should have:
\begin{equation*}
\sum_j \pi^{\text{sim}}_j q^{\text{sim}}_{kj} \approx \frac{|i:(A^{\text{sim}})^i_k=1}{1000}
\end{equation*}
that is, the value $\sum_j \pi^{\text{sim}}_j q^{\text{sim}}_{kj}$ should approximately match the empirical frequency of an affirmative answer to question $k$. We thus defined:
\begin{equation*}
\pi^{\text{sim}} = \argmin{\pi}\left\{\left(\sum_j \pi^{\text{sim}}_j q^{\text{sim}}_{kj} - \frac{|\{i:(A^{\text{sim}})^i_k=1\}|}{1000}\right)^2  \text{ s.t. } \pi \in \Delta_{32} \right\}
\end{equation*}
To avoid computational 0-1 errors, we clamped all elements of $\hat{q}^{\text{sim}}$ and $\pi^{\text{sim}}$ to $[1 \times 10^{-3},1 - 1 \times 10^{-3}]$.

We determined two block structures: a fine block structure used for simulation, and a coarser block structure used for evaluations (since we generally need blocks to be of size at least $r=32$). To find the block structure, we computed correlation between each pair of observed answers and hierarchically clustered questions, adjusting for the number of observations (using the function \texttt{cor.test} in R) to a maximum block size of 12. We then computed the matrices $\Sigma_{b,j}$ as the empirical correlation matrices of questions in blocks $b$ for the coarse CoD group to which $d_j$ belonged. 

We then combined these smaller blocks into three larger blocks of sizes at least 32, which were used in the calculation of imputation error, coherence, and two- and three- way agreement ($\hat{I}$, $\hat{C}$, $\hat{R}_2$ and $\hat{R}_3$). We simulated between 500 and 15000 samples ($n$) per simulation run, and simulated $\approx 30,000$ datasets, until standard errors were sufficiently small as to be negligible. 

\subsection{Evaluations}

\subsubsection{Differentiation of correct and incorrect probbases}

For the evaluation of whether estimated objective functions could differentiate perturbed and correct probbases, we considered three levels of pertubation:
\begin{enumerate}
\item A `random' probbase for which $\hat{q}_{kj} \iidsim U(0,1)$
\item A `perturbed' probbase for which $(\hat{q}-q^{\text{sim}})_{kj} \iidsim N\left(0,\left(\frac{1}{10}\right)^2 \right)$
\item A `slightly perturbed' probbase for which for 20 index pairs $(k,j)$ chosen uniformly at random, $(\hat{q}-q^{\text{sim}})_{kj} \sim N\left(0,\left(\frac{1}{10}\right)^2 \right)$ and $\hat{q}_{kj}=q^{\text{sim}}_{kj}$ otherwise. 
\end{enumerate}
For each simulation run, we simulated either 50 or 100 randomly-perturbed probbases according to each of the above levels (using 50 to reduce per-run time for large $n$). We then evaluated each estimated objective function for the real probbase and for each randomly-perturbed probbase, and computed the quantile of the value for the real probbase amongst the values for the randomly-perturbed probbases. To draw Figure~\ref{fig:pertubation} and similar supplementary figures, we used a LOESS fit with default parameters in $R$ (degree 2, smoothing parameter $\alpha = 0.75$). 

\subsubsection{Recovery of unknown probbase entries}

For evaluating recovery of probbase entries, we firstly chose a number of probbase entries to try and recover between 2 and 20 uniformly at random. We then chose this many probbase entries uniformly at random amongst values of $q^{\text{sim}}$ in $(0.05,0.95)$, and treated these entries as unknowns. 

For each estimated objective function, we then performed multivariate optimisation using the limited memory bound-constrained Broyden -  Fletcher - Goldfarb - Shanno algorithm~\citep{byrd95} to find the values of these unknowns which minimised the estimated objective. We then recorded the absolute deviation between the true values $q^{\text{sim}}$ and the estimated values. We plotted the relationship between $n$ and mean absolute error using a LOESS fit with degree 2 and smoothing parameter $\alpha=0.75$. 

\subsubsection{Identification of perturbed probbase entries}

Finally, we chose 20\% of entries of $q^{\text{sim}}$ uniformly at random, and perturbed them by adding independent Gaussian noise to each entry with standard deviation $1/10$. We recorded the difference between each perturbed probbase entry and true probbase entry as a target. 

In each simulation, we then considered each element of the perturbed probbase individually, and treated the single element as unknown. We then minimised the estimated objective over this single element, and recorded the absolute difference between the value minimising the objective and the value in the perturbed probbase as a predictor. 

We assembled all predictor-target pairs across all simulations, and assessed the potential to predict whether the target exceeded 10\% using a receiver-operator characteristic curve. 

\section{Supplementary Figures}

\begin{figure}[H]
\centering
    \begin{subfigure}[t]{0.45\textwidth}
    \includegraphics[width=\textwidth]{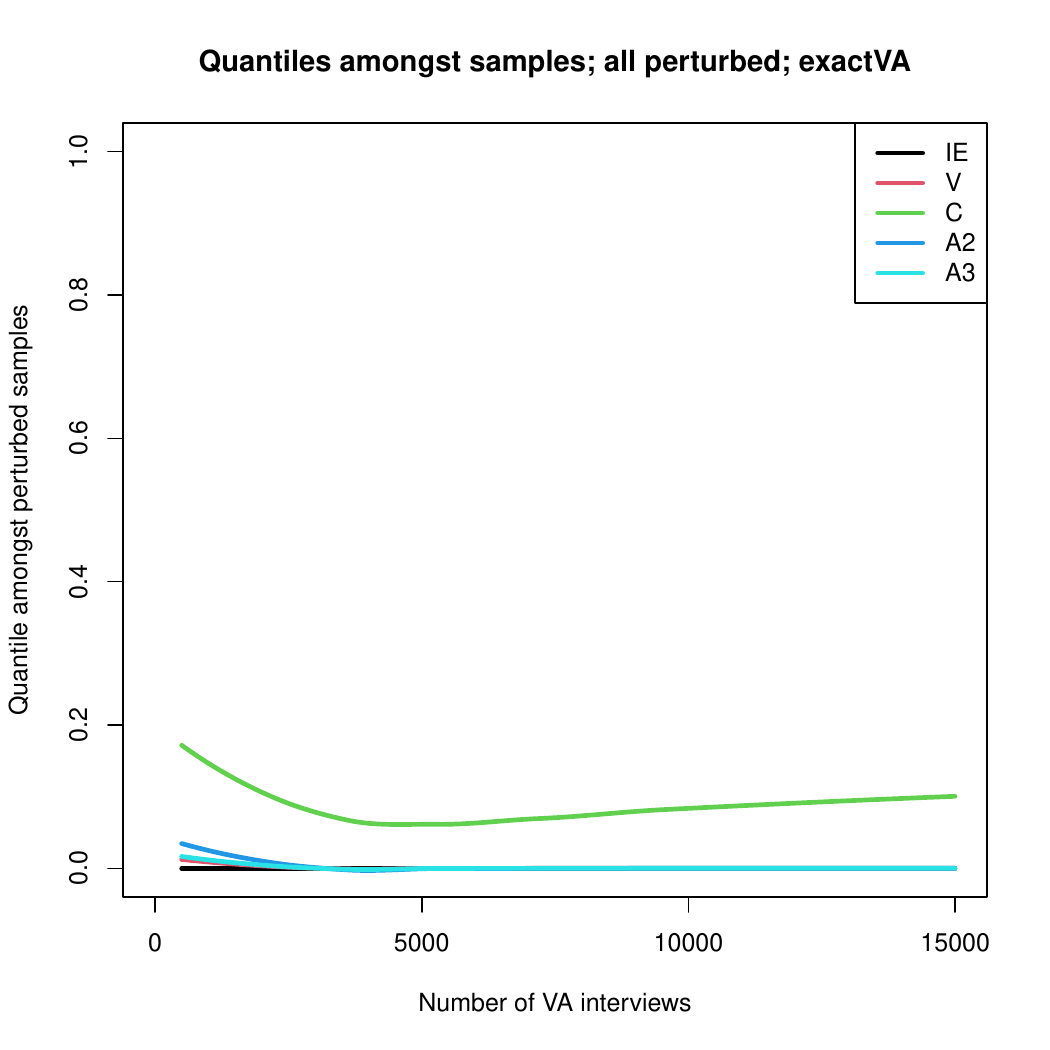}
    \end{subfigure}
    \begin{subfigure}[t]{0.45\textwidth}
    \includegraphics[width=\textwidth]{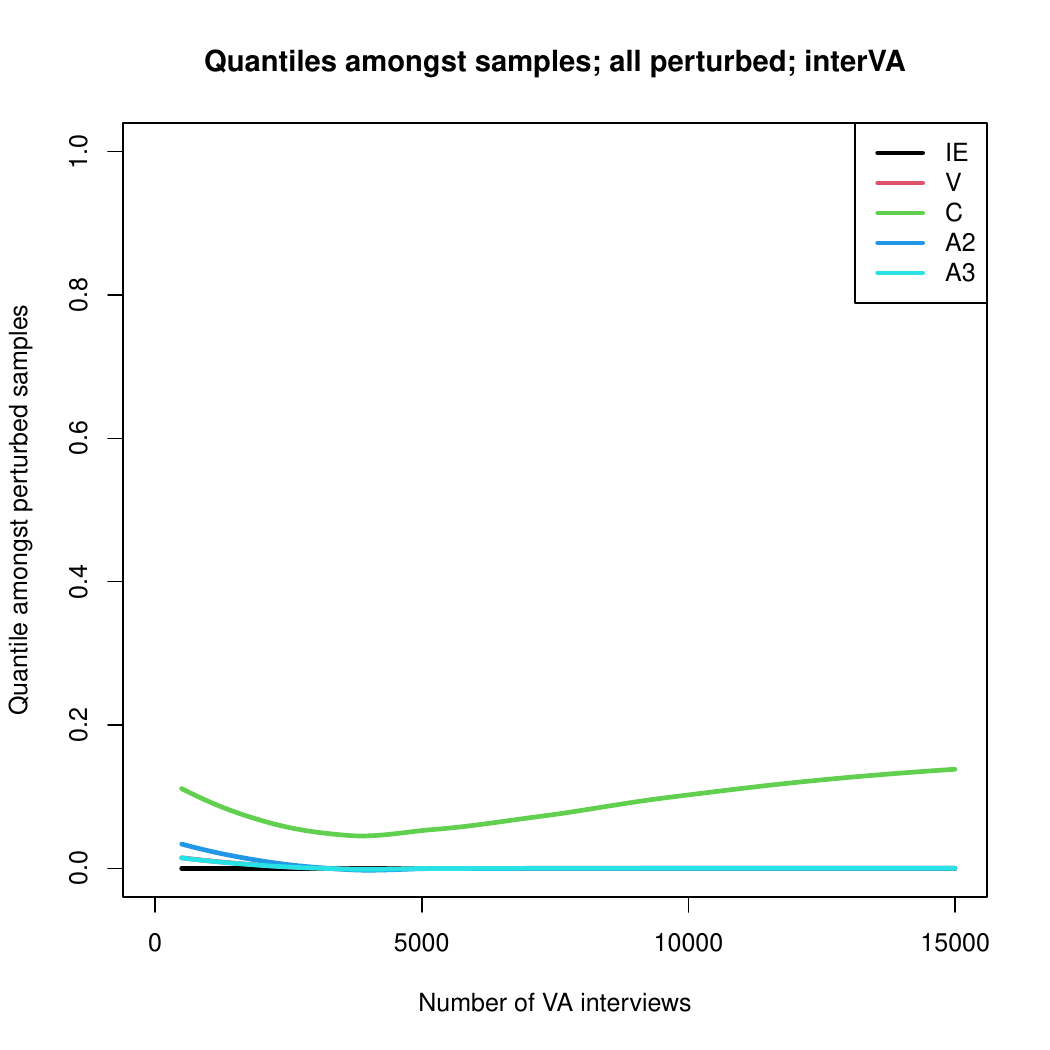}
\end{subfigure}
\caption{Differentiation between true probbase and globally perturbed probbases ($N(0,(1/100)^2)$ noise added to all elements) using estimated objective functions (Imputation error, IE: $\hat{I}$, viability, V: $\hat{N}$, coherence, C: $\hat{C}$, two-way agreement, A2: $\hat{R}_2$ and three-way agreement, A3: $\hat{R}_3$). Leftmost plot shows computations for oracle VA algorithm; rightmost plot for InterVA algorithm. Lines show estimated mean (LOESS) quantile of test function evaluated on true probbase amongst test function evaluated on perturbed probbases, when using a database with total number of samples given by the value on the x-axis. Pointwise standard errors are less than the widths of the lines. Most quantiles are close to zero. }
\label{supp_fig:pertubation_moderate}
\end{figure}

\begin{figure}[H]
\centering
\begin{subfigure}{\textwidth}
    \begin{subfigure}[t]{0.45\textwidth}
    \includegraphics[width=\textwidth]{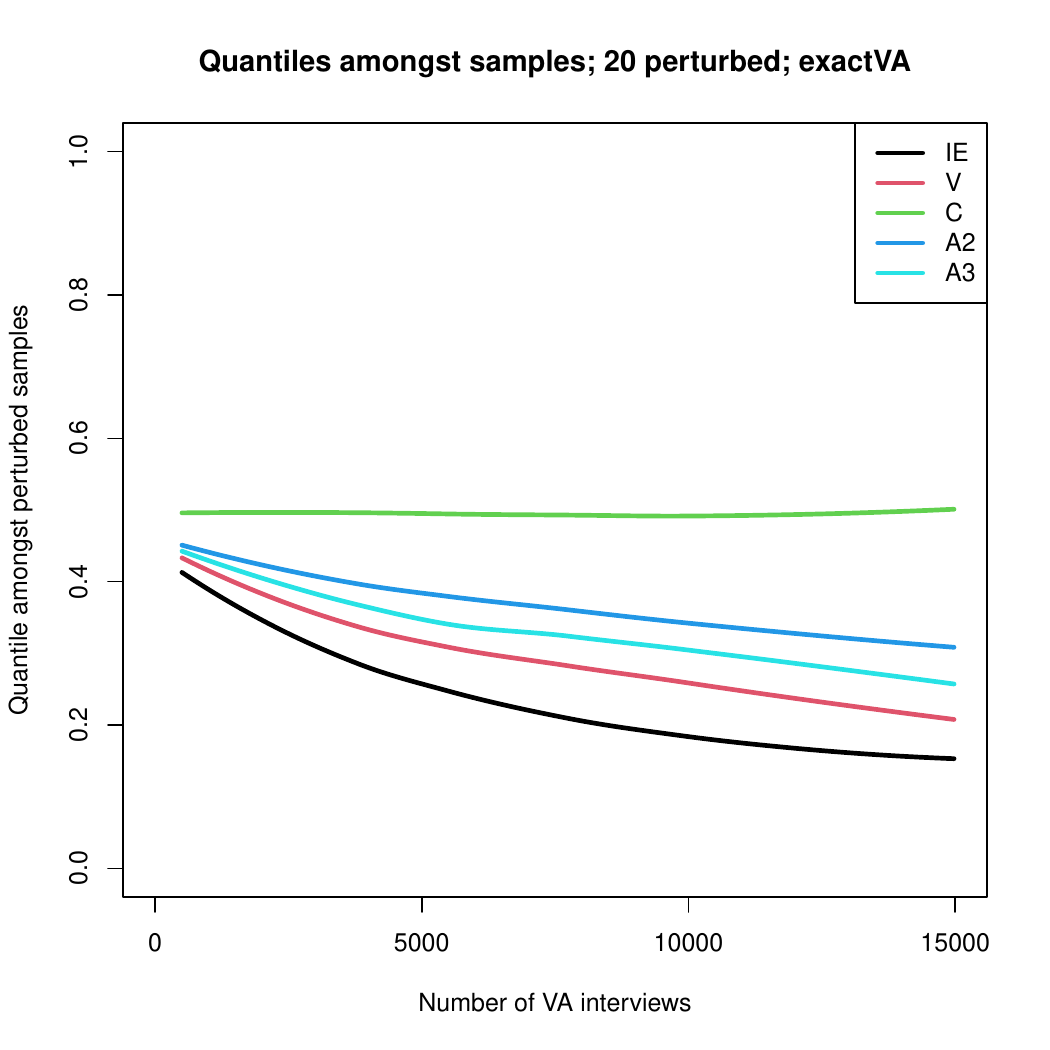}
    \end{subfigure}
    \begin{subfigure}[t]{0.45\textwidth}
    \includegraphics[width=\textwidth]{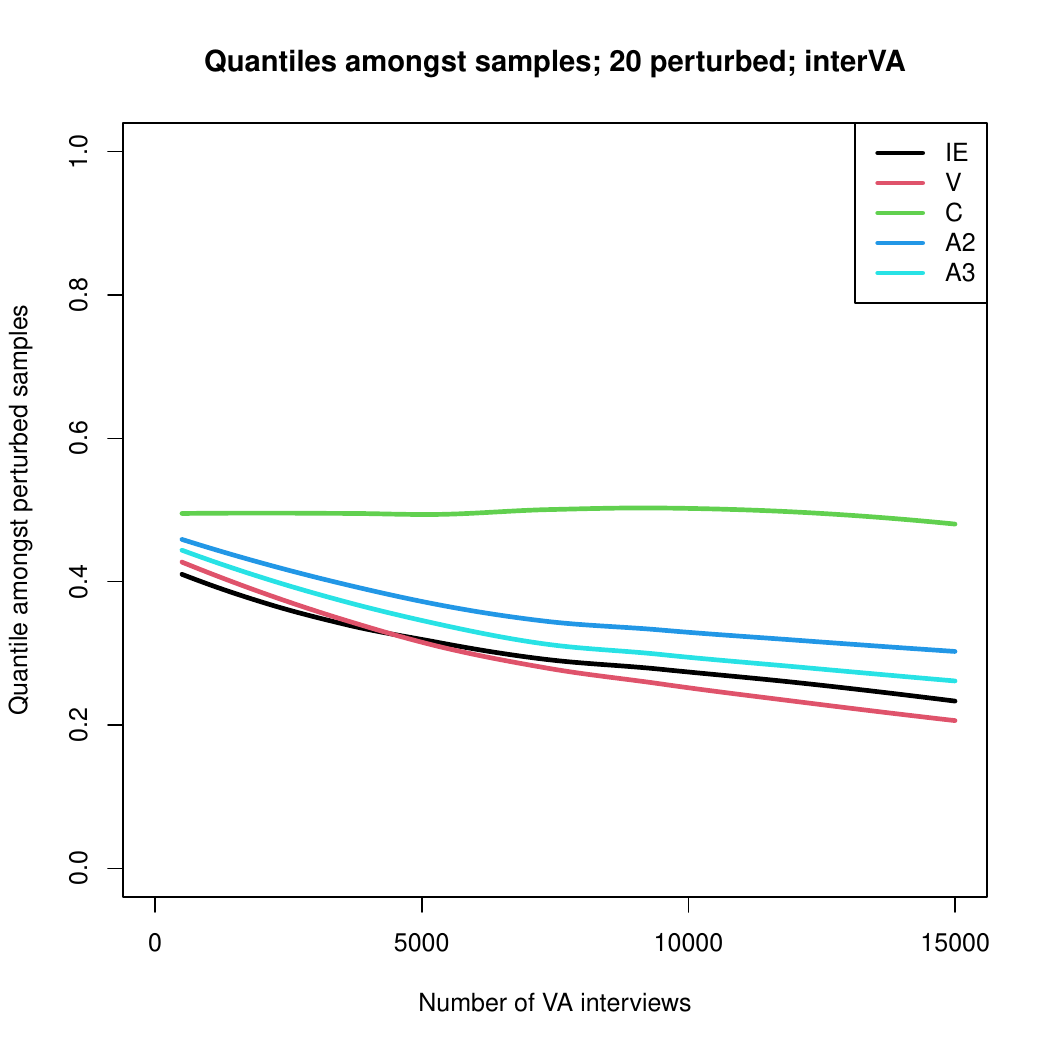}
\end{subfigure}
\end{subfigure}
\begin{subfigure}{\textwidth}
    \begin{subfigure}[t]{0.45\textwidth}
    \includegraphics[width=\textwidth]{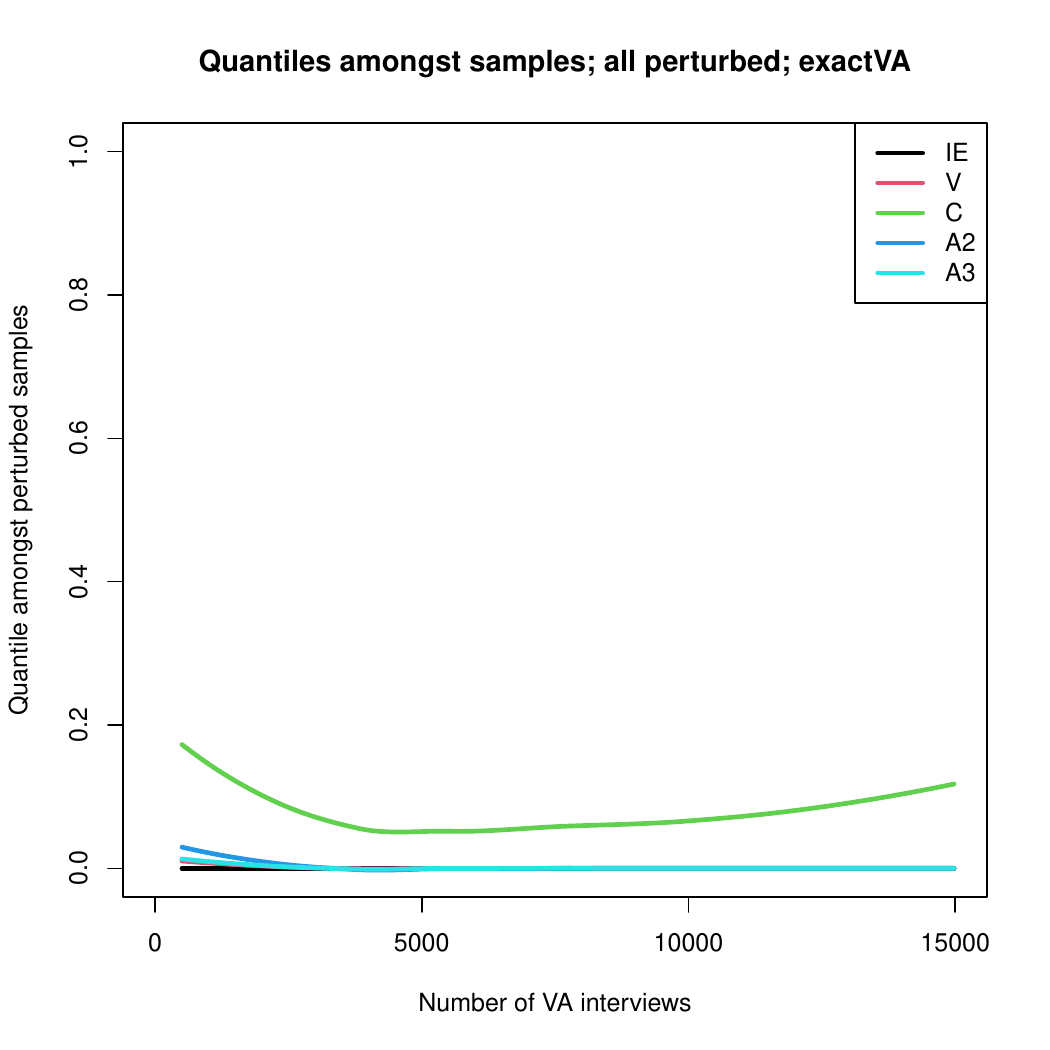}
    \end{subfigure}
    \begin{subfigure}[t]{0.45\textwidth}
    \includegraphics[width=\textwidth]{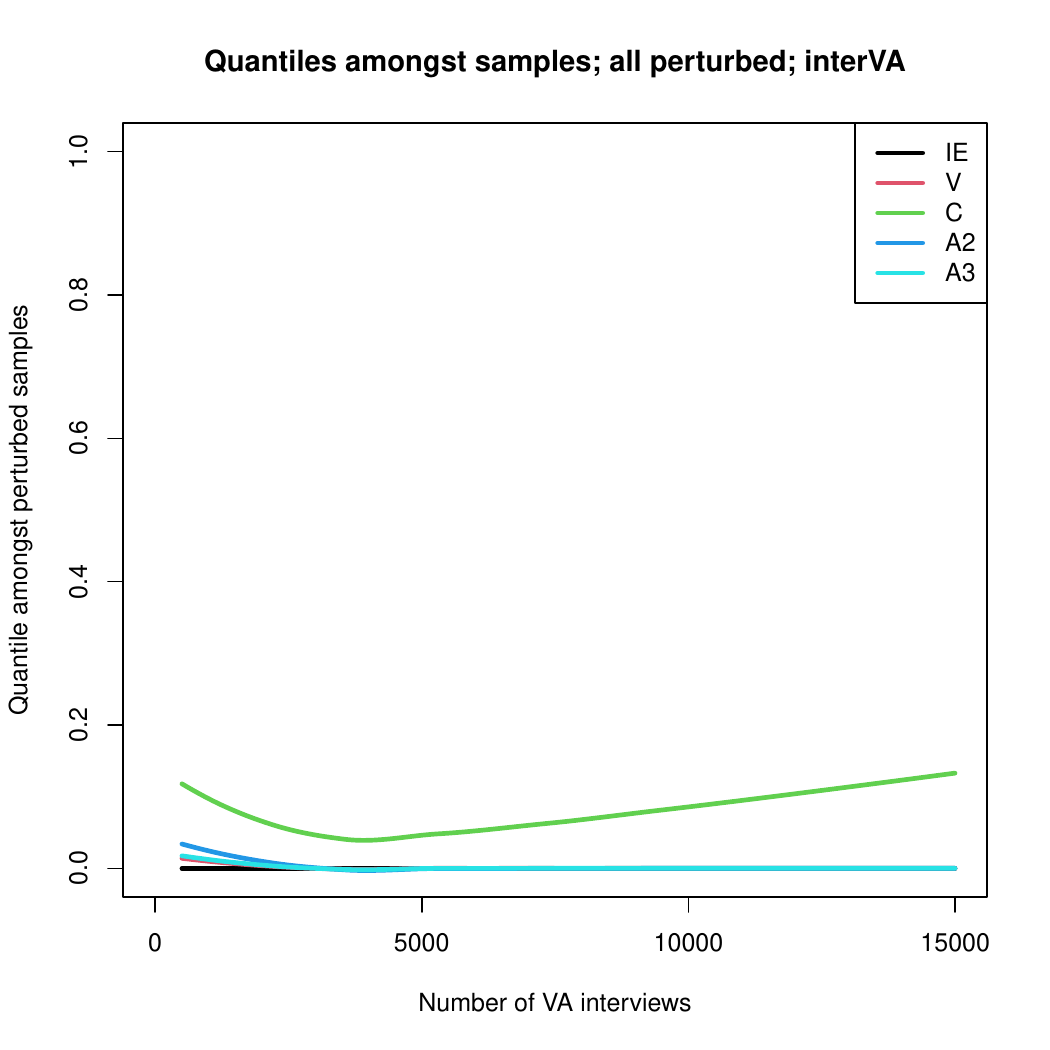}
\end{subfigure}
\end{subfigure}
\caption{Differentiation between true probbase and slightly or globally perturbed probbases ($N(0,(1/100)^2)$ noise added to twenty uniformly-randomly chosen or all elements) using estimated objective functions (Imputation error, IE: $\hat{I}$, viability, V: $\hat{N}$, coherence, C: $\hat{C}$, two-way agreement, A2: $\hat{R}_2$ and three-way agreement, A3: $\hat{R}_3$) on simulations for which the sampling distribution does \textbf{not} satisfy assumption~\ref{asm:cond_indep}. Leftmost plots show computations for oracle VA algorithm; rightmost plot for InterVA algorithm. Lines show estimated mean (LOESS) quantile of test function evaluated on true probbase amongst test function evaluated on perturbed probbases, when using a database with total number of samples given by the value on the x-axis. Pointwise standard errors are less than the widths of the lines. Mean quantiles are essentially indistinguishable from Figure~\ref{fig:pertubation} and Supplementary Figure~\ref{supp_fig:pertubation_moderate}}
\label{supp_fig:pertubation_noci}
\end{figure}

\begin{figure}[H]
\centering
    \begin{subfigure}[t]{0.45\textwidth}
    \includegraphics[width=\textwidth]{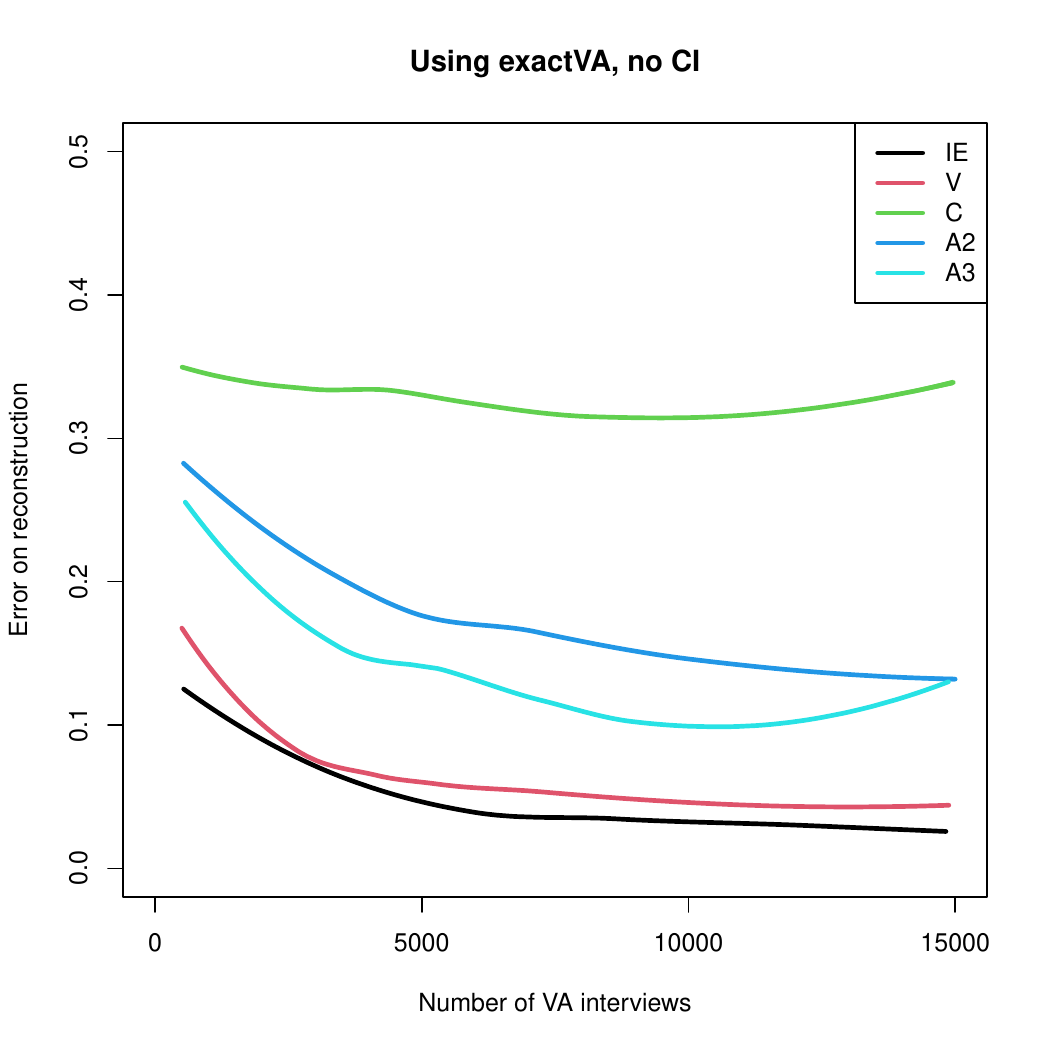}
    \end{subfigure}
    \begin{subfigure}[t]{0.45\textwidth}
    \includegraphics[width=\textwidth]{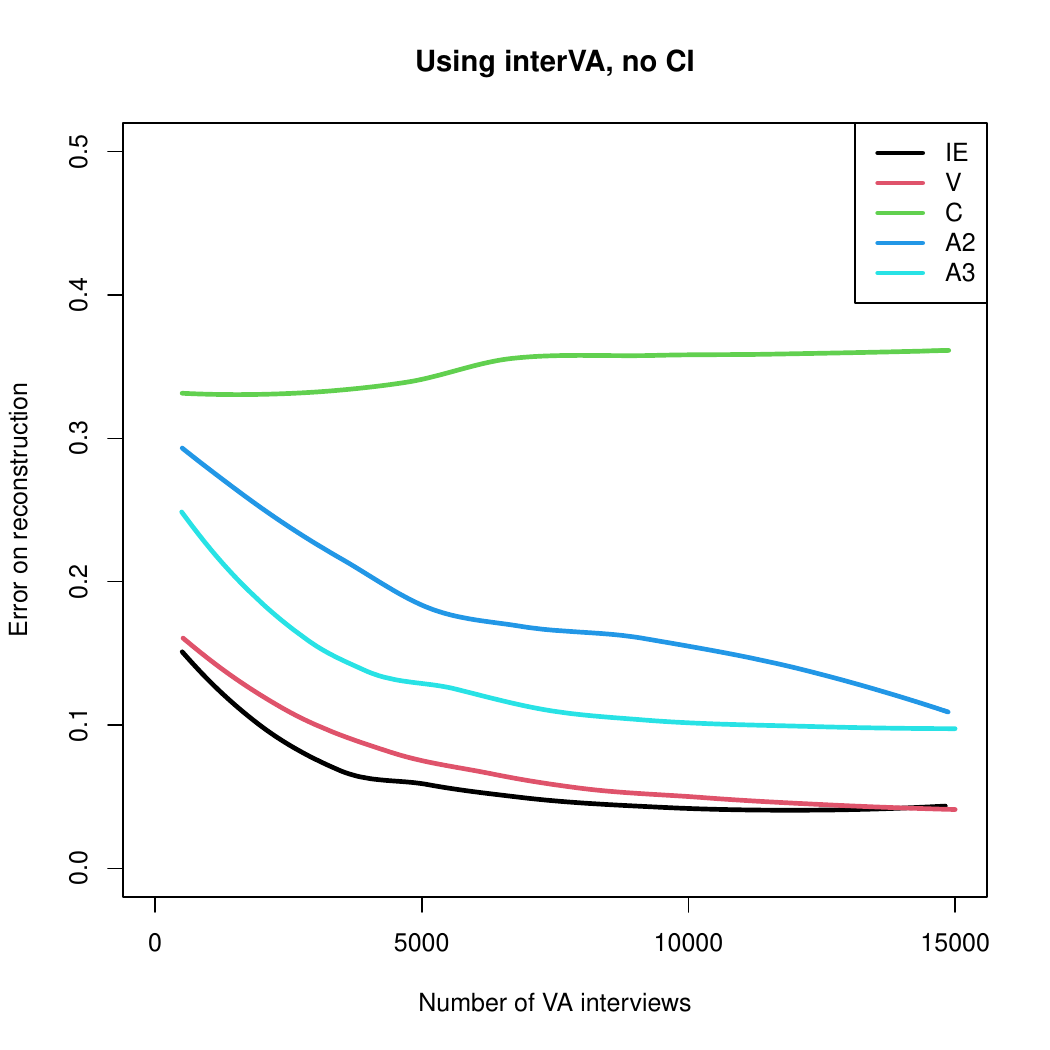}
\end{subfigure}
\caption{Reconstruction of 2-20 missing probbase values by minimisation of estimated objective functions (Imputation error, IE: $\hat{I}$, viability, V: $\hat{N}$, coherence, C: $\hat{C}$, two-way agreement, A2: $\hat{R}_2$ and three-way agreement, A3: $\hat{R}_3$) on simulations for which the sampling distribution does \textbf{not} satisfy assumption~\ref{asm:cond_indep}. The leftmost plot shows computations for an oracle VA algorithm; the rightmost plot for the InterVA algorithm. Lines show mean absolute difference between recovered values and true values when using a database with total number of samples given by the value on the x-axis. Pointwise standard errors are less than the widths of the lines. Mean absolute errors are essentially indistinguishable from Figure~\ref{fig:reconstruction}.}
\label{supp_fig:reconstruction_noci}
\end{figure}

\begin{figure}[H]
\centering
    \begin{subfigure}[t]{0.45\textwidth}
    \includegraphics[width=\textwidth]{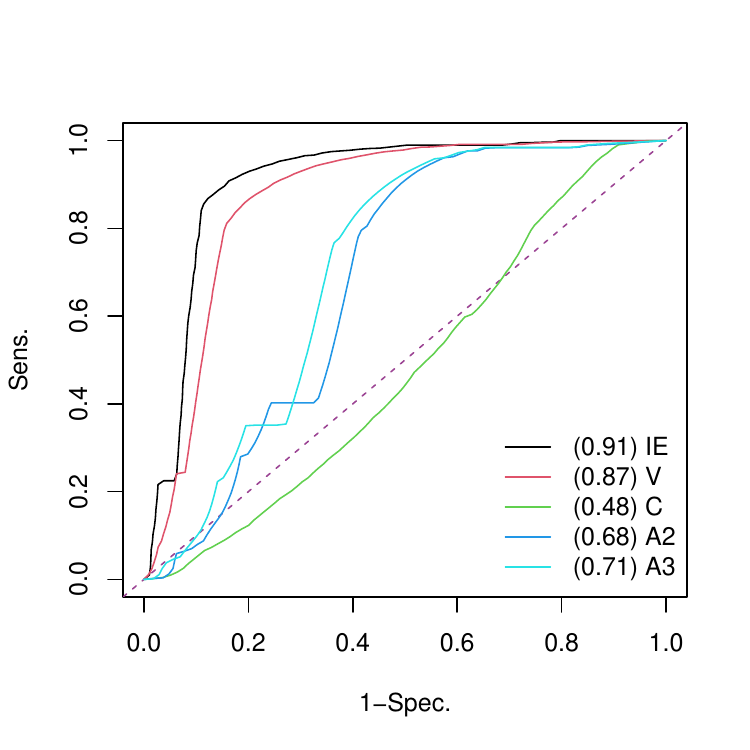}
    \end{subfigure}
    \begin{subfigure}[t]{0.45\textwidth}
    \includegraphics[width=\textwidth]{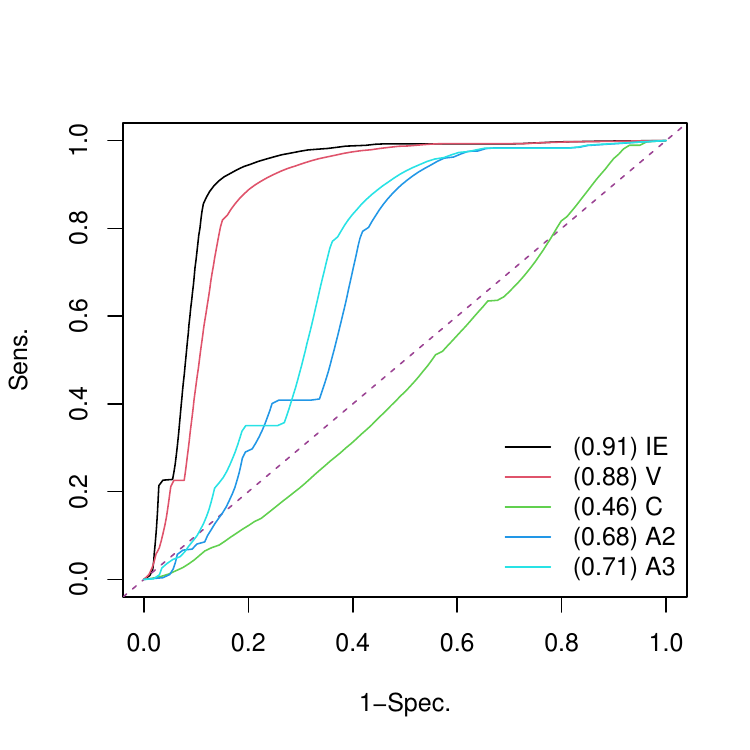}
\end{subfigure}
\caption{Search for probbase elements perturbed by at least 10\% (a randomly-chosen 20\% of elements) on simulations for which the sampling distribution does \textbf{not} satisfy assumption~\ref{asm:cond_indep}. Curves show ROC plots using test functions (Imputation error, IE: $\hat{I}$, viability, V: $\hat{N}$, coherence, C: $\hat{C}$, two-way agreement, A2: $\hat{R}_2$ and three-way agreement, A3: $\hat{R}_3$). The leftmost plot shows computations for an oracle VA algorithm; the rightmost plot for the InterVA algorithm. Bracked values in legend show AUROC. ROC curves are essentially indistinguishable from those in Figure~\ref{fig:search}.}
\label{supp_fig:search_noci}
\end{figure}

\clearpage

\bibliography{va}

\end{document}